\documentstyle[sprocl,psfig,epsf,twoside]{article}

\def\vp{{\bf p}}

\def\al{\alpha}
\def\bea{\begin{eqnarray}}
\def\eea{\end{eqnarray}}

\newcommand{\ab}{{\bar a}}
\newcommand{\bb}{{\bar b}}
\newcommand{\half} {{\textstyle {1\over 2}}}
\newcommand{\third}{{\textstyle {1\over 3}}}
\newcommand{\twothirds}{{\textstyle {2\over 3}}}

\newcommand{\eighth}{{\textstyle {1\over 8}}}
\newcommand{\quarter}{{\textstyle {1\over 4}}}
\newcommand{\e}{ {\rm e} }

\def\bp{{\bf p}}

\let\al=\alpha
\let\be=\beta
\let\ga=\gamma

\let\de=\delta
\let\De=\Delta

\let\eps=\epsilon

\let\La=\Lambda

\let\th=\theta

\let\<=\langle
\let\>=\rangle

\def\dsp{\displaystyle}

\let\ad=\dagger

\newcommand{\psib}{{\bar\psi}}
\newcommand{\psid}{{\psi^\ad}}

\newcommand{\da}{{\dot a}}

\newcommand{\dc}{{\dot c}}

\newcommand{\bk}{{\bf k}}

\newcommand{\xiy}{\xi^y}
\newcommand{\xiz}{\xi^z}

\newcommand{\cc}{ {\rm c.c.} }

\newcommand{\beq}{\begin{equation}}
\newcommand{\eeq}{\end{equation}}

\newcommand{\FF} {{\cal F}}

\def\slash#1{#1\!\!\!/}
\newcommand{\muslash}{\slash{\mu}}

\newcommand{\pslash}{\slash{p}}

\begin{document}


\textheight=6.98truein   
\thispagestyle{plain}
\vspace*{0.85in}
\centerline{\LARGE\bf CHAPTER 35}  

\vspace*{0.21in} 
\noindent \rule[0.5pt]{4.5in}{2pt} \\  
\noindent \rule[8.5pt]{4.5in}{0.5pt}  
\vspace*{0.25in}

\begin{center}
\huge\bf   The Condensed Matter Physics of QCD
\end{center}

\vspace{2cm}

\begin{center}
{\Large\bf  K.~Rajagopal}
\vspace{0.7cm}
\end{center}

\begin{center}
{\Large\bf  F.~Wilczek}
\end{center}

\newpage
\pagestyle{myheadings}  
\markboth{\small \em Handbook of QCD / Volume 3}{\small \em 
The Condensed Matter Physics of QCD}
\title{THE CONDENSED MATTER PHYSICS OF QCD}

\author{KRISHNA RAJAGOPAL AND FRANK WILCZEK}

\address{Center for Theoretical Physics, Massachusetts Institute of 
Technology\\
Cambridge, MA USA 02139}


\maketitle\abstracts{Important progress 
in understanding the behavior of hadronic matter at 
high density has been achieved recently, by adapting
the techniques of condensed matter theory.  At asymptotic
densities, the combination of asymptotic freedom and BCS
theory make a rigorous analysis possible.
New phases of matter with remarkable properties are
predicted.  They provide a theoretical laboratory
within which chiral symmetry breaking and confinement
can be studied at weak coupling.  They may also play
a role in the description of neutron star interiors.  
We discuss the phase diagram of QCD 
as a function of temperature and density,
and close with a look
at possible astrophysical signatures.}

\vspace{1cm}

\tableofcontents

\newpage
  
\section{Introduction and Summary}

In this article we shall discuss the behavior of QCD at high density. We
shall mainly be concerned, more precisely, with the regime of very large
baryon
number density and relatively low temperature.  The appeal of the subject
should be obvious.
It provides the answer to a child-like question: What happens to matter, as
you squeeze it harder and harder?   Moreover, this regime may be realized
in  neutron
star interiors and in the violent events associated with collapse of
massive stars or  collisions of neutron stars, so it is important for
astrophysics.   Finally, we may
hope to gain insight into QCD at moderate or low density --- the realm of
ordinary matter and of terrestrial experiments --- by extrapolating from
above.

What is less obvious, but turns out (with certain qualifications) to be
true, is that the high density regime of QCD is tractable theoretically.
Heuristic
arguments to this effect, invoking asymptotic freedom,\cite{AF} 
go back to the
earliest days of the subject.\cite{CollinsPerry}   
High density brings in a large energy
scale, the chemical
potential $\mu$, and one might hope that the relevant coupling to describe
the dynamics is $g(\mu)$.  This becomes small as $\mu \rightarrow \infty$,
and
suggests the possibility of a perturbative treatment.  This naive
expectation, however, does not stand up to critical scrutiny.  As we shall
discuss at length below,
perturbation theory around the naive ground state  (free quark Fermi spheres)
encounters infrared divergences.  Furthermore, the naive perturbative
ground state is
unstable.   Therefore, straightforward perturbative treatment of QCD at high
density fails.

Fortunately, related difficulties have been met and overcome
previously, in the theory of superconductivity.    There we learn that
arbitrarily weak
attractive interactions can change the ground state qualitatively.  In the
true ground state there is an effective mass for photons --- the Meissner
effect --- and
energy gaps for charged excitations.   These phenomena remove potential
infrared divergences, and render the perturbation theory around the true
ground state
regular (nondegenerate).

We can readily adapt  the methods of superconductivity theory to QCD.  It
is also instructive to consider, in this connection, variants of QCD with
different numbers
of colors and flavors, and different spectra of quark masses.  A rich and
highly structured theory emerges, displaying calculable, highly non-trivial
dependence on
all these variables.

The central result of the analysis is the identification of condensates in
diquark channels, analogous to the Cooper pairs of electrons in ordinary
superconductors.
This is the phenomenon of color 
superconductivity.\cite{Barrois,BailinLove,ARW1,RappETC}

Compared to ordinary superconductivity, color superconductivity, though it
appears superficially to be more complex mathematically, is in a profound
sense
simpler and more directly related to fundamentals.  
Ordinary
superconductivity takes place in solids and the accurate
effective interactions are
determined by band structure and other complicated 
effects. 
Furthermore, ordinary
superconductivity in a metal involves electron pairing,
and the fundamental interaction between electrons (the
screened Coulomb
interaction) is repulsive.  
The effective attraction near the Fermi surface that leads to
superconductivity arises in classic superconductors only as a subtle
consequence of retarded
phonon interactions, and in the cuprate superconductors through some mechanism
yet unknown.  In color superconductivity, by contrast, the attractive
interaction can
arise already from the primary, strong, interactions.  
This has two consequences. First, the accurate form of these
interactions can be calculated from first principles, using asymptotic
freedom.  This makes calculations at high enough density robust.
Second, at accessible densities, where the strong
interactions are much stronger than the electromagnetic interactions,
we expect the color superconductors themselves
to be robust in the sense that the ratio of their 
gaps and critical temperatures to the Fermi energy should be quite
large.

In QCD with three colors and three flavors,
we find an improved ground state at high density, based on color
superconductivity, around which weak-coupling perturbation theory is valid.
In particular, all the colored degrees of freedom acquire
gaps. Thus,
the improved ground state differs qualitatively from the naive
one.  

The resulting predictions regarding the low-energy spectrum and
dynamics are
striking.  Color symmetry and chiral symmetry are spontaneously broken.
The spectrum of elementary excitations is quite different from
that found in naive perturbation theory.  Nominally massless quarks and
gluons become massive, new massless collective modes appear, and various
quantum
numbers get modified.    All the elementary excitations carry integral
electric charges.\cite{CFL}
Altogether, one finds an uncanny resemblance between the
properties one computes at asymptotic densities, directly from the
microscopic Lagrangian, and the properties one expects to hold at low
density,  based on the
known phenomenology of hadrons.    In particular, the traditional
``mysteries'' of confinement and chiral symmetry breaking are fully
embodied in a controlled,
fully microscopic, weak-coupling (but nonperturbative!) calculation, that
accurately describes a directly 
physical, intrinsically interesting regime.\cite{CFL,SW1}

Though some of the ideas go back a decade or more,\cite{Barrois,BailinLove} 
the full power of color
superconductivity to provide a rigorous foundation for the investigation of
high density
QCD has only become apparent relatively recently, and the subject is
developing rapidly.  In this survey we shall emphasize what we see as the
most fundamental
ideas that have appeared in the field to date, and attempt to identify some
significant challenges for the future.   Although we shall supply extensive
references we
will not attempt to catalogue all the very latest results, nor snapshot the
developing state of the art in its technical and quantitative aspects.

In Section 2, we shall discuss in detail the high-density behavior of a
slightly idealized version of real-world QCD, in which we imagine there are
three flavors of
massless quarks.  This is the case for which the clearest and most
beautiful results emerge, based on the phenomenon of 
color-flavor locking.
In Section 3, we
shall discuss what changes must be made, in order to take into account the
realistic quark spectrum.   We shall also briefly discuss QCD variants with
unrealistic
numbers of colors and flavors, which shed additional light on the theory.
In Section 4, we shall discuss
in greater depth the theoretical analysis which underlies the preceding
sections, and the
techniques it relies upon.     
In Section 5, we shall synthesize the preceding discussion, to produce a
tentative sketch
of the phase diagram for high-density QCD.    
Finally in Section 6 we shall discuss possible
applications to astrophysics, including both existing results and some
avenues that we believe might reward further investigation.

\section{Color-Flavor Locking}

In this Section we shall analyze the high-density, zero-temperature behavior
of a slight idealization of QCD,
in which the masses of the
$u$, $d$ and $s$ quarks are set to zero, and those of the $c$, $b$
and $t$ quarks to
infinity.  This idealization gives rise to an especially clear and
beautiful form of the theory.  Also, as we shall discuss in Sections 3
and 5, the
analysis applies with only very
minor modifications to an important phase of real-world QCD.

In this Section our discussion
will be broadly conceptual.   The formal and algorithmic underpinnings are
spelled out in more detail in Section 4, and of course in the
primary literature.  In particular, we focus in this section
on a presentation of the physical properties of dense quark
matter in the idealized three-flavor world, deferring discussion
of quantitative calculations of the magnitude of the gap
at the Fermi surface as much as possible to Section 4.

Let us briefly describe the foundational argument justifying
the weak-coupling approximation at high density, 
leaving a more detailed and precise discussion to Section 4.
The relevant degrees of freedom in cold,
dense quark matter are those which involve quarks with momenta
near the Fermi surface. At high density, when
the Fermi momentum is large, the QCD gauge coupling $g(\mu)$ is small.
When a pair of quasi-particles scatter, the typical momentum transfer
is of order $\mu$ and the interaction is therefore weak.
The exception, of course, is the case of scattering by
a small angle. 
In QCD at asymptotically high densities, the long-range 
magnetic gauge interactions
are unscreened in the absence of superconductivity,
which raises the possibility of infrared problems in small
angle scattering. The Meissner effect induced by the superconducting
condensate itself can provide the requisite screening, however,
regulating the putative collinear divergence and 
guaranteeing that the calculation is truly a weak-coupling
calculation, at least at truly asymptotic density.
Quantitative calculations 
support this qualitative conclusion, although it turns out
that 
dynamical screening is more important than the Meissner
effect in the regulation of small angle scattering.

\subsection{Form of the Condensate}

Because of the infinite degeneracy among pairs
of quarks with equal and opposite momenta at the Fermi surface,
an attractive interaction between quarks, even if arbitrarily weak,
renders
the Fermi surface unstable to the formation of 
a condensate of quark Cooper pairs.  Creating a pair
costs no free energy at the Fermi surface, while the attractive
interaction results in a free energy benefit.
The consequence of an attractive interaction between
opposite-momentum modes near the Fermi surface is
therefore the formation of a condensate in the zero temperature
ground state.   We expect those quark quasiparticles
which interact with the condensate to acquire
an energy gap, and we expect a Meissner effect to occur for 
all gauge bosons except those which see the condensate as neutral.

Single gluon exchange, the QCD
analogue of the Coulomb interaction between two quarks, is attractive 
if the quarks are antisymmetric in color and the pair is 
therefore in the 
color ${\bf\bar 3}$ channel. At weak-coupling, this interaction
dominates and 
this argument suffices to guarantee condensation
in the color ${\bf\bar 3}$ channel.  
The instanton interaction
is also attractive in the ${\bf \bar 3}$ channel,
which may be relevant at stronger coupling.
At any coupling,
attraction in the ${\bf \bar 3}$ channel
is quite reasonable intuitively,
for in this channel the total color flux is reduced
as one brings the quarks together.

Thus, according to the preceding discussion, we should expect
the formation of a condensate in a color ${\bf \bar 3}$ channel.  
The form of the condensate can be
anticipated by analyzing the renormalization of couplings toward the
Fermi surface starting from realistic
microscopic couplings, as we describe in Section 4.  
This indicates the most likely flavor and spin channels for
condensate formation.  Ultimately, one must compare the energies of
different candidate ground states, constructed along the lines
indicated by BCS, and choose the most favorable.

An analysis of this sort indicates that the true ground state contains
nonzero condensates approximately of the form\cite{CFL}
\begin{equation}
\label{CFLcondensate}
\langle \psi^{a\alpha}_{iL} (\vec p) \psi^{b\beta}_{jL} (-\vec p) \epsilon_{ab}
\rangle ~=~
-\langle \psi (\vec p)^{a\alpha}_{iR} \psi^{b\beta}_{jR}
(-\vec p)\epsilon_{ab}\rangle ~=~
\Delta(p^2) \epsilon^{\alpha\beta A}\epsilon_{ijA}\ .
\end{equation}
We have explicitly displayed
color ($\al,\be$), flavor ($i,j$),
and spinor ($a,b$) indices. 
The $A$-index is summed and therefore links color and flavor.
We have used a two-component spinor notation;
note that properly the right-helicity
fields should involve dotted spinors.  The important information conveyed
by the
spinors is that the condensation does
not violate rotational invariance.  The relative minus sign between
left-helicity and right-helicity
condensates signifies that the ground state is a scalar, rather than a
pseudoscalar, so
that parity is unbroken.  The
magnitude of the condensate depends on the magnitude of the 3-momentum
$\vec{p}$.  In weak coupling,
it is largest for $|\vec{p}|$ near the nominal (free-particle) Fermi surface.

Many different treatments have shown that a condensate
of the form (\ref{CFLcondensate}) is the dominant condensate
in three-flavor QCD.\cite{CFL,RappETC2,SchaeferPatterns,ShovWij,EHHS}  
The essential physical argument
that favors this pattern is that, by 
leaving the maximal
unbroken symmetry group, this pattern allows
quarks of all three colors and all three flavors to pair.\cite{CFL} 
Less 
symmetric condensates in which not all quarks pair do not  lower
the free energy as much as the ``maximal'' 
choice (\ref{CFLcondensate}).\cite{SchaeferPatterns,EHHS} 

In reality, condensation in the color ${\bf \bar 3}$ channel
(\ref{CFLcondensate}) induces a small but nonzero condensate
in the color $\bf 6$ channel even if the interaction
is repulsive in this channel,\cite{CFL}  because this additional condensation
breaks no further symmetries.\cite{ABR2+1} This means that the
right hand side of (\ref{CFLcondensate}) 
is slightly more complicated and should, in fact, be
written in terms of two gap parameters $\kappa_1$ and $\kappa_2$,
as $\kappa_1(p^2)\delta^\alpha_a \delta^\beta_b +\kappa_2(p^2)\delta^\alpha_b
\delta^\beta_a$. The pure color ${\bf \bar 3}$ 
condensate displayed in (\ref{CFLcondensate})
has $\kappa_2=-\kappa_1$. Using (\ref{CFLcondensate}) is a good approximation
because the induced color $\bf 6$ condensate is much smaller than
the dominant color ${\bf \bar 3}$ condensate mandated by the attraction
in this channel.\cite{CFL,SchaeferPatterns,ShovWij}

We can now explain the  term ``color-flavor locking''.
Writing 
$\epsilon^{\alpha\beta A}\epsilon_{abA}=\delta^\alpha_a\delta^\beta_b-\delta^\alpha_b\delta^\beta_a$, we 
see that the condensates (\ref{CFLcondensate}) involve
Kronecker delta functions that
link color and flavor indices.  These condensates
transform  nontrivially
under separate color and flavor transformations.   Neither color
transformations nor flavor
transformations, separately, are valid symmetries of the ground state.
However, the delta functions {\it do} remain
invariant if we simultaneously rotate both color and flavor.  Thus these
symmetries are locked together.

Pairing which locks two previously unrelated symmetries is
not a new phenomenon.
The condensate for the color-flavor locked state of QCD 
perhaps most directly resembles that
found in the B phase of superfluid helium 3.   In an ordinary nonrelativistic
system, orbital and spin rotations are separate symmetries.
In the B phase, however, a condensate of pairs
of atoms formed at the Fermi surface 
transforms nontrivially under spin and orbital rotations
but is invariant under simultaneous rotations of both: it locks
spin and orbital rotations.  Another analogy 
is to the physics
of chiral symmetry breaking in the QCD vacuum. There, the
formation of a condensate of left-handed quarks and right-handed
antiquarks locks the $SU(3)_L$ and $SU(3)_R$ flavor symmetries.


All the main qualitative properties of the color-flavor locked state are
direct consequences of
(\ref{CFLcondensate}), properly interpreted.  Elucidation
of these consequences
will occupy us for the remainder of the Section.

\subsection{Symmetry Breaking}

An aspect of (\ref{CFLcondensate}) that might appear troubling at first
sight is its lack of gauge invariance.
This actually turns out to be a profound advantage.

There are powerful general arguments that local gauge invariance cannot be
broken.\cite{Elitzur}  
Indeed, local gauge
invariance is really a tautology, stating the redundancy of 
variables.  Yet its ``breaking'' is
central to two of the most successful theories in physics, namely BCS
superconductivity theory and the
standard model of electroweak interactions.  In BCS theory we postulate a
nonzero vacuum expectation
value for the Cooper pair field, which is electrically charged.  In the
electroweak standard model we
postulate a nonzero value for the Higgs field, which violates both the
weak isospin $SU(2)$ and the
hypercharge $U(1)$ gauge symmetries.

In each case, we should interpret the condensate as follows.  We are
working in a gauge theory at weak
coupling.  It is then very convenient to fix a gauge, because after we have
done so --- but not before! --- the
gauge potentials in which we perturb will make only small fluctuations
around zero.  Of course at the end
of any calculation we must restore the gauge symmetry, by averaging over
the gauge fixing parameters.
Only gauge-invariant results will survive this
averaging.  However, in the intermediate steps,
within a fixed gauge, one can capture important correlations that
characterize the ground state by
specifying the existence of nonzero condensates relative to the ambient
gauge.  In superconductivity, the essence of the physics is the correlation
in the fermionic wave function which describes the 
Cooper pairs, and the resulting modification
of the dispersion relations which describe the excitation
spectrum.  In particular, the gap in the spectrum of fermionic excitations
at the Fermi surface is a gauge invariant quantity.
Describing this physics within a fixed gauge
as a condensate which ``breaks'' the gauge symmetry
is a convenient fiction.

In the standard electroweak model one employs a nonzero vacuum expectation
value for a Higgs doublet
field $\langle \phi^a \rangle~=~ v \delta^a_1$, which is not gauge
invariant.  One might be tempted to 
use the magnitude of its absolute square, 
which is gauge invariant, as an
order parameter to characterize
the symmetry breaking. However, $\langle \phi^\dagger \phi \rangle$ never
vanishes, whether or not any
symmetry is broken.  In fact there can be no 
gauge invariant order parameter for the
electroweak phase transition, since it
has long been known 
that one can, by allowing the $SU(2)$ gauge couplings
to become large, go over into a
``confined'' regime, encountering no sharp phase transition along the 
way.\cite{FradkinShenker}
The absence of massless gauge
bosons and of long-range forces is the essence of the Meissner-Anderson-Higgs 
effect --- and it is also the essence of
confinement!

So evidently, when used with care, the notion of spontaneous
gauge symmetry breaking can
be an {\it extremely} 
convenient fiction.  In particular, by forging a connection
with
superconductivity and condensate
formation, it brings the universality class of confinement down to earth,
and makes it accessible to weak coupling methods.
These condensates
need not
break any true (i.e. global) symmetries.
If a global symmetry {\it is} broken, some
combination of the 
condensates themselves is a gauge invariant physical observable,
and not just a convenient fiction.

With this discussion in mind, let us consider the consequences of
(\ref{CFLcondensate}) for symmetries.\cite{CFL} 
(See also Ref.~16
in which this ordering was considered at zero density.)

The exact microscopic symmetries of QCD with three massless flavors are
\begin{equation}
G_{\rm microscopic} ~=~ SU(3)_{\rm color} \times SU(3)_L \times SU(3)_R
\times U(1)_B ~,
\label{Gmicroscopic}
\end{equation}
where the first factor is the (local) vector color gauge symmetry, the
second and third factors are global chiral flavor
symmetries, and the fourth factor is baryon number.  (For
the present, we set the electromagnetic coupling constant to zero,
and therefore neglect the fact that the  $U(1)$ subgroup
of $SU(3)_L \times SU(3)_R$ which describes
electromagnetism is in fact a local symmetry.)

The color-flavor locked phase (\ref{CFLcondensate}) 
features two condensates, one involving left-handed quarks
alone and one involving right-handed quarks alone.  The
former locks $SU(3)_L$ flavor rotations to $SU(3)_{\rm color}$:
the condensate is not symmetric under either alone, 
but is symmetric under simultaneous $SU(3)_{L+{\rm color}}$
rotations.  Similarly, the condensate involving right-handed
quarks alone locks  $SU(3)_R$ flavor rotations to $SU(3)_{\rm color}$.
As a consequence, 
of all the symmetries in
$G_{\rm microscopic}$, only the subgroup
\begin{equation}
G_{\rm CFL} ~=~ SU(3)_{{\rm color}+L+R} \times Z_2
\end{equation}
leaves the correlated ground state (\ref{CFLcondensate}) invariant.
The color and chiral flavor symmetries are broken, by color-flavor locking,
down to the (global) vector ``diagonal'' symmetry, that
makes equal transformations in all three sectors -- color, left-handed
flavor, and right-handed flavor.   Baryon number symmetry is
broken down to a discrete $Z_2$ symmetry under which all quark
fields are multiplied by $-1$.  Even though the condensates
in (\ref{CFLcondensate}) do not {\it appear} to lock $SU(3)_L$
to $SU(3)_R$, they manage to do so by locking both to $SU(3)_{\rm color}$.
Color-flavor locking, therefore, provides a mechanism
by which chiral symmetry can be broken.  


The spontaneous breaking of chiral symmetry is a familiar phenomenon in
zero-density QCD.  Here, at high density, it occurs by a rather different
mechanism.   In zero-density QCD  chiral symmetry breaking is due to
condensation of left-handed quarks with right-handed
antiquarks.   The pairing of opposite helicities, of course, breaks chiral
symmetry.  Here we have only pairing of left-handed quarks with
left-handed quarks, and right-handed quarks with right-handed quarks.
Nevertheless chiral symmetry is broken indirectly, as we have described.  

Now we must mention explicitly a general formal consequence of symmetry
breaking, that we glided over
earlier.  That is, the form of the condensate in (\ref{CFLcondensate}) is not
unique.  There are states with equally good energy where the
correlated ground state is subjected to the action of any transformation in
$G_{\rm microscopic}$.
This action will in general (except for
elements of
$G_{\rm CFL}$) alter the form of the condensate.
So there is a manifold of distinct vacua associated with the broken
symmetries.   By interpolating among these vacua, with
small space-time gradients, one produces low-energy excitations around a
fixed, reference ground state such as the one specified by
(\ref{CFLcondensate}).

The significance of such states is familiar from other contexts.  When the
broken symmetry is a global symmetry, they lead to
Nambu-Goldstone particles.  When the broken symmetry is a local gauge
symmetry, the Meissner-Anderson-Higgs mechanism is triggered, and the
would-be Nambu-Goldstone particles
become the longitudinal parts of massive gauge fields.

Here the local color symmetry has been broken completely.  Thus all the
gluons acquire mass.   This result has a major positive
consequence for the logical status of our analysis.  It removes the
possibility of infrared divergences associated with gluon exchange.
Similarly, the existence of an energy gap for 
all the quarks removes the other
potential source of infrared divergences, from integration over
low-energy excitations around the Fermi surface.   Altogether, then,
weak coupling
perturbation theory {\it around the correct, condensed ground
state\/} is free of the difficulties that appeared around the naive ground
state.


The spontaneous violation of baryon number is perhaps
less familiar to a particle physicist. On first hearing,
one might think this is a  dramatic or even catastrophic prediction, since we
know that baryonic matter in the Universe is stable over very long periods
of time.    That is obviously too naive
an interpretation, however, since in the
theory of ordinary superconductors we deal with electron pairing without
worrying over the violation of lepton number, and in the
theory of helium superfluids we deal with condensates of atoms or diatoms,
which formally violate both baryon and lepton number.
Finally, and (as we shall see below) most directly
to the point, ordinary nuclear matter
is a superfluid in which nucleon-nucleon pairing violates baryon
number symmetry.

In all these cases, if we are dealing with a finite sample of the
superconductor or superfluid, there is no true violation of the
conservation laws.  Indeed, we may draw a surface surrounding the sample,
and apply Gauss' law to the equation of current conservation
(modified, within the sample, to include the condensate) in the usual way
to see that changes in the bulk quantum numbers are
accompanied by appropriate compensating fluxes.   The correct
interpretation of the formal ``violation'' of these symmetries is that there
can be large fluctuations 
and easy transport of the corresponding quantum
numbers within the sample.   These are precisely the
phenomena of superconductivity and superfluidity.  The mathematical
connection between broken symmetry and super-transport is
quite direct: As we have discussed, the symmetry breaking
order parameter can point in any one of a manifold of ``directions'';
the supercurrents are carried by spatial variation 
in the ``direction'' of the condensate; that is, the phenomenon of
superfluidity is a direct manifestation of the Nambu-Goldstone
mode associated with the spontaneous breaking of a global
symmetry; similarly, superconductivity is a direct manifestation
of the mode which becomes the longitudinal component of
a massive gauge boson and is thus a direct manifestation
of the breaking of a gauge symmetry.

As already mentioned, the standard Higgs mechanism in the electroweak
sector of the standard model has no gauge-invariant order
parameter.  With color-flavor locking the situation is more fortunate,
because global as well as gauge symmetries are broken.  Physically,
this implies that there are sharp differences between the color-flavor
locked phase and the quark-gluon plasma phase (in which
all symmetries of the QCD lagrangian are unbroken),  
so that any passage between them
must be marked by one or more phase transitions.   In fact, it is a simple
matter to abstract gauge invariant order parameters, which
have a strict meaning valid at any coupling, from our primary, gauge
variant condensate at weak coupling.   For instance, to form a
gauge invariant order parameter capturing chiral symmetry breaking we may
take the product of the left-handed version of
(\ref{CFLcondensate}) with the right-handed version and saturate the color
indices, to obtain
\begin{equation}
\label{chiralOP}
\langle \psi^\alpha_{Li}\psi^\beta_{Lj}{\bar \psi}^k_{R\alpha} 
{\bar \psi}^l_{R\beta}
\rangle \sim
\langle \psi^\alpha_{Li}\psi^\beta_{Lj}\rangle \langle 
{\bar \psi}^k_{R\alpha} {\bar \psi}^l_{R\beta} \rangle \sim
\Delta^2 \epsilon_{ijm}\epsilon^{klm}\ .
\end{equation}
Likewise we can take a product of three copies of the condensate and
saturate the color indices, to obtain a gauge invariant order
parameter for the baryon-number violating superfluid order parameter.
These secondary order parameters will survive gauge unfixing
unscathed.  Unlike the primary condensate, from which they were
derived, they are more than convenient fictions.

If we turn on a common mass for all the quarks, the chiral $SU(3)_L\times
SU(3)_R$ flavor symmetry of $G_{\rm microscopic}$ will be
reduced to the diagonal $SU(3)_{L+R}$.   If we turn on unequal masses, the
symmetry will be even less.  In any case, however, the $U(1)$
of baryon number a good microscopic symmetry, and  the corresponding
six-quark order parameter remains a strict signature of the
color-flavor locked phase, distinguishing it both from the 
quark-gluon plasma phase, and
from some other states of quark matter we shall encounter
in Section 3.

As it stands the order parameter (\ref{chiralOP}) is not quite the usual one,
but roughly speaking its square.  It leaves invariant an
additional $Z_2$, under which only the left-handed quark fields change sign.
Actually this $Z_2$ is not a legitimate symmetry of the full
theory, but suffers from an anomaly.   So we might expect that the usual
chiral order parameter is induced by the anomalous
interactions that violate the axial baryon number symmetry of the classical
Lagrangian.  To put this another way, because axial baryon number
is not a symmetry of QCD, 
once chiral symmetry is broken
by color-flavor locking
there is no symmetry argument
precluding the existence of an ordinary chiral condensate. 
Indeed,
instanton effects do induce a nonzero $\langle \bar \psi_R\psi_L \rangle$
because the instanton-induced interaction is a six-fermion operator which 
can be written as a product of $\bar\psi_R\psi_L$ and the
operator in (\ref{chiralOP}) which already has a nonzero
expectation value,\cite{CFL} 
but this turns out to be a small effect.\cite{RappETC2,SchaeferPatterns}

At weak coupling, we can be more specific about these matters.  The most
important interactions near the Fermi surface, quantitatively,
arise from gluon exchange.   These are responsible for the primary
condensation.  The instanton interaction is much less important
asymptotically because the gauge fields which make
up the instantons themselves are screened, the effects of instantons
are intrinsically smaller for more energetic
quarks, and because the instanton-induced interaction
involves six fermion fields, and hence (one can
show) becomes irrelevant upon renormalization toward the Fermi surface.
The instanton interaction is qualitatively important, however,
because it represents the leading contribution to axial baryon number
violation.  It is only such $U(1)_A$ violating interactions that
remove the degeneracy among states with different relative phases between
the left- and right-handed condensates in
(\ref{CFLcondensate}).   In the absence of intrinsic $U(1)_A$ breaking, the
spontaneous violation of this symmetry in the color-flavor
locked phase would be accompanied by the existence of a pseudoscalar
$SU(3)_{{\rm color}+L+R}$ singlet Nambu-Goldstone bosons.   Since the
intrinsic violation of this symmetry is parametrically small, the
corresponding boson will not be strictly massless, but only very light.
Quantum fluctuations in this light $\eta'$-field, 
among other things, will keep the
conventional chiral symmetry breaking order parameter small
compared to (\ref{chiralOP}) at high density.

\subsection{Elementary Excitations}

The physics of 
the excitations in the 
CFL phase has been the focus of much recent 
work.\cite{CFL,SW1,ABR2+1,SW2,RappETC2,Zahed,effectiveCFL,gapless,SchaeferPatterns,ShovWij,EHHS,SonStephMesons,RWZ,HongLeeMin,ManuelTytgat,RSWZ,Zarembo,BBSMeson,RischkeMeissner,HongEMMass,SchaeferKCond,Nowak,MSW_CFL,ManuelTytgat2,CasalbuoniGattoNardulli,HongHongPark}
There are three sorts of elementary excitations.  They are the modes
produced directly by the fundamental quark and gluon fields, and
the collective modes associated with spontaneous symmetry breaking.   These
modes can be classified under the unbroken
$SU(3)\times Z_2$ symmetry, and the unbroken rotation and parity
symmetries.

The quark fields of course produce spin 1/2 fermions.  Some of these are
true long-lived quasiparticles, since there are no lighter states
of half-integer spin that they might decay into.  With the conventions we
have been using, as expressed in (\ref{CFLcondensate}), the
quark fields are triplets and antitriplets under color and flavor,
respectively.  Thus they decompose into an octet and a singlet under the
diagonal $SU(3)_{{\rm color}+L+R}$.  
There is an energy gap for production of pairs
above the ground state.  More precisely, there are two gaps: a
smaller one for the octet, and a larger one for the singlet.\cite{CFL}
The dispersion 
relations describing these fermionic quasiparticle excitations in
the CFL phase have been described in some 
detail.\cite{ABR2+1,gapless,CasalbuoniGattoNardulli}


The gluon fields produce an $SU(3)_{{\rm color}+L+R}$ 
octet of spin 1 bosons.  As
previously mentioned, they acquire a common mass by the
Meissner-Anderson-Higgs mechanism. The quantitative expressions
for the 
masses of these vector
mesons which have been computed at weak coupling~\cite{SonStephMesons,TwoFlavorMeissner1,RischkeMeissner,CasalbuoniGattoNardulli} and
in an instanton-liquid model.\cite{TwoFlavorMeissner2}

The fermionic excitations have a gap; the vector mesons
have mass; but, the Nambu-Goldstone bosons are massless.
These bosons form a pseudoscalar octet associated with chiral
symmetry breaking, and a scalar singlet associated with baryon number
superfluidity.  The octet, but not the singlet, will be lifted from
zero mass if the quarks are massive.  Finally there is the parametrically
light, but never strictly massless, pseudoscalar singlet associated
with $U(1)_A$ breaking.

The Nambu-Goldstone bosons arising from chiral
symmetry breaking in the CFL phase
are Fermi surface excitations in which the orientation
of the left-handed and right-handed diquark condensates oscillate
out of phase in flavor space.
The effective field theory describing
these oscillations has been 
constructed.\cite{effectiveCFL,SonStephMesons,Zarembo}  
Up to two derivatives, it is given by
\begin{equation}\label{Leff}
{\cal L}_{\rm eff}=\frac{f_\pi^2}{4}{\rm Tr}\left(\partial_0\Sigma
\partial_0\Sigma^\dagger + v_\pi^2 \partial_i\Sigma
\partial^i\Sigma^\dagger \right) -c \left( {\rm det}M\,
{\rm Tr}(M^{-1}\Sigma)+ {\rm h.c.}\right)\ .
\end{equation}
The  Nambu-Goldstone boson field matrix $\Sigma$ is a color
singlet and transforms under $SU(3)_L\times SU(3)_R$ as 
$\Sigma\rightarrow U_L \Sigma U^\dagger_R$ as usual.
$M={\rm diag}(m_u,m_d,m_s)$ is the quark mass matrix.
The construction of $\Sigma$ from rotations of the CFL condensates
can be found in Refs.~19,21:
one first finds the 19 putative Nambu-Goldstone bosons
expected when $G_{\rm microscopic}$ is broken to $G_{\rm CFL}$;
one then identifies 8 of these which become the longitudinal
parts of massive vector bosons; the remaining ten
are the octet described by (\ref{Leff}), the singlet $\eta'$ and
the superfluid mode.  See Refs.~19,21
for the singlet terms in the effective Lagrangian.
The higher derivative
terms in the effective Lagrangian have also been analyzed.\cite{Zarembo}
It has also been suggested that once
higher derivative interactions are introduced, the effective theory 
may support Skyrmions, in which the Nambu-Goldstone
boson field configuration has nonzero winding number.\cite{Zahed}
These solitons have energies comparable to the gap,\cite{HongHongPark} and
are an alternative description of the gapped
fermionic excitations in the CFL phase, in the same
sense that baryons can alternatively be described as Skyrmions
of the vacuum pion field.

The masses of the pseudoscalar mesons which are the 
pseudo-Nambu-Goldstone bosons associated with chiral symmetry
breaking 
can be obtained from ${\cal L_{\rm eff}}$
of (\ref{Leff}).\cite{SonStephMesons} For example,
\begin{equation}
m^2_{\rm \pi^\pm}=\frac{2c}{f_\pi^2}m_s(m_u+m_d)\ ,
\ \ \ \ \ \ \ \ \ \ 
m^2_{\rm K^\pm}=\frac{2c}{f_\pi^2}m_d(m_u+m_s)\ .
\end{equation}
Thus, the kaon is lighter than
the pion, by a factor of $m_d/(m_u+m_d)$.\cite{SonStephMesons} 
Note that the effective Lagrangian is quadratic in $M$.
This arises because $L_{\rm eff}$ respects the $Z_2$ symmetry
under which only the left-handed quarks change sign.\cite{CFL}  As 
we discussed in the previous section, this is almost a symmetry
of the CFL phase: it would be a symmetry if instanton effects
could be neglected.\cite{CFL} However, instanton effects
generate a nonzero, but small, ordinary $\langle\bar\psi_R \psi_L\rangle$ 
condensate, which breaks the $Z_2$,\cite{CFL,RappETC2,SchaeferPatterns}
and results in a contribution to the meson $m^2$ which
is linear in $M$ and which may be 
numerically significant.\cite{ManuelTytgat}
The induced $\langle\bar\psi_R \psi_L\rangle$ 
was significantly overestimated in 
Refs.~\cite{RappETC2,SchaeferPatterns}, however.\cite{SchaeferKCond}
Making the already small $\langle\bar\psi_R \psi_L\rangle$ even smaller
will significantly reduce the $Z_2$-violating contributions
to the meson masses.


If we were describing pions in vacuum, or pions in nuclear matter,
the only way to obtain the coefficients in the effective
theory would be to measure them in an experiment or, if
possible, to calculate them on the lattice.  Indeed in any
theory with strong interactions, the purpose of writing
an effective theory for the low energy degrees of freedom
is to express the predictions for many low energy processes
in terms of a few parameters, which must be obtained from
experiment.  In
the color-flavor locked phase, however, 
the full theory is weakly coupled 
at asymptotically high densities. In this regime, therefore, 
the coefficients
$f_\pi^2$, $v_\pi^2$ and $c$ are calculable from
first principles using weak coupling methods!
Up to possible logarithmic corrections, the result 
is\cite{SonStephMesons,RWZ,HongLeeMin,ManuelTytgat,RSWZ,BBSMeson,ManuelTytgat2,CasalbuoniGattoNardulli}
\begin{equation}
f_\pi^2=\frac{21-8\log 2}{36\pi^2}\mu^2\ ,\ \ \ \ 
v_\pi^2=\frac{1}{3}\ ,\ \ \ \ c=\frac{3\Delta^2}{2\pi^2}\ .
\end{equation}
The electromagnetic\cite{HongEMMass,ManuelTytgat2} and
nonzero temperature\cite{ManuelTytgat2} corrections to these
quantities have also been calculated.  

Quantitatively, (see Section 4 for a discussion of estimates
of $\Delta$) we estimate that the lightest pseudoscalar meson,
the kaon, has a mass in the range of 5 to 20 MeV at $\mu=400$ MeV,
and becomes lighter still at higher densities.
There are two reasons why the Nambu-Goldstone bosons are so
much lighter in the CFL phase than in the vacuum. First,
their mass$^2$ is proportional to $m_{\rm quark}^2$
rather than to $m_{\rm quark}$, as at zero density.
In addition, there is a further suppression by a factor
of $\Delta/\mu$, which arises because
the Nambu-Goldstone bosons are collective excitations
of the condensates formed from particle-particle
and hole-hole pairs near the Fermi surface, whereas
the quark mass term connects particles with 
antiparticles, far from the Fermi surface.\cite{HongLeeMin}

In QCD with unequal quark masses, at very high
densities the CFL phase is much as we have described it,
except that the gaps associated with $\langle us \rangle$,
$\langle ds \rangle$ and  $\langle ud \rangle$ pairing 
will differ slightly,
the Fermi momenta for the different quark flavors
may differ slightly,\cite{RajagopalWilczekNeutrality}
and there 
may therefore be a small electron chemical potential $\mu_e$.
(We discuss lower densities, where the differences between Fermi
momenta become comparable to $\Delta$, in Section 3.)
Because the strange quarks are less
numerous due to their greater mass, it would seem that the requirement of 
overall charge neutrality can only be satisfied if $\mu_e>m_e$
and there is a nonzero density of electrons.  Sch\"afer
has recently noted, however, that if $\mu_e>m_{K^-}$, charge
neutrality will be achieved by the formation of a condensate
of $K^-$ bosons, rather than by the formation of a small Fermi sphere
of electrons.\cite{SchaeferKCond}  The CFL condensate rotates
in the $K^-$ direction, relative to the direction favored by
the quark mass term in the effective Lagrangian.  This costs
energy, but because it introduces a negative electric charge
it is free-energetically favored, if $\mu_e>m_{K^-}$.

The formation of a kaon condensate in the CFL phase 
changes none of its symmetries. In this sense, it is
a less dramatic effect than the formation of a kaon condensate
in nuclear matter made of neutrons and protons 
only:~\cite{KaonCondensation} there,
kaon condensation breaks $U(1)_S$. This symmetry is already
broken in the CFL phase.  Kaon condensation in the CFL
phase is more akin to kaon condensation in hypernuclear matter
made up of equal measures of all the octet baryons, 
in which $U(1)_S$ is already
broken by hyperon-hyperon pairing.  We shall elaborate
much further on this connection in Section 2.5.

\subsection{The Modification of Electromagnetism}

It is physically significant, and proves extremely instructive, to consider
the effect of color-flavor locking on the electromagnetic
properties of high-density hadronic matter.

To do this, we 
consider coupling in the appropriate additional 
$U(1)_{\rm EM}$ gauge
field $A_\mu$, representing the photon.   This couples to
$u,d,s$ quarks with strength ${2\over 3}e, -{1\over
3}e,-{1\over 3}e$, respectively.  
Evidently this $U(1)_{\rm EM}$ symmetry is
broken by the condensate (\ref{CFLcondensate}), through the terms pairing
differently-charged quarks.   Were this the complete story,
the color-flavor locked phase would be an electromagnetic
superconductor.   The truth is far different, however.

The situation is
analogous to what occurs in the electroweak sector of the standard model.
There, the Higgs field condensate breaks both the original weak
$SU(2)$ and the hypercharge $U(1)$.  However, one linear combination of these
symmetries leaves the condensate invariant,  and remains a
valid gauge symmetry of the ground state.  Indeed, this is how we identify
electromagnetism within the standard model.

Here we must similarly consider the possibility of mixing broken color
$SU(3)$ and (original) electromagnetic $U(1)_{\rm EM}$ generators to find
a valid residual symmetry.  
Indeed, we should expect this to occur, by the following argument.
In QCD with three flavors, 
$U(1)_{\rm EM}$ is a subgroup of $SU(3)_{L+R}$. When
we neglected electromagnetism, we found that in
the color-flavor locked phase $SU(3)_{L+R}$ is broken
but $SU(3)_{\rm color +L+R}$ is an unbroken global symmetry.
We therefore expect that gauging 
a $U(1)$ subgroup of $SU(3)_{L+R}$ must correspond, in the CFL phase,
to gauging a $U(1)$ subgroup of 
the unbroken $SU(3)_{\rm color +L+R}$.

Once we are alerted to this possibility, it is
not difficult to identify the appropriate combination of
the photon and gluons which remains unbroken.\cite{CFL,ABRflux}   
The unbroken $U(1)$ is generated by 
\begin{equation}
\tilde Q = Q + \eta T_8
\end{equation}
with $\eta=1/\sqrt{3}$. $Q$ is the conventional electromagnetic
charge generator and $T_8$ is associated with one of the gluons.
In the representation of the quarks, 
\begin{equation}
\begin{array}{rcl@{\,\,}ll}
Q &=& &\mbox{diag}(\twothirds,-\third,-\third) &
\mbox{in flavor $u,d,s$ space,} \\[2ex]
T_8 &=& \frac{1}{\sqrt{3}} & \mbox{diag}(-2,1,1) &
\mbox{in color $r,g,b$ space}. \\
\end{array}
\end{equation}
As is conventional, we have taken ${\rm tr}(T_8T_8)=2$.
By construction, the $\tilde Q$-charges of all the Cooper pairs
in the condensate (\ref{CFLcondensate}) are zero. (For example,
with these conventions, red up quarks pair only with green
down or blue strange quarks, and both these pairs have
$\tilde Q=0$ in sum.)
The condensate
is $\tilde Q$-neutral, the $U(1)$ symmetry generated by $\tilde Q$
is unbroken, and the associated $\tilde Q$-photon will remain massless.
To see exactly 
which gauge field remains unbroken, look at the covariant
derivative of the condensate,
\begin{equation}
D_\mu\langle q^\alpha_a q^\beta_b\rangle = 
\Bigl(\partial_\mu + eA_\mu Q + g G^8_\mu T_8\Bigr)
\langle q^\alpha_a q^\beta_b\rangle \,.
\end{equation}
The kinetic term $|D\langle q^\alpha_a q^\beta_b\rangle|^2$
will give a mass to one gauge field
\begin{equation}
A^X_\mu = \frac{-\eta e A_\mu + g G^8_\mu}{\sqrt{ \eta^2 e^2 + g^2}}
= -\sin\al_0 A_\mu + \cos\al_0 G^8_\mu\,,
\end{equation}
while the orthogonal linear combination 
\begin{equation}\label{rot:Aprime}
A^{\tilde Q}_\mu = \frac{g A_\mu + \eta e G^8_\mu}{\sqrt{ \eta^2 e^2 + g^2}}
= \cos\al_0 A_\mu + \sin\al_0 G^8_\mu
\end{equation}
is the $\tilde Q$-photon which remains massless.
That is, $B^{\tilde Q}$ satisfies the ordinary Maxwell
equations while $B^X$ experiences a Meissner effect.
The denominators arise from keeping the gauge field kinetic terms
correctly normalized, and we have defined the 
angle $\al_0$ which specifies the unbroken $U(1)$ via
\begin{equation}\label{rot:alpha0}
\cos\al_0 = \frac{g}{\sqrt{ \eta^2 e^2 + g^2}}\ .
\end{equation}
The mixing angle $\alpha_0$ is the analogue of the Weinberg
angle in electroweak theory, in which the 
presence of the Higgs condensate causes the $A_\mu^Y$ and the third
$SU(2)_W$ gauge boson to mix to form the photon, $A_\mu$, and 
the massive $Z$ boson.   
At accessible densities the gluons are strongly coupled
($g^2/(4\pi) \sim 1$), 
and of course the photons are weakly coupled
($e^2/(4\pi) \approx 1/137$), so $\al_0\simeq \eta e/g$ 
is small.
The ``rotated photon''
consists mostly of the usual photon, with only a small
admixture of the $T_8$ gluon.


Let us now consider the charges of 
all the elementary excitations which we
enumerated in Section 2.3.  For reference, 
the electron couples to $A^{\tilde Q}_\mu$
with charge 
\begin{equation}
\tilde e ~=~ {eg\over \sqrt{\eta^2e^2 + g^2}}~.
\end{equation}
which is less than $e$ because the electron couples only
to the $A_\mu$ component of $A^{\tilde Q}_\mu$.
Now in computing the $\tilde Q$-charge of the quark with color and flavor
indices $\alpha,a$ we must take the appropriate combination
from 
$$\frac{e(-{2\over 3}g, {1\over 3}g,{1\over3}g) + g({2\over 3}e, -{1\over
3}e,-{1\over 3})}{\sqrt{\eta^2 e^2+g^2}}\ .$$   
One readily perceives
that the possible values are $0, \pm \tilde e$.  Thus, in units
of the electron charge, the
quarks carry integer $\tilde Q$-charge!   Quite remarkably, high-density QCD
realizes a mathematically consistent gauge theory version of the old vision
of Han and Nambu: the physical quark excitations have
integer electric charges that depend on an internal color quantum number!

Similarly, the gluons all have $\tilde Q$-charges $0,\pm \tilde e$.
Indeed, they have the $\tilde Q$-charges one would expect for
an octet of massive vector bosons.
The Nambu-Goldstone bosons arising from the breaking
of chiral symmetry, of course, have the same
charge assignments as the familiar $\pi$, $K$ and $\eta$ octet
of pseudoscalars.  The baryon superfluid mode is 
$\tilde Q$-neutral.
In the color-flavor locked phase, we conclude, 
all the elementary excitations
are integrally charged.\footnote{We shall see in Section 3 
that in two-flavor QCD, in
which color-flavor locking does not occur, the color
superconducting condensate which forms also leaves
a $\tilde Q$-photon massless.  The only difference
relative to the CFL phase is that $\eta=-1/2\sqrt{3}$.
(However, the $\tilde Q$-charges
of the excitations are not all integral in this theory.)}  
This is a classic aspect of confinement,
here embodied in a controlled, weak-coupling framework.


It is fun to consider how a chunk of our color-flavor locked material would
look.  If the quarks were truly massless, then so would be some
charged Nambu-Goldstone bosons, and one might expect a rather unusual
``bosonic metal'', in which the low-energy electromagnetic
response is dominated by these modes.  As we mentioned
in Section 2.3,  turning on 
quark masses gives small masses to the Nambu-Goldstone
bosons.  Electromagnetic radiative corrections
further lift the masses of the charged Nambu-Goldstone bosons.
As a result, the color-flavor locked material becomes a transparent
insulator, with no charged excitations at zero
temperature.  Altogether, a chunk of color-flavor locked
material would resemble a diamond: an ordinary light wave incident
upon it would be partially reflected, but some fraction would
be admitted as a $\tilde Q$-light wave; the $\tilde Q$-light
would travel through the transparent ``diamond'', and would 
partially emerge and be partially internally reflected on the far side.
Turning on small but unequal quark masses mars
the diamond somewhat:~\cite{RajagopalWilczekNeutrality} 
maintaining overall electric charge neutrality
in this case will require either a nonzero electron density
or a condensate of charged kaons,\cite{SchaeferKCond} 
and in either case, $\tilde Q$-light 
will have charged excitations off which to scatter.

Although a quantitative calculation of
light reflecting off the facets of a CFL-diamond
has not yet been done, 
the effect of a chunk of color superconducting quark matter (whether
in the CFL phase or in the less symmetric phase in which only
up and down quarks pair)
on a static magnetic field has been
described in complete detail.\cite{ABRflux} 
Some fraction of an externally
applied ordinary magnetic field penetrates the superconductor
in the form of a $\tilde Q$-magnetic field, while some fraction
of the ordinary magnetic field is expelled by the Meissner effect.
The fraction of the field which is expelled depends both
on $\alpha_0$ and on the shape of the chunk 
color superconducting quark matter, but it is small when
$\alpha_0$ is small, as in nature. Most of the flux
is admitted, as $\tilde Q$-flux.  This $\tilde Q$-magnetic field
satisfies Maxwell's equations and is not restricted to flux tubes.

\subsection{Quark-Hadron Continuity}

The universal features of the color-flavor locked state: confinement,
chiral symmetry breaking leaving a vector $SU(3)$ unbroken, and
baryon number superfluidity, are exactly what one expects to
find in nuclear matter in three-flavor QCD.\cite{SW1}  
Perhaps this is not immediately obvious in the case of baryon
number superfluidity, but let us recall that pairing phenomena, which
would go over into neutron 
superfluidity and proton superconductivity in nuclear matter, 
are very well established in ordinary
nuclei.  
In three-flavor QCD,
there are good reasons\cite{JaffeH} 
to think that the pairing interaction in
the flavor singlet dibaryon channel (the so-called $H$-dibaryon 
channel)
would be quite attractive in three-flavor QCD, and support a robust baryon
number superfluidity. Thus, the symmetries of the color-flavor
locked phase are precisely those of nuclear matter in three-flavor
QCD, perhaps better referred to as hypernuclear matter.\cite{SW1}

Furthermore, there is an uncanny resemblance
between the low-lying spectrum computed from first principles
for QCD at asymptotically high density, and what one 
expects to find in hypernuclear matter, in a world with
three degenerate quark flavors.
It is hard to
resist the inference that in this theory,
there need be no phase transistion between nuclear
density and high density.\cite{SW1}  
There need be no sharp boundary between hypernuclear matter,
where 
microscopic caculations are difficult but the convenient degrees of freedom
are ``obviously'' hadrons, and the asymptotic high-density
phase, where weak-coupling (but non-perturbative) calculations are
possible, and the right degrees of freedom are elementary quarks and
gluons, together with the collective Nambu-Goldstone modes.  We call this
quark-hadron continuity.\cite{SW1}  Perhaps the least surprising
aspect of this, by now, is the continuity between the pseudoscalar mesons
at nuclear density and those at asymptotically high densities,
since in both regimes these are present as a result of the
breaking of the same symmetry.
It might seem more   shocking that a quark
can go over continuously into, or ``be'', a baryon, since baryons are
supposed to contain three quarks, but remember that in the
color-flavor locked phase the quarks are immersed in a diquark condensate,
and so a one-quark excitation can pick two quarks up from (or lose
to quarks to) 
the condensate at will.  The difference between one and
three is negotiable.
What about the gluons? 
Within the color-flavor locked phase, similarly, they are quite
literally the physical vector mesons.  They are massive, as we 
have discussed, and have the right quantum numbers.
Thus the original vision of
Yang and Mills -- who proposed non-abelian gauge theory as a model of
$\rho$ mesons -- is here embodied.

Note that the hypothesis of continuity between hypernuclear
and dense quark matter certainly does not preclude 
quantitative change. Far from it.  
The dispersion relation for a fermion --- whether
a quark in the CFL phase or a baryon in the hypernuclear phase ---
is characterized by a gap at the Fermi surface and by a 
gap at zero momentum, i.e. a mass.  
As a function
of increasing density, gaps at the hyperon Fermi surfaces
due to hyperon-hyperon pairing evolve continuously
to become the gaps at the quark Fermi surfaces which
characterize the color-flavor locked phase.\cite{ABR2+1}
During this evolution, the gaps are thought to increase
significantly.  In contrast, the gaps at zero momentum 
decrease dramatically with increasing density as they
evolve from being of order the hyperon masses in hypernuclear
matter to being of order the current quark masses at
asymptotically high densities.   

Note that in order for
quark-hadron continuity to be realized, $U(1)_{EM}$ must
not be broken by hyperon-hyperon pairing.\cite{ABR2+1}
At every point during the evolution of the theory as a function 
of increasing density, there will be an unbroken $U(1)$ and a massless
$\tilde Q$-photon and the excitations will be integer charged.
As the density is increased, however, the definition of 
the $\tilde Q$-photon in terms of the vacuum photon and gluon
fields rotates.  The ordinary photon rotates
to become the $\tilde Q$-photon of the CFL phase, defined in
(\ref{rot:Aprime}).
Turning to the massive vector
bosons, in hypernuclear matter there is both
an octet and a singlet. The singlet must become much
heavier than the octet as a function of increasing density,
since in the low energy description of the 
color-flavor locked phase one finds the octet alone.
We see that if quark-hadron continuity is realized in QCD 
with three degenerate quarks, it requires various
quantititative (but continuous) changes.
What is remarkable is that it
is even possible to imagine watching all the physical
excitations of the theory  
evolving continuously as one dials the density up
and goes from a strongly coupled hadronic world to 
a weak-coupling world of quarks and gluons.

If the quarks are massless, the Nambu-Goldstone bosons are
massless in both hypernuclear and CFL quark matter, and in between.
Once nondegenerate quark masses are introduced, 
however, the evolution of the
Nambu-Goldstone masses as a function of increasing density
becomes more intricate, as the kaon must go from
being heavier than the pion to being lighter.

Note, finally, that the whole story becomes further
complicated once the strange quark is made as heavy
as in nature.\cite{ABR2+1,SW2}  
Although the color-flavor locked phase
is certainly obtained at asymptotically densities, where
quark masses are neglectable, the
nuclear matter phase, made of neutrons and
protons only, is not continuously connectable with 
the color-flavor locked phase.  If quark-hadron continuity
is to be realized in the phase diagram of nature, what
must happen is that, as a function of increasing density,
one first goes from nuclear matter to hypernuclear matter,
with sufficiently high density that all the hyperons have
similar Fermi surfaces.  This first stage must involve
phase transitions, as the symmetries of hypernuclear matter
differ from those of ordinary nuclear matter. Then, as 
the density is increased further, the hypernuclear matter
may evolve continuosly to become CFL quark matter, with
pairing among hyperons becoming CFL pairing among quarks.

We now have a description of the properties
of the CFL phase and its excitations, in which much
can be described quantitatively if the value of the gap $\Delta$
is known.  
We describe estimates of $\Delta$ in Section 4.
First, however, we give a full description of the less symmetric variants
of color superconductivity which arise in QCD with 
$N_f\neq 3$. 
Already, however, in our idealized world (in which we either have three
degenerate quarks or such high densities that the quark mass
differences can be neglected) let us pause to marvel at our
theoretical good fortune. 
The color-flavor locked phase is
a concrete realization of the idea of complementarity:
the same phase of a gauge theory can be 
described simultaneously
as one in which the gauge symmetry is spontaneously broken
and as one in which color is confined.\cite{FradkinShenker}
This means that it provides us with a weak-coupling
laboratory within which we can study a confined phase from
first principles at weak coupling.
It is furthermore a phase of QCD wherein the physics of
chiral symmetry breaking --- indeed all the parameters
of the chiral effective Lagrangian and all known 
or conjectured phenomena of
the pseudoscalar meson sector, including
kaon condensation --- are amenable to controlled,
weak-coupling calculation.

\section{2SC and Other Variants}

\subsection{Two Flavors}

In the previous Section, we have described quark
matter in QCD with
three degenerate flavors of light quarks.  
Nature is less symmetric, and in order to
bracket nature we now describe the color superconducting
phase in QCD with two flavors of light quarks.
Pairs of quarks cannot be color
singlets, and in QCD with two flavors of massless quarks
the Cooper pairs form in the (attractive) color ${\bf \bar 3}$ 
channel.\cite{Barrois,BailinLove,ARW1,RappETC}
The resulting 
$\langle \epsilon_{\alpha\beta 3}\epsilon^{ij}\psi^\alpha_i \psi^\beta_j\rangle$ condensate 
picks a color direction (in this case the 3 or blue direction),
creates a gap $\Delta$ at the Fermi surfaces of
quarks with the other two
out of three colors (red and green),
and breaks $SU(3)_{\rm color}$ to an 
$SU(2)_{\rm color}$ subgroup, giving mass to five of
the gluons by the Anderson-Higgs mechanism. The
masses of these vector bosons have been computed
in the weak-coupling theory valid at asymptotically high
densities,\cite{TwoFlavorMeissner1} and in an instanton liquid
model.\cite{TwoFlavorMeissner2}
Axial color is not a symmetry of the QCD action, but at asymptotically
high densities where the QCD coupling $g$
is weak, explicit axial color breaking is also weak.  As a result,
the pseudoscalar excitations of the condensate which would be Goldstone
bosons arising from  
axial-$SU(3)_{\rm color}$ to axial-$SU(2)_{\rm color}$ breaking if
$g$ were zero may be rather light.\cite{MSW_2SC}

In QCD with two flavors, the Cooper pairs are $ud-du$
flavor singlets and the global flavor symmetry 
$SU(2)_L\times SU(2)_R$ is intact. There
is also an unbroken global symmetry which plays the
role of $U(1)_B$. Thus, no global symmetries are broken
in this 2SC phase.  This means that as a function
of increasing density, there must be a phase transition
between the hadronic and 2SC phases, at which chiral
symmetry is restored.
This phase transition
is first order\cite{ARW1,bergesraj,PisarskiRischke1OPT,CarterDiakonov}
since it involves
a competition between chiral condensation and
diquark condensation.\cite{bergesraj,CarterDiakonov}

Because no global symmetries are broken, there are no light
scalar degrees of freedom in the 2SC phase.  The 2SC phase
is not a superfluid.  The potentially light degrees of freedom are
the gapless blue quarks, and the gauge bosons associated
with unbroken gauge symmetries.  The gluons 
associated with the unbroken $SU(2)_{\rm color}$ will
exhibit strong dynamics on long enough length scales; this
aspect of the infrared physics of the 2SC phase is not under
perturbative control.  Thus, whereas in the CFL phase we expect
that any physical quantity can be  obtained from a controlled
weak-coupling calculation at sufficiently high density,
this claim cannot be made for the 2SC phase.
Note that the gapless blue quarks are neutral under $SU(2)_{\rm color}$.
As in the CFL phase, there is a massless
abelian gauge boson, formed from a suitable linear
combination of the photon and one of the eight original gluons.
The two blue quarks have charges $0$ and $1$ under this
unbroken, but rotated, $\tilde Q$-electromagnetism.

We expect that the blue quarks, left ungapped by the primary
2SC condensate, will find some way to pair in a higher
angular momentum channel.  Indeed, the instanton interaction
pairs these quarks in a $J=1$ condensate which breaks
rotational invariance.  Early work suggested that the
associated gap is of order keV,\cite{ARW1} but this
estimate should be revisited.

It is interesting that both the 2SC and CFL phases satisfy
anomaly matching constraints, even though
it is not yet completely clear whether this
must be the case when Lorentz invariance
is broken by a nonzero density.\cite{AnomalyMatching}
It is not yet clear how high density 
QCD with larger numbers of flavors,\cite{SchaeferPatterns} 
which we discuss below, satisfies
anomaly matching constraints.
Also, anomaly matching in the 2SC phase requires that the
up and down quarks of the third color remain ungapped; 
this requirement must, therefore,
be modified once these quarks pair to form a 
$J=1$ condensate, breaking rotational invariance.\cite{ARW1}

\subsection{Two$+$One Flavors}

Nature chooses two light quarks and one middle-weight
strange quark. If we imagine beginning
with the CFL phase and increasing $m_s$, how
do we get to the 2SC phase? This question
has been answered in Refs.~13,17.
A nonzero $m_s$ weakens those condensates which
involve pairing between light and strange quarks.
The CFL phase requires
nonzero $\langle us \rangle$ and $\langle ds \rangle$
condensates; because these condensates
pair quarks with differing Fermi momenta 
they can only exist if the resulting gaps (call them $\Delta_{us}$ and 
$\Delta_{ds}$) are 
larger than of order $m_s^2/2\mu$, the
difference between the $u$ and $s$ Fermi momenta in
the absence of pairing.  This means that as
a function of increasing $m_s$ at fixed $\mu$
(or decreasing $\mu$ at fixed $m_s$) there
must be a first order unlocking phase transition.\cite{ABR2+1,SW2}
The argument can be phrased thus: the 2SC and CFL phases
must be separated by a phase transition, because
chiral symmetry is broken in the CFL phase but not
in the 2SC phase; suppose this transition were second order;
this would require $\Delta_{us}$ and 
$\Delta_{ds}$ to be infinitesimally small but nonzero just
on the CFL side of the transition; however, these gaps
must be greater than of order $m_s^2/2\mu$; 
a second order phase transition is therefore a logical
impossibility, either in mean field theory or beyond;
the transition must therefore be first order.  
Note that the $m_s$ that appears in these estimates
is a density dependent effective strange quark mass,
somewhat greater than the current quark mass.

Putting in reasonable numbers for
quark matter which may arise
in compact stars, for $m_s=200-300$~MeV and $\mu=400-500$~MeV
we find that the CFL phase is obtained if the interactions
are strong enough to generate a gap $\Delta$ which is 
larger than about $40-110$~MeV,
while the 2SC phase is obtained if $\Delta$ is smaller.
As we shall see in the next section, $\Delta\sim 40-110$~MeV
is within the range of current estimates and
present
calculational methods are therefore not 
precise enough to determine whether quark matter with
these parameters is in the CFL or 2SC phases.
At asymptotically high densities, however, the CFL
phase is necessarily favored.

\subsection{One Flavor}

The remaining variant of color superconductivity
which may arise in nature is that
involving pairing between quarks
of a single flavor.  If $m_s$ is too large for CFL pairing 
(i.e. $m_s$ greater than of order $\sqrt{2\mu\Delta}$)
but is nevertheless less than $\mu$ itself, the 2SC
phase will include a nonzero density of strange quarks.
Although, by hypothesis, these strange quarks do
not pair with the light quarks, they can pair with
each other, forming a $\langle ss \rangle$ condensate
with angular momentum $J=1$.  This pairing has
been analyzed in Ref.~47. (See also
Ref.~48.) 
Interestingly, even though the Cooper pairs have $J=1$,
the condensate does not break rotational invariance.\cite{Schaefer1Flavor}
The Cooper pairs are in the color ${\bf \bar 3}$ channel,
and the condensate locks color and spatial rotation:
it leaves unbroken a global symmetry assocated  with
simulataneous color and spatial rotation.
The resulting gap is much smaller than the $J=0$ 2SC and CFL
gaps. It 
has been estimated to be of order
hundreds of keV,\cite{Schaefer1Flavor} although applying
results of Ref.~49 suggests a somewhat smaller gap, around
10 keV.

\subsection{Four or More Flavors}

We end this section with brief mention of four variants
which are unphysical, but nevertheless instructive:
QCD with more than three light flavors, QCD with two colors,
QCD with many colors,
and QCD with large isospin density and zero baryon density. 

Dense quark matter in QCD with more than three flavors was
studied in Ref.~10. The main result
is that the color-flavor locking phenomenon persists:
Condensates form which lock color rotations to flavor
rotations, and the $SU(N_f)_L\times SU(N_f)_R$ group
is broken down to a vector subgroup.  Unlike with
$N_f=3$, however, the unbroken group is not the 
full $SU(N_f)_{L+R}$ which is unbroken in the vacuum.
In the case of $N_f=4$, for example, one finds  
$SU(4)_L\times SU(4)_R\rightarrow O(4)_{L+R}$
while in the case of $N_f=5$,
$SU(5)_L\times SU(5)_R\rightarrow SU(2)_{L+R}$.\cite{SchaeferPatterns}
For $N_f=4,5$ as for $N_f=3$, chiral symmetry is broken in
dense quark matter.  However, because the unbroken vector
groups are smaller than $SU(N_f)_V$, there must be a phase
transition between hadronic matter and dense quark 
matter in these theories.\cite{SchaeferPatterns}

If $N_f$ is a multiple of three, the order parameter 
takes the form of multiple copies of the $N_f=3$ order
parameter, each locking a block of three flavors to 
color.\cite{SchaeferPatterns} 
All quarks are gapped in this phase, as in the $N_f=3$ 
CFL phase. For
$N_f=6$, the resulting symmetry breaking pattern 
is $SU(6)_L\times SU(6)_R\rightarrow SU(3)_{L+R}\times U(1)_{L+R}
\times U(1)_{L-R}$.\cite{SchaeferPatterns} 
The unbroken $SU(3)_{L+R}$ is a simultaneous
rotation of both three flavor blocks for $L$ and $R$ and a 
global color rotation.  Note that the unbroken $U(1)$'s are
subgroups of the original $SU(6)$ groups: they correspond
to vector and axial flavor rotations which rotate one
three flavor block relative to the other.  
Note that for $N_f=6$, unlike for $N_f=3,4,5$,
chiral symmetry is not completely
broken at high density: an axial $U(1)$ subgroup remains unbroken.
As the primary condensate we have just described
leaves no quarks ungapped, there is no 
reason to expect the formation of any subdominant
condensate which could break the unbroken chiral symmetry.
Both because of this unbroken chiral symmetry and because
the unbroken vector symmetry differs from that of the vacuum, 
there must be a phase
transition between hadronic matter and dense quark
matter in QCD with $N_f=6$.\cite{SchaeferPatterns}

\subsection{Two Colors}

The simplest case of all to analyze is QCD with two colors
and two flavors.  The condensate is antisymmetric in color
and flavor, and is therefore a singlet in both color
and flavor.  Because it is a singlet in color, dense
quark matter in this theory is not a color superconductor.
Although the condensate is a singlet under
the ordinary $SU(2)_L\times SU(2)_R$ flavor group, it nevertheless
does break symmetries because the symmetry 
of the vacuum in QCD with $N_f=N_c=2$ is enhanced to $SU(4)$.
One reason why $N_c=2$ QCD is interesting to study at nonzero
density is that it provides an example where
quark pairing can be studied on the lattice.\cite{HandsMorrison} 
The $N_c=2$ case has also
been studied analytically in Refs.~6,51; 
pairing in this
theory is simpler to analyze because 
quark Cooper pairs are color singlets. We refer the reader
to these references for details.

\subsection{Many Colors}

The $N_c\rightarrow \infty$
limit of QCD is often one in which hard problems become
tractable. However, the ground state of $N_c=\infty$ QCD
is a chiral density wave, not a color superconductor.\cite{DGR}
At asymptotically high densities, 
color superconductivity persists up
to $N_c$'s of order thousands~\cite{ShusterSon,PRWZ} before being
supplanted by the phase described in Ref.~52.  At any finite
$N_c$, color superconductivity occurs at 
arbitrarily weak coupling whereas
the chiral density wave does not.
For $N_c=3$, color superconductivity is 
still favored over the chiral density wave (although not by much)
even if the interaction 
is so strong that the color superconductivity gap is 
$\sim \mu/2$.\cite{RappCrystal}

\subsection{QCD at Large Isospin Density}

The phase of $N_c=3$ QCD with nonzero 
isospin density ($\mu_I\neq 0$) and zero baryon density ($\mu=0$) {\it can}
be simulated on the lattice.\cite{SonStephIsospin}  
The sign problems that plague simulations at $\mu\neq 0$ do not
arise for $\mu_I\neq 0$.
Although not physically realizable, physics with
$\mu_I\neq 0$ and $\mu=0$  is very interesting
to consider because phenomena arise which are similar
to those occurring at large $\mu$ and, in this context,
these phenomena are accessible to numerical ``experiments''.
Such lattice simulations 
can be used to
test calculational methods which have also been applied at large $\mu$,
where lattice simulation is unavailable.
At low isospin density, this theory describes a dilute
gas of Bose-condensed pions. 
Large $\mu_I$ physics features large Fermi surfaces for down quarks
and anti-up quarks, Cooper pairing of down
and anti-up quarks,  and a gap
whose $g$-dependence is as in (\ref{eq:SonResult}), albeit
with a different 
coefficient of $1/g$ in the exponent.\cite{SonStephIsospin}  
This condensate has the same quantum numbers as the pion
condensate expected at much lower $\mu_I$,
which means
that a hypothesis of continuity between hadronic --- in this
case pionic --- and quark matter as a function of $\mu_I$. 
Both the dilute pion gas limit and the asymptotically
large $\mu_I$ limit can be treated analytically;
the possibility of continuity between these two
limits can be tested on the lattice.\cite{SonStephIsospin}
The transition from a 
weak coupling superconductor with condensed Cooper pairs
to a gas of tightly bound bosons which form a Bose condensate
can be studied in a completely controlled fashion.

\section{Calculational Methods}

In this Section we review the basic theoretical methods used to analyze
color superconductivity.   To keep
the discussion within appropriate bounds, we shall concentrate on the
methods used to characterize the
ground state.  The derivation of effective theories for the low-energy
dynamics,  electromagnetic
response, and transport properties is of course predicated on
identification of the ground state,
but presents many additional points of interest as we have
already seen in Section 2.  In that Section, we deferred
all discussion of methods by which the gap $\Delta$, which
characterizes the ground state, is calculated.  
Much effort has gone into calculating
the magnitude of the gaps in the 2SC and CFL 
phases,\cite{BailinLove,ARW1,RappETC,CFL,ABR2+1,SW2,bergesraj,Hsu1,SW0,Son,CarterDiakonov,AKS,RappETC2,PisarskiRischke,Hong,HMSW,SW3,rockefeller,Hsu2,ShovWij,EHHS,Vanderheyden,BBS,RajagopalShuster,Manuel} and in
this section we face up to the challenge of describing what 
has been learned.

It would be ideal if the calculation of
the gap were within the scope of
lattice gauge theory as is, for example, the calculation
of the critical temperature on the vertical axis of the phase diagram.
Unfortunately, lattice methods relying
on importance sampling have to this
point been rendered exponentially 
impractical at nonzero baryon density by the 
complex action at nonzero $\mu$.  
Various lattice methods {\it can} be applied for $\mu\neq 0$
as long as $T/\mu$ is large enough;~\cite{Philipsen} 
so far, though, none have proved applicable at temperatures
which are low
enough that color superconductivity occurs.
As we saw in the previous section, lattice simulations
are possible in two-color QCD and in QCD at large isospin density.
Finally, sophisticated algorithms have recently 
allowed theories which are simpler than QCD but which
have as severe a fermion sign problem as that in QCD at nonzero 
chemical potential to be simulated.\cite{MeronCluster}
All of this bodes well for the future.

To date, in the absence of suitable lattice methods, 
quantitative analyses of color superconductivity 
have followed two distinct strategies.
The
first approach is utilitarian and semi-phenomenological, emphasizing the
use of  simplified models.  This
will occupy us in Sections 4.1-4.3.  The overarching theme here is to define
models which incorporate the
salient physical effects, yet are tractable using known mathematical
techniques of
quantum many-body theory.  Free parameters within a model of
choice are chosen to give reasonable 
vacuum physics.
Examples include analyses in which the
interaction between quarks is replaced simply by four-fermion
interactions with the quantum numbers of
the instanton interaction~\cite{ARW1,RappETC,bergesraj}
or of one-gluon exchange,\cite{CFL,ABR2+1}
random matrix models,\cite{Vanderheyden} and more sophisticated
analyses done using
the instanton liquid model.\cite{CarterDiakonov,RappETC2,RappCrystal}
Renormalization group methods have also been used to
explore the space of all possible 
effective four-fermion interactions.\cite{Hsu1,SW0}
These methods yield results which are in qualitative 
agreement: the favored condensates are as described
in Sections 2 and 3; 
the gaps range between several tens of MeV up to of order 
$100$~MeV;
the associated critical temperatures (above which the 
diquark condensates vanish)
can be as large as about $T_c\sim 50$~MeV.
This agreement between different models reflects the fact
that what matters most is simply the strength of the attraction
between quarks in the color ${\bf \bar 3}$ channel, and by
fixing the parameters of the model interaction to fit, say,
the magnitude of the  vacuum chiral condensate, one ends up
with attractions of similar strengths in different models.

The second, more ambitious approach is fully
microscopic.   Such an
approach is feasible, for high-density QCD, due to asymptotic freedom.
Several important results have
been obtained from the microscopic approach, perhaps most notably the
asymptotic form of the gap.\cite{Son}
Very significant challenges remain, however.    It is not really known, for
example, how to calculate
corrections to the leading term in a systematic way.     We review the
microscopic approach in
Sections 4.4-4.5.

These approaches have complementary virtues -- simplicity versus rigor,
openness to
phenomenological input versus quantitative predictive power
at asymptotically high density.  Fortunately,
they broadly
agree in their main conclusions as to the patterns of symmetry breaking and the
magnitude of the gap at accessible densities of order $\mu=400-500$~MeV.

\subsection{Renormalization Toward the Fermi Surface}

The
adequacy of weak coupling for describing QCD at high density is by no means
obvious.   Our strategy will be to adopt it as a working
hypothesis, bring out its consequences, and see whether we find a
consistent picture.   In doing this, we must consider how to work from the
fundamental interactions in
the perturbative, no-particle state (``Fock vacuum''), which are assumed to
be simple, to the effective
interactions in the dense medium.

To begin, let us focus on the quark degrees of freedom. In line with
our announced strategy, we begin by approximating the
ground state with filled Fermi spheres for all the quarks,
with Fermi momentum $p_F$. The low-energy
excitations then include states where some modes just below the
nominal Fermi surfaces are vacant (hole modes) and some modes just above
are occupied (particle modes).  There is a continuum of such
states.  Moreover, states containing pairs of particle or hole modes with
three-momenta $(\vec p, -\vec p)$, for various values of the
direction of
$\vec p$ but a common $|\vec p|$, are all degenerate.
We
are therefore faced with a system whose excitations have
a large density
of states (proportional to the area of the Fermi surface)
which can be excited at arbitrarily small free energy cost
(for pairs whose $|\vec p|$ is arbitrarily close to $p_F$.)
In the absence of interaction, there would be nothing
more to say.  The effect of interactions, however,
is to allow all pairs
$(\vec p, -\vec p)$ with a common $|\vec p|$ to scatter into
one another, consistent with momentum conservation.
Thus,
we are perturbing a system with a continuum of excitations
with arbitrarily low
energies.
In this situation, with large numbers of nearly degenerate states,
straightforward perturbation theory can fail, even for weak coupling.

The Wilsonian renormalization group is often an appropriate tool for
analyzing problems of this sort.  Following this approach, one
attempts to map the original problem onto a problem with fewer degrees of
freedom, by integrating out the effect of the higher-energy
(or, in a relativistic theory, more virtual) modes.  Then one finds a new
formulation of the problem, in a smaller space, with new
couplings.  In favorable cases the reformulated problem is simpler than the
original, and one can turn around and solve it. 

This use of the renormalization group is in the same spirit
as that appropriate for the analysis of long wavelength
physics at a second order phase transition, or the long
wavelength physics of the QCD vacuum (a.k.a chiral perturbation theory.)
In traditional perturbative QCD
one runs
the logic backwards.   The fundamental short-distance theory, with nothing
integrated out, is
simple and weakly coupled.   When one integrates out highly virtual modes,
one finds that QCD becomes more strongly coupled.   
Thus, renormalization group methods are used in two
distinct ways in vacuum QCD: first, they inform us how the 
fundamentally simple theory of quarks and gluons comes to look
complicated at low energy and, second, once 
rephrased in terms of the correct effective degrees
of freedom relevant at low energies, they allow us to make
predictions for low energy physics.  
Renormalization group
methods are used in two analogous ways at high density. In this Section
we shall use them to inform us at what free energy scale the 
theory based on the original quark degrees of freedom breaks down,
signalling the formation of a gap and the need to switch
to new degrees of freedom.  These new degrees of freedom
include the gapped quasiparticles, which can then be integrated out.
If
the gapped phase features 
massless or light Nambu-Goldstone boson excitations
of the condensate then, as we have seen in Section 2, 
we can use renormalization group methods to make predictions for the 
physics of these effective 
degrees of freedom at momenta well below the gap.

In both chiral perturbation
theory of the QCD vacuum and the Wilsonian effective
theory appropriate at high density, one integrates out modes
with large free energy, keeping only those with    smaller
and smaller free energy.  In the vacuum, free energy is
just energy and this corresponds to keeping modes
with momenta closer and closer to zero. At high density,
however, the modes with small free energy are those
with momenta near the Fermi surface, and it is on
these modes that the renormalization group focuses.
The application
of Wilsonian methods to physics at a Fermi surface was
pioneered by Shankar\cite{Shankar} and Polchinski.\cite{Polchinski}

Asymptotic freedom plays a foundational role in the analysis of
QCD at asymptotically high density.  It allows us to make
it plausible that the
fundamental couplings among the active modes, near the Fermi surface, are
generically weak.  Indeed, these modes involve large energy
and momenta, and so scattering events which displace them by a fixed finite
angle involve large momentum transfer, whereas scattering
events with small angle should not much alter the physical properties of
the state. This intuitive argument will be refined, and justified,
in Section 4.4.

Granting that the fundamental coupling is weak, let us
consider the consequences of integrating out modes whose 
free energy is between $\epsilon$ and $\delta\epsilon$
on the effective interactions among the
remaining modes, which are those within a band in momentum
space centered on the Fermi surface which have free energy 
less than $\delta\epsilon$.  
We will obtain renormalized couplings among
the remaining modes, due to one-loop diagrams  in which
external legs are within $\delta\epsilon$ of the Fermi
surface while loop momenta are between 
$\delta\epsilon$ and $\epsilon$.  This 
approach was first applied to high density QCD
in Ref.~57
and was generalized in Ref.~58.

The first result of this analysis is that, in general,
four-fermion, six-fermion and higher-order interactions
are all suppressed as we approach the Fermi surface.\cite{Shankar,Polchinski} 
This fixed point corresponds to Landau Fermi liquid theory. The
only, and crucial, exceptions are those four-fermion operators
that involve
involve particles or holes with equal and opposite three momenta.   
These allow pairs of particles on the Fermi surface to scatter, 
as mentioned above.
For
couplings $G_\eta$ of this kind we find
\begin{equation}
\label{rgfermi}
{dG_\eta \over d\ln \delta} ~=~ -\kappa_\eta G_\eta^2~.
\end{equation}
Here $\eta$ is a catch-all label that might include color, flavor, angular
momentum. The effective couplings will
depend on all these indices.   In general we will have a matrix equation,
with couplings between the different channels (different
values of $\eta$).  
But to bring out the main point, it is enough to
consider the simplest possible case, with one channel, as we have done
in  (\ref{rgfermi}). In this single channel calculation, 
a simple calculation shows that $\kappa$ is positive.
(Note that we have chosen a notation in which
attractive interactions correspond to $G>0$.)  
This evolution equation is quite simple to integrate, and we find
\begin{equation}
\label{rgfsoln}
{1\over G_\eta(\delta)} - {1\over G_\eta (1)} ~=~ \kappa_\eta \ln \delta~.
\end{equation}
Thus we
see that the qualitative behavior of $G_\eta(\delta)$ depends
on the sign of $G_\eta(1)$.  If this is negative, then as $\delta
\rightarrow 0$ one finds that $G_\eta(\delta) \rightarrow 0$ from
below.  Repulsive interactions are thus irrelevant.
On the other hand, if $G_\eta(1)>0$ (indicating an attractive
coupling) then $G_\eta(\delta)$ grows, and formally it diverges
when
\begin{equation}
\label{divergence}
\delta ~=~ \exp\left( -\frac{1}{\kappa_\eta G_\eta (1)}\right).
\end{equation}
Of course once the effective coupling becomes large the working equation
(\ref{rgfermi}), which neglects higher-order corrections, is no
longer trustworthy.    All we may legitimately infer from (\ref{divergence})
is that an attractive interaction in any channel, however weak,
will, by renormalization toward the Fermi surface, induce a strong
effective coupling between particles with equal and opposite momenta,
on opposite sides of the Fermi surface.

What we have done here is simply to rephrase Cooper's discovery of the
pairing instability in modern language,
suitable for generalization.   Now let us apply it to four-fermion
interactions of the type we expect to
occur in QCD.

We will restrict ourselves to massless
QCD, a spherical Fermi surface, and local operators invariant under
the appropriate chiral symmetry.   The cases of three and two flavors are
rather different, and we will
analyze them separately in turn.

Following Ref.~57, we take the basic
four-fermion interactions
in $N_f=3$ QCD to be proportional to the operators
\begin{eqnarray}
\label{nf3_ops}
O^0_{LL} &=& (\bar\psi_L\gamma_0\psi_L)^2\,, \hspace{1cm}
O^0_{LR} \;=\; (\bar\psi_L\gamma_0\psi_L)(\bar\psi_R\gamma_0\psi_R)\,, \\[2ex]
O^i_{LL} &=& (\bar\psi_L\gamma_i\psi_L)^2\,,  \hspace{1cm}
O^i_{LR} \;=\; (\bar\psi_L\vec\gamma\psi_L)
       (\bar\psi_R\vec\gamma\psi_R)\,. \nonumber
\end{eqnarray}
Each of these operators comes in two color structures,
for example color symmetric and color anti-symmetric
\beq
(\bar\psi^a\psi^b)(\bar\psi^c\psi^d)
  \left(\delta_{ab}\delta_{cd}\pm \delta_{ad}\delta_{bc}\right).
\eeq
Nothing essentially new emerges upon considering superficially
different isospin structures, or different Dirac matrices. All
such structures can be reduced to linear combinations of the
basic ones (\ref{nf3_ops}), or their parity conjugates, by
Fierz rearrangements. In total, then, we need to consider eight
operators.

As discussed above, these operators are renormalized by quark-quark scattering
in the vicinity of the Fermi surface. This means that both
incoming and outgoing quarks have momenta $\vec p_1, \vec p_2\simeq
\pm \vec p$ and $\vec p_3, \vec p_4\simeq\pm \vec q$ with $|\vec p|,
|\vec q| \simeq p_F$. We set the external frequency to be zero.
A graph with vertices $\Gamma_1$ and $\Gamma_2$ then gives \cite{Hsu1}
\beq
\label{loop}
G_1 G_2 I\; (\Gamma_1)_{i'i}(\Gamma_1)_{k'k}
  \left[ -(\gamma_0)_{ij}(\gamma_0)_{kl}-\frac{1}{3}
          (\vec\gamma)_{ij}(\vec\gamma)_{kl}\right]
 (\Gamma_2)_{jj'}(\Gamma_2)_{ll'}
\eeq
with $I=\frac{i}{8\pi^2}\mu^2\log(\delta)$.
We will denote the density of states on the
Fermi surface by $N=\mu^2/(2\pi^2)$ and the logarithm of the scale
as $t=\log(\delta)$. The
renormalization group does not mix $LL$ and $LR$ operators,
nor different color structures. This means that
the evolution equations contain at most $2\times 2$ blocks.  A simple
calculation
now yields\cite{Hsu1}
\bea
\label{nf2_evol}
\frac{d(G^{LL}_0+G^{LL}_i)}{dt} &=& -\frac{N}{3}
   (G^{LL}_0+G^{LL}_i)^2 \,,\\
\frac{d(G^{LL}_0-3G^{LL}_i)}{dt} &=& -N
   (G^{LL}_0-3G^{LL}_i)^2 \,,\\
\frac{d(G^{LR}_0+3G^{LR}_i)}{dt} &=& 0\,,\\
\frac{d(G^{LR}_0-G^{LR}_i)}{dt} &=& -\frac{2N}{3}
   (G^{LR}_0-G^{LR}_i)^2\,.
\eea
In this basis the evolution equations are already diagonal. The coupling
$G_1=G^{LL}_0+G^{LL}_i$ evolves as
\bea
 G_1(t) &=& \frac{1}{1+(N/3)G_1(0)t}\,,
\eea
with analogous results for the other operators. Note that the evolution
starts at $t=0$ and moves towards the Fermi surface as $t\to-\infty$.
If the coupling is attractive at the matching scale, $G_1(0)>0$, it
will grow during the evolution, and reach a Landau pole at $t_c=3/(N
G_1(0))$. The location of the pole is controlled by the initial value
of the coupling and the coefficient in the evolution equation. If the
initial coupling is negative, the coupling decreases during the
evolution. The second operator in (\ref{nf2_evol}) has the largest
coefficient and will reach the Landau pole first, unless the initial
value is very small or negative.  In that case, subdominant operators can come
to cause the leading instability.

The form of the operators that diagonalize the
evolution equations is readily understood.\cite{SW0}  
First, order
the operators according to the size of the coefficient in the
evolution equations
\bea
\label{O_dom}
O_{dom} &=& (\bar\psi_L\gamma_0\psi_L)^2 -
               (\bar\psi_L\vec\gamma\psi_L)^2\,,\\
O_{sub,1} &=& (\bar\psi_L\gamma_0\psi_L)(\bar\psi_R\gamma_0\psi_R)
  -\frac{1}{3} (\bar\psi_L\vec\gamma\psi_L)(\bar\psi_R\vec\gamma\psi_R)\,,\\
O_{sub,2} &=& (\bar\psi_L\gamma_0\psi_L)^2
  + \frac{1}{3} (\bar\psi_L\vec\gamma\psi_L)^2\,,\\
\label{O_mar}
O_{mar}   &=& (\bar\psi_L\gamma_0\psi_L)(\bar\psi_R\gamma_0\psi_R)
  + (\bar\psi_L\vec\gamma\psi_L) (\bar\psi_R\vec\gamma\psi_R)\,.
\eea
Upon Fierz rearrangement
we find
\bea
O_{dom} &=&   2(\psi_LC \psi_L)(\bar\psi_L C\bar\psi_L)\,,\\
O_{sub,1} &=& \frac{1}{3}(\psi_L C\vec\gamma\psi_R)
 (\bar\psi_RC\vec\gamma\psi_L) + \ldots \,,\\
O_{sub,2} &=& \frac{4}{3}(\psi_L C\vec\Sigma\psi_L)
 (\bar\psi_LC\vec\Sigma \bar\psi_L) \,,\\
O_{mar}   &=&  \frac{1}{2}(\psi_LC\gamma_0\psi_R)
 (\bar\psi_RC\gamma_0\psi_L) + \ldots \,.
\eea
This demonstrates that the linear combinations in
(\ref{O_dom}-\ref{O_mar}) correspond to simple structures in the
quark-quark channel.  (It also means that it might have been more
natural to perform the whole calculation directly in a basis of diquark
operators.)

The full structure of the $(LR)$ operators is 
$$O_{sub,1},O_{mar}=
(\psi C\gamma\tau_{S,A}\psi)(\bar\psi C\gamma\tau_{S,A}\bar\psi)+
(\psi C\gamma\gamma_5\tau_{A,S}\psi)(\bar\psi C\gamma\gamma_5
\tau_{A,S}\bar\psi)\ ,
$$ 
where $\tau_{S,A}$ are symmetric, respectively
anti-symmetric,
flavor generators. Note that because these two structures have different
flavor symmetry, the flavor structure cannot be factored
out.

The dominant operator corresponds to the scalar diquark
channel, while the subdominant operators contain vector diquarks.
Note that one cannot decide which color channel is preferred
from the evolution equation alone.
To decide that question, we can appeal to
the fact that ``reasonable'' interactions,
including specifically one gluon exchange,
will be attractive in the color anti-symmetric
but repulsive in the color symmetric channel. Indeed, it is the color
anti-symmetric configuration that minimizes the total color flux
(and hence field energy) emanating from the quark pair.  If the color wave
function is anti-symmetric, the dominant operator fixes the
isospin wave function to
be anti-symmetric as well.  

The form of the dominant operator indicates the existence
of potential instabilities, but does not itself indicate
how they are resolved.  The renormalization group analysis
tells us the scale at which the dominant couplings get
strong but, strictly speaking, tells us nothing more.
In Sections 4.2 and 4.3, we shall see how to learn more,
and in particular to confirm that the resolution of
the instability is that a gap forms in the dominant 
attractive channel.  Because the renormalization group
methods do not allow us to directly describe the formation
of the gap (although they indicate its magnitude) they
cannot be used to confirm that color-flavor locking is favored.
For this, a variational calculation is required as we shall
describe in Section 4.2.

The dominant four-fermion operator in three-flavor
QCD does not distinguish
between scalar and pseudoscalar diquarks.  Indeed,
for $N_f \geq 3$ all four-quark operators
consistent with chiral symmetry exhibit an accidental
axial baryon symmetry, under which scalar and pseudoscalar diquarks are
equivalent.
For $N_f=3$ this
degeneracy is lifted by six-fermion 
operators,\cite{CFL,RappETC2,SchaeferPatterns}
which are irrelevant operators in the sense of the
renormalization group.
In the
microscopic theory, these operators will be induced by instantons.  Their
formal irrelevance
does not imply that they are negligible physically, especially since they
are the leading terms which
break
the residual axial baryon number symmetry.

Anticipating that the instability  
indicated by 
renormalization toward the Fermi
surface is stabilized by the 
formation of a gap, we conclude that upon further 
renormalization the couplings cease to run.
We therefore expect that, instead of running to zero, the
instanton coupling remains at some finite value.
Instantons have important physical effects in
the condensed phase, even for
$N_f=3$,\cite{RappETC2,SchaeferPatterns} 
as already mentioned above.  Specifically, they cause
quark-antiquark
pairs to condense, by inducing the ``normal'' chirality-violating order
parameter (familiar at zero density) from the diquark-antidiquark
condensate (which arose, in turn, as
a secondary consequence of the  primary diquark condensation).  
The resulting quark-antiquark condensate is very
small.\cite{RappETC2,SchaeferPatterns}  The instanton interaction also
lifts the  degeneracy between the
scalar and pseudoscalar diquark condensates, favoring
the scalar.

We turn now to QCD with two flavors.
This requires us to take into account additional
operators. At first hearing it might seem odd that with fewer 
flavors we encounter more possible interactions. It occurs because for
$N_f =2$, but not for larger values, two quarks of the same chirality
can form a chiral $SU(2) \times SU(2)$ singlet. Related to this, for
$N_f=2$ we have $U(1)_A$ violating {\it four}-fermion operators.
In the microscopic theory, these operators will be induced by instantons.

The new operators are
\bea
\label{inst_ops}
 O_S &=& \det_f (\bar\psi_R\psi_L), \hspace{1cm}
 O_T \;=\; \det_f (\bar\psi_R\vec\Sigma\psi_L)\ .
\eea
Both operators are determinants in flavor space. For quark-quark
scattering, this implies that the two quarks have to have different
flavors. The fact that the flavor structure is fixed implies that
the color structure is fixed, too. For a given $(qq)$ spin, only
one of the two color structures contributes. Finally, both quarks
have to have the same chirality, and the chirality is flipped
by the interaction.

These considerations determine
the structure of the evolution equations.
Two left handed quarks can interact via one of the instanton operators,
become right handed, and then rescatter through an anti-instanton,
or through one of the $U(1)_A$ symmetric $RR$ operators. The result
will be a renormalization of the $LL$ vertex in the first case, and
a renormalization of the instanton in the second. 
Note that the flavor structure will
always remain a determinant. Even though instantons generate all
possible Dirac structures in (\ref{nf3_ops}), the color-flavor structure
is more restricted.

Evidently,
instantons do not affect the evolution of the $LR$ couplings at all.
The evolution equations of the $LL$ couplings are modified to become
(henceforth we drop the
subscript $LL$):~\cite{SW0}
\bea
\label{nf3_evol}
\frac{dG_0}{dt} &=& \frac{N}{2} \Bigg\{
  -G_0^2 + 2G_0G_i - 5G_i^2 - K_S^2 + 2K_S K_T - 5 K_T^2 \Bigg\}\,, \\
\frac{dG_i}{dt} &=& \frac{N}{2} \Bigg\{
 \frac{1}{3}G_0^2 - \frac{10}{3}G_0G_i + \frac{13}{3}G_i^2
 +\frac{1}{3}K_S^2 - \frac{10}{3} K_S K_T + \frac{13}{3}K_T^2 \Bigg\}\,, 
\hspace{0.4cm}\\
\frac{dK_S}{dt} &=& \frac{N}{2} \Bigg\{
  2\left(-G_0 + G_i \right)K_S
 +2\left( G_0 - 5 G_i \right)K_T \Bigg\}\,, \\
\frac{dK_T}{dt} &=& \frac{N}{2} \Bigg\{
  \frac{2}{3}\left( G_0 - 5 G_i \right)K_S
 +\frac{2}{3}\left( -5G_0 +13 G_i \right)K_T \Bigg\}~.
\eea
These equations can be uncoupled in the form
\bea
\label{nf3_evol_2}
\frac{dG_1}{dt} &=& - \frac{N}{3} \left( G_1^2+K_1^2 \right)\,, \\
\frac{dK_1}{dt} &=& - \frac{2N}{3}\; G_1 K_1\,, \\
\frac{dG_2}{dt} &=& -  N \left( G_2^2+K_2^2 \right)\,, \\
\frac{dK_2}{dt} &=& - 2N \; G_2 K_2 \,,
\eea
where $G_1=G_0+G_i$, $K_1=K_S+K_T$ and $G_2=G_0-3G_i$, $K_2=K_S-3K_T$.
The equations for $G$, $K$ uncouple even further. We have
\bea
\label{nf3_evol_3}
\frac{d(G_2+K_2)}{dt} &=& -N \left( G_2 +K_2 \right)^2 \,,\\
\frac{d(G_2-K_2)}{dt} &=& -N \left( G_2 -K_2 \right)^2 \,,
\eea
as well as the analogous equation for $G_1$, $K_1$.

The differential
equations are now easily solved,\cite{SW0} leading to
\bea
\label{nf3_flow}
 G_2(t) &=& \frac{1}{2}\left\{ \frac{1}{a+Nt} + \frac{1}{b+Nt}\right\}, \\
 K_2(t) &=& \frac{1}{2}\left\{ \frac{1}{a+Nt} - \frac{1}{b+Nt}\right\},
\eea
together with the analogous results for $G_1,K_1$. Here,
$a,b=(G_2(0)\pm K_2(0))^{-1}$.  We see that $G_2$
and $K_2$ will grow and eventually reach a Landau pole if either
$a$ or $b$ is positive. The location of the pole is determined
by the smaller of the values, $t_c=-a/N$ or $t_c=-b/N$. The
same is true for $G_1$ and $K_1$, but the couplings evolve
more slowly, and the Landau pole is reached later.

At this level a number of qualitatively
different scenarios are possible, depending on the
sign and relative magnitude of $G(0)$ and $K(0)$.\cite{SW0} (Henceforth
we drop all subscripts.) If $G(0)$ and $K(0)$
are both positive then they will both grow, and the location of the
nearest
Landau pole is determined by $G(0)+K(0)$.  The asymptotic ratio of
the two couplings is 1.
If $G(0)$ and $K(0)$ are both negative, and the
magnitude of $G(0)$ is bigger than the magnitude of $K(0)$, then
the evolution drives both couplings to zero. These are the
standard cases.  Attraction leads to an instability, and repulsive
forces are suppressed.

More interesting cases arise when the sign of the two couplings is
different.  The case $G(0),K(0)<0$ and
$|K(0)|>|G(0)|$ is especially weird.\cite{SW0}
Both $G(0),K(0)$ are repulsive, but the evolution
drives $G(0)$ to positive
values.  Both couplings reach a Landau pole, and near the pole
their asymptotic
ratio approaches minus one. Similarly, we can have a negative $G(0)$ and
positive $K(0)$ with $K(0)>|G(0)|$. Again, the evolution will drive
$G(0)$ to positive values.\cite{SW0}

The dominant
and sub-dominant instanton operators are
\bea
O_{dom} &=&\det_f\left[ (\bar\psi_R\psi_L)^2 -
  (\bar\psi_R\vec\Sigma\psi_L)^2 \right]\,,\\
O_{sub} &=& \det_f\left[(\bar\psi_R\psi_L)^2 +\frac{1}{3}
  (\bar\psi_R\vec\Sigma\psi_L)^2\right]\,.
\eea
Upon Fierz rearrangement, we find
\bea
O_{dom} &=& 2(\psi_LC\tau_2\psi_L)(\bar\psi_R C\tau_2\bar\psi_R)\,, \\
O_{sub} &=& \frac{2}{3}(\psi_LC\tau_2\vec\Sigma\psi_L)
 (\bar\psi_R C\tau_2\vec\Sigma\bar\psi_R)\,,
\eea
corresponding to scalar and tensor diquarks. Both operators are
flavor singlet. Overall symmetry then fixes the color wave functions,
anti-symmetric $\bar 3$ for the scalar, and symmetric 6 for the tensor.
The dominant pairing induced by instantons is in the scalar diquark
channel, the only other attractive channel is the tensor.
All this neatly confirms the scenario discussed in 
Refs.~5,6 and Section 3.1.

Just as we found for $N_f \geq 3$, there is an appealing heuristic
understanding for the amazingly simple behavior of the evolution
equations, obtained by focussing on the diquark channels.
Instantons distinguish between scalar diquarks with positive and
negative parity. $G+K$ corresponds to the positive parity operator
$(\psi C\gamma_5\psi)$ and $G-K$ to the negative parity
$(\psi C \psi)$.  The asymptotic approach of $G/K\to 1$, then
corresponds to the fact that scalar
diquark condensation is favored over pseudoscalar diquark
condensation. This is always the case if $K(0)>0$.

We can also 
understand the strange case $G(0),K(0)<0$ and $|K(0)|>|G(0)|$.
In this case the interaction for scalar diquarks is repulsive,
but the interaction in the pseudoscalar channel is attractive
and leads to an instability. Note that this can only happen
if we have the ``wrong'' sign of the instanton interaction, {\it
i.e}. for $\theta = \pi$.
Similarly, we can understand why the asymptotic ratio of the molecular
(instanton-anti-instanton) and direct instanton couplings approaches
$G/K=\pm 1$. Instantons induce a repulsive interaction
for pseudoscalar diquarks. During the evolution, this coupling
will be suppressed, whereas the attractive scalar interaction
grows. But this means that in the pseudoscalar channel, the repulsive
(instanton) and attractive (molecular) forces have to cancel in the
asymptotic limit, so the effective couplings become equal.

To match all the instanton coupling constants
to microscopic QCD would require control
of the instanton density, the relevant value
of $\alpha_s$, and the screening mechanism, and does not seem practicable.
Some simple qualitative conclusions may be inferred, however.
{}From the form of the instanton vertex we can fix the ratio
of the two instanton-like couplings, 
$K_T/K_S=1/(2N_c-1)$.\cite{SVZ_80b,SS_98} 
This emphasizes that the tensor channel
coupling is expected to be small.

Also, both one gluon exchange
and ideas based on instantons (e.g., the instanton liquid model)  yield
$G^{LL}_0>0,-G^{LL}_i>0$.
Thus the favored
scenario is that both instanton and $U(1)_A$ symmetric
couplings flow at the same rate.  In any case, the leading instability
occurs in a scalar diquark channel.

The simplified analysis presented here is incomplete, in that
we have retained only instantaneous s-wave interactions.   Realistic
interactions,
such as those generated by gluon exchange or instantons, are momentum- and
frequency-dependent.  Nevertheless the simplified analysis is valuable, in
that it indicates the internal
quantum numbers of the channels likely to be most unstable.  Moreover,
other things being equal, s-wave
instabilities seem likely to dominate, since they add constructively over
the fermi surface.   (This intuition
is borne out by variational calculations 
calculations of gaps and pairing energies associated
with higher angular momentum condensates,
which turn out to be very small.\cite{ARW1,Schaefer1Flavor,BowersLOFF})

\subsection{Model Hamiltonian by Variational Methods}

The renormalization group determines the running
of couplings from the matching point downwards, towards the Fermi surface.
The catastrophic growth of an attractive interaction we have
encountered indicates an
instability of the system, 
indicates
in what channel(s) the instability is most pronounced, and
indicates the energy scale at which the instability develops.
At the scale at which the instability sets in, however, 
the perturbative renormalization procedure breaks down.
The great
achievement of Bardeen, Cooper, and Schrieffer (BCS)\cite{BCS}
was to demonstrate how
Fermi-surface instabilities
of this type are resolved by pairing, to produce a superconducting ground
state.  This ground state is characterized by a gap $\Delta$,
whose magnitude does indeed turn out to be of the same order
as the energy scale indicated by the renormalization group 
analysis.  
The instability arose in a renormalization
group analysis done in terms
of fermionic excitations (about the naive ground state with no
pairing) which have zero free energy at the Fermi surface.
In the gapped state, which the BCS analysis determines is
formed,
the resolution of the instability
is simply that the correct fermionic degrees of freedom are 
quasiparticle excitations which have a gap in their spectrum.
In this section we outline the BCS argument that the resolution
of the instability is a paired state, and give the associated
calculation of $\Delta$.  

As an aside, note that once
$\Delta$ is known and the ground state has been characterized,
if (as in the CFL phase) there are bosonic excitations 
of this ground state with energy less than $\Delta$, a renormalization
group analysis framed in terms of these new degrees of freedom
can be employed in their description. The results of such analyses were
presented in
Section 2.3.

The original BCS theory was founded on a variational calculation using a
model Hamiltonian.  Their model
Hamiltonian retained only an idealized form of the effective attraction
near the Fermi surface, that
encouraged condensation in the s-wave, spin-singlet channel.
The pioneering work of Nambu and Jona-Lasinio,\cite{OriginalNJL} 
who first applied the ideas
behind BCS theory to particle--anti-particle pairing in 
quantum field theories describing elementary particle interactions,
followed a similar approach.  
We shall follow the literature in referring to
models of this class
as NJL models, but shall apply them \`a la BCS, to quark-quark 
pairing.

In QCD there are many more fermion species (colors and
flavors) than in conventional superconductors, and the form of the
condensate is a major focus of
interest.   Renormalization
group calculations, as presented in 
Section 4.1, provide important
guidance.  The
dominant operator indicates the existence of potential instabilities that
are amplified by renormalization,
but does not indicate  how they are resolved.  
To decide the most
favorable form of pairing, we
must compare the energies associated with different possibilities.  For
this qualitative purpose, and of
course to estimate the quantitative relationship between the size of the
pairing gap and the microscopic
parameters, we still require detailed dynamical calculations.  
NJL models provide
an appropriate, tractable starting point.\cite{ARW1,RappETC}
One chooses a simple four-quark interaction,  
which provides attraction in the channel which
the renormalization group analysis indicates is dominant.
One first normalizes the parameters in such a toy model to give
reasonable vacuum physics, and then uses the 
model to estimate the gap.
Here we shall exemplify the use of such models
by using them to estimate the size of the gaps in
the 2SC and CFL phases and to argue that the CFL condensate
is indeed the favored pattern of condensation in three-flavor
QCD.  

We use an NJL model
with free energy 
\begin{equation}\label{freeenergy}
\Omega = \int d^3x\, \psib(x) (
\nabla\!\!\!\!/ - \mu\ga_0) \psi(x) + H_I\ ,
\end{equation}
where the interaction Hamiltonian 
\begin{equation}\label{Hinteraction}
H_I = \frac{3}{8} G \int d^3x\, \FF\,
 \left(\psib(x) \ga_\mu T^A \psi(x)\right)\, \left(\psib(x) 
\ga^\mu T^A \psi(x)\right)\ 
\end{equation}
describes a four-fermion interaction
with the color, flavor, and spinor structure of single-gluon exchange.

In microscopic QCD the interactions become weak at high momentum.  Here
this feature is
caricatured with a form factor $\FF$.  In detail, when we expand $H_I$ in
momentum modes
we are instructed to include
a form factor $F(p)$ on each leg of the interaction vertex.
Examples which have been used in the literature include
power-law and smoothed-step 
profiles for $F$:~\cite{ARW1,CFL,ABR2+1}
\beq\label{ham:F}
F(p) = \Bigl( {\Lambda^2\over p^2+\La^2} \Bigr)^\nu,\qquad
{\rm or}\quad
F(p)= \dsp \left( 1 + \exp \Bigl[  {p-\La \over w} \Bigr] \right)^{-1}.
\eeq
One of the important results found in these investigations is
that if, for any given $F(p)$, the coupling constant $G$ is chosen
to yield a reasonable vacuum chiral condensate (say with
a constituent quark mass of 400 MeV) then the resulting
superconducting gap is reasonably insensitive to the choice
of $F(p)$.  
Varying the scale
$\Lambda$ and the shape of the form factor while tuning the coupling to
keep some physical property of the vacuum
fixed affects the gap very little.
This robustness indicates that the magnitude
of the gap is largely determined just by the strength of the interaction
(normalized with respect to a physical observable) and does
not depend sensitively on other details.
For this reason, and following Ref.~13,49  we
make the simplifying choice of replacing the smooth form 
factor $F(p)$ by a step function: $F(p)=1$ for $p<\Lambda$
and $F(p)=0$ for $p>\Lambda$.  

We begin with two flavors and follow the analysis
presented in 
Refs.~5,49. We seek a condensate of the 2SC form
by 
making the BCS ansatz 
\beq\label{BCS:ansatz}
\begin{array}{rcl}
|\Psi\> &=& A^\ad_L A^\ad_R |0\>\,,\\[2ex]
A^\ad_L &=& 
  \dsp\prod_{\vp, \alpha, \beta} 
    \left( \cos \theta_L(\vp) 
    + \epsilon^{\alpha\beta3} \e^{i \xi_L(\vp)} 
    \sin \theta_L(\vp)\, a^\ad_{Lu\alpha}(\vp) 
    \, a^\ad_{Ld\beta}(-\!\vp) \right)\,, 
   \\[4ex]
A^\ad_R &=& \mbox{as above,~} L \to R\,,
\end{array}
\eeq
for the ground state wave function. Here, $\alpha$ and $\beta$ are color
indices, $u$ and $d$ label flavors explicitly, and $a^\ad$ 
is the particle creation operator 
(for example, $a^\ad_{Ld\al}$ creates a left-handed down quark with color
$\al$).  The Cooper pairs described by this ansatz are
evidently antisymmetric in color; that they are also antisymmetric
in flavor follows from the anticommutation relations satisfied
by the creation operators.
The $\theta$'s and $\xi$'s are
the variational parameters of our ansatz: they are 
to be chosen to minimize
the free energy of this BCS state.  
Note that we really should have included pairing among antiparticles
in our ansatz also. However, this doubles the length of the
equations and makes little difference in the end.  We therefore
leave the antiparticles out until the end of the derivation, and restore
them in the gap equation itself.
Note also that $|0\>$ is the no-particle state.  At nonzero chemical
potential, in the absence of any interactions, the ansatz
(\ref{BCS:ansatz}) describes filled Fermi seas if 
$\theta_L(\vp)=\theta_R(\vp)=\pi/2$ for $|\vp|<\mu$, and 
$\theta_L(\vp)=\theta_R(\vp)=0$ otherwise.  Throughout this
section, we assume $\mu$ is the same for all flavors of quarks.
The consequences of relaxing this assumption are described
in Ref.~49 and sketched 
in Section 6.

Explicit computation demonstrates that the condensate
\beq\label{BCS:cond}
\Gamma_L \equiv -\frac{1}{2}\< \Psi | \epsilon_{ij} \epsilon_{\alpha \beta 3} 
  \,\psi^{i\alpha}({\bf r}) C L \,\psi^{j\beta}({\bf r}) 
  | \Psi\> 
\eeq
is nonvanishing. Here, $C = i \gamma^0 \gamma^2$ and $L =
(1-\gamma_5)/2$ is the usual left-handed projection operator.
$\Gamma_L$ can be expressed in terms of the variational parameters 
as:
\beq\label{BCS:gammas}
\Gamma_L =  \dsp\frac{4}{V} \dsp\sum_{\vp} 
  \sin \theta_L(\vp) \cos \theta_L(\vp) \e^{i(\xi_L(\vp)-\phi(\vp))} \ .
\eeq
Here $V$ is the spatial
volume of the system and
the dependence on the azimuthal angle $\phi$ follows from
our use of the spinor conventions described in 
Refs.~5,7,79.  
The expression for $\Gamma_R$ is the same
as that in (\ref{BCS:gammas}) except
that $\phi(\vp)$ is replaced by $\pi-\phi(\vp)$.  
In Eq.~(\ref{BCS:gammas})
and throughout, $(1/V)\sum_\vp$ becomes $\int d^3p/(2\pi)^3$
in an infinite system.  Our simplified choice of form factor
corresponds to restricting the range of this integral
to $|\vp|<\Lambda$.

We must now minimize the expectation value of the free
energy (\ref{freeenergy}) in the state $|\Psi\rangle$:
\begin{equation}
\langle \Psi|\Omega|\Psi \rangle 
= 4\dsp\sum_{\vp}\left(|\vp|-\mu\right)\sin^2 \theta_L(\vp) 
+ 4\dsp\sum_{\vp}\left(|\vp|-\mu\right)\sin^2 \theta_L(\vp)
+\langle H_I \rangle
\end{equation}
where
\begin{equation}\label{gap:HI}
\langle H_I \rangle = -\frac{GV}{2}\left(|\Gamma_L|^2+|\Gamma_R|^2\right)\ 
\end{equation}
in terms of the condensates which are in turn 
specified in terms of the variational
parameters by (\ref{BCS:gammas}).  We
have neglected the contribution to $\langle\Omega\rangle$ made 
by the third color quarks, which
do not pair.

Upon variation with respect to 
the $\xi$'s, we find that they must be chosen
to cancel the azimuthal phases $\phi(\vp)$ in 
(\ref{BCS:gammas}). 
In this way, we obtain maximum coherence
in the sums over $\vp$, giving the largest possible magnitudes for 
the condensates and gap parameters.  We have
\beq\label{gap:phases}
\xi_L(\vp) = \phi(\vp) + \varphi_L\,, \hspace{0.3in} 
\xi_R(\vp) = \pi-\phi(\vp) + \varphi_R\,,
\eeq
where $\varphi_L$ and $\varphi_R$ are arbitrary $\vp$-independent
angles.  
These constant phases do not affect the free 
energy --- they correspond to the Goldstone bosons for the
broken left-handed and right-handed baryon number symmetries --- and
are therefore not constrained by the variational procedure.  
For convenience, we set $\varphi_L = \varphi_R = 0$ 
and obtain condensates
and gap parameters that are purely real.  

The relative phase $\varphi_L - \varphi_R$ 
determines how the condensate transforms under a parity
transformation. Its value determines whether
the condensate is scalar, pseudoscalar,
or an arbitrary combination of the two.
Because single gluon exchange cannot change the handedness of a
massless quark, the left- and right-handed 
condensates are not coupled in the 
free energy $\Omega$.
Our choice of 
interaction Hamiltonian
therefore allows an arbitrary choice of $\varphi_L - \varphi_R$.
A global $U(1)_A$ transformation changes $\varphi_L - \varphi_R$,
and indeed this is a symmetry of our toy model.
If we included $U(1)_A$-breaking interactions in
our Hamiltonian, to obtain a more complete description of QCD,
we would find that the free energy depends
on $\varphi_L - \varphi_R$, and thus selects a 
preferred value. For example, had we taken $H_I$ to be the
two-flavor instanton interaction as in Refs.~5,6, 
the interaction energy would appear as 
$\Gamma_L^{*}\Gamma_R + \Gamma_L \Gamma_R^{*}$ instead
of as in (\ref{gap:HI}). This would
enforce a fixed phase relation $\varphi_L-\varphi_R=0$,
favoring the parity conserving $\langle\psi C\gamma_5 \psi\rangle$
condensate.\cite{ARW1,RappETC}

We now apply the variational method to determine the angles
$\theta(\vp)$ in our trial wavefunction, by minimizing the free
energy: $\partial \< \Omega\> / \partial \theta(\vp) = 0$.
Everything is
the same for left and right condensates so we hereafter drop the $L$
and $R$ labels.  Upon variation with respect to $\theta(\vp)$,
we obtain
\beq\label{gap:tan2th}
\tan 2\theta(\vp) = \frac{ \Delta}{|\vp|- \mu}\,,
\eeq
where $\Delta_{L,R}=G\Gamma_{L,R}$. With the $\theta$ angles
now expressed in terms of $\Delta$, we can 
use the wave function $|\Psi\rangle$ to obtain expressions
for the quasiparticle dispersion relation:
\begin{equation}
E(\vp)=\sqrt{\left(|\vp|-\mu\right)^2+\Delta^2}\ .
\end{equation}
Here, $E$ is the free energy cost of removing a pair and 
replacing it with either and up quark with momentum $\vp$
or a down quark with momentum $-\vp$.  This result
confirms that $\Delta$ is the gap in the spectrum
of fermionic excitations.  Note, however, that in the 2SC
phase the third color (``blue'') quarks remain gapless in the
ansatz within which we are working. As mentioned in
Section 3, it is likely that they form an angular momentum $J=1$
condensate with a gap which is many orders of magnitude
smaller than the gap $\Delta$ for the red and green quarks.

Substituting the expression (\ref{gap:tan2th}) for the $\theta$ angles 
into the expression (\ref{BCS:gammas}) for the $\Gamma$ condensate,
and using the relation $\Delta = G\Gamma$, we 
obtain a self-consistency equation for $\Delta$:
\beq\label{BCSgap}
1 = \frac{2G}{V}\sum_\vp \left\{\frac{1}{\sqrt{(|\vp|-\mu)^2+\Delta^2}}
+ \frac{1}{\sqrt{(|\vp|+\mu)^2+\Delta^2}}
\right\}\ .
\eeq
The first term on the right-hand side of the gap equation (\ref{BCSgap})
yields a logarithmic
divergence at the Fermi surface if $\Delta$ is small.
This term 
is the contribution to the gap equation from particles
and holes, and the logarithmic divergence is the manifestation
of the BCS instability. Because the right hand side diverges for
$\Delta\rightarrow 0$, there must be a solution to the gap equation
with $\Delta\neq 0$ even for an arbitrarily small coupling $G$, 
as long as $G$ is positive (attractive).
The second term, with $(|\vp|+\mu)$ in the denominator,
is the contribution from antiparticles, which
we have suppressed above and restored here.  Note that even though
the same value of $\Delta$ appears in both the antiparticle and 
particle/hole contributions, the antiparticle contribution to
the gap equation is small because its denominator is everywhere
greater than $\mu$.  

We shall rederive the gap equation (\ref{BCSgap}) diagrammatically
in Section 4.3. The variational method we have used above
has the virtue of providing us in addition with an expression for
the corresponding wave function itself.

In order to use this gap equation
to estimate $\Delta$, we need to fix the coupling $G$.
To do this, we
evaluate the vacuum chiral gap (or constituent quark mass) and
set it to $M=400$~MeV.
In vacuum, $M$ satisfies the gap equation
\beq\label{Ch:gap2}
1=\frac{8G}{V}\sum_\vp {1\over\sqrt{|\vp|^2 + M^2}}\ ,
\eeq
which may be derived along very similar lines to the above,
working at zero chemical potential and postulating
a chiral symmetry breaking condensate in which quarks
and anti-quarks pair in the standard fashion.  
Note that for $M\rightarrow 0$, the right-hand side of the chiral
gap equation has no logarithmic divergence; no BCS Fermi surface
instability.  This means that $M\neq 0$ only if $G$ is above
some threshold.  Let us choose
$\Lambda=800$~MeV and fix $G$ such that $M=400$~MeV.
Upon solving the BCS gap equation (\ref{BCSgap}) 
we then find that $\Delta=106$~MeV when $\mu=400$~MeV
and $\Delta=145$~MeV when $\mu=500$~MeV.

Note that if, instead, we had used the four-fermion interaction with
the quantum numbers of the instanton vertex, as in Ref.~5,
we would have obtained {\it precisely} the same  results.
That is, had we
made the same choice for the form factor and for 
$\Lambda$, and then chosen the
coupling constant for the instanton vertex such that 
the vacuum constituent quark mass is 400 MeV, 
it turns out that the resulting $\Delta$ is exactly
that which 
we have obtained using the single-gluon exchange interaction.

Varying $\Lambda$, or using a smooth form factor instead
of a sharp cutoff, yields similar results for $\Delta$,
ranging from about 50~MeV to somewhat more than 100~MeV.\cite{ARW1,bergesraj}
The authors of Ref.~6 also use the instanton
interaction, and also normalize it such that $M=400$~MeV in vacuum, but they
model the fact that instanton effects decrease at
nonzero density by introducing a density-dependent
coupling constant. They find gaps ranging from 20~MeV
to 90~MeV.  More sophisticated treatments of the
instanton interaction, including form factors obtained
from suitable Fourier transforms of instanton profiles,
tend to yield larger gaps, perhaps
as large as 200 MeV.\cite{CarterDiakonov,RappCrystal}

The models we are discussing in
this subsection can only be used for qualitative guidance,
but it is very pleasing to see so many of them agreeing
on the order of magnitude of the gap.

The 2SC phase has also been studied at nonzero temperature 
using the instanton interaction.\cite{RappETC,bergesraj}
The critical temperature above which the condensate 
vanishes (in mean field theory) turns out to be within
a few percent of that expected from BCS theory: 
$T_c=0.57 \Delta(T=0)$.\cite{bergesraj}
The standard BCS result
is valid in the $\Delta/\mu\rightarrow 0$ limit, and the
interaction in QCD is strong enough that $\Delta/\mu\sim 1/4$
at the densities we are discussing, so a few percent discrepancy
is to be expected.  Note that based on our estimate
of the gap, the critical temperature we estimate
for the color superconducting phase is of order many tens of MeV.
This makes it almost certain that this phase is inaccessible
in heavy ion collisions, which are much hotter. Although
$T_c$ is low relative to heavy ion collision temperatures,
color superconductors are in three senses high temperature
superconductors. First, $T_c\sim 10^{12}$ Kelvin. Second, 
and more seriously, $T_c$ can easily be
of order $10\%$ of the Fermi energy.
This makes these materials high temperature superconductors by any
definition, even relative to the cuprates.  This robustness
of the phenomenon of color superconductivity in the face
of nonzero temperature directly reflects the strength
of the attraction between quarks in QCD. Third,
$T_c$ is much hotter than the temperature in a neutron
star which is more than a few seconds old, making it 
clear that if quark matter exists in the core of a neutron
star, it must be a color superconductor.

We turn now to QCD with three flavors, and derive a gap
equation appropriate for the CFL phase.  We use the 
same single gluon exchange interaction Hamiltonian $H_I$
as above.  This time, however, we present a derivation
which follows the ideas of Bogoliubov and Valatin.\cite{Bogoliubov}
We focus from the beginning
only on the condensate of left-handed quarks. (As above,
since we do not include any $U(1)_A$ breaking interaction
the left- and right-handed condensates can be rotated one
into the other.  Introducing instanton effects, done
in the CFL phase in Refs.~9,10, favors
the Lorentz scalar combination.)

We rewrite the interaction Hamiltonian in terms
of Weyl spinors, keeping only the left-handed fields:
\beq\label{HI}
H_I = \frac{G}{2}\int d^3x\, \FF \,
(3 \de^\al_\de \de^\ga_\be - \de^\al_\be \de^\ga_\de)
(\psid^i_{\al\da}\psid^j_{\ga\dc} \eps^{\da\dc}
 \psi^\be_{ib}\psi^\de_{jd} \eps^{bd})\,,
\eeq
We have explicitly displayed color ($\al,\be\ldots$), flavor ($i,j\ldots$),
and spinor ($a,\da\ldots$) indices, and rewritten the color
and flavor generators.

We now make the mean-field ansatz
that in the true ground state $|\Psi\rangle$ at
a given chemical potential, 
\beq\label{Col:mf}
\begin{array}{rcl}
\<\Psi| \psid^i_{\al\da}\psid^j_{\ga\dc} \eps^{\da\dc} |\Psi \>
 &=&
\dsp {1\over G} P^i_\al{}^j_\ga\,, \\[3ex]
P^i_\al{}^j_\ga &=&
  \third(\De_8 + \eighth\De_1)\de^i_\al\de^j_\ga
+ \eighth\De_1 \de^i_\ga \de^j_\al\ ,
\end{array}
\eeq
where the numerical factors have been chosen so that
$\De_1$ and $\De_8$, which parameterize $P$, will turn
out to be the gaps for singlet and octet quark excitations,
respectively (see below).  This ansatz takes the color ${\bf\bar 3}$ form
(\ref{CFLcondensate}) if $\Delta_1=-2\Delta_8$.  Solutions
to the gap equation turn out to be within a few percent
of satisfying this, meaning that there is only a small admixture
of the color ${\bf 6}$ condensate.
In the mean field approximation, 
the interaction Hamiltonian becomes
\beq
\begin{array}{rcl}
H_I &=& \dsp \half \int d^3x\, \FF\, Q_\be^i{}_\de^j \,
\psi^\be_{ib}\psi^\de_{jd} \eps^{bd} + \cc\,, \\[2ex]
Q_\be^i{}_\de^j &=&
 \De_8 \de^i_\de\de^j_\be + \third(\De_1-\De_8) \de^i_\be \de^j_\de\ .
\end{array}
\eeq
Replacing indices $i,\be$ with a single color-flavor index $\rho$,
we can simultaneously diagonalize the $9\times 9$ matrices
$Q$ and $P$, and find that they have two eigenvalues,
\beq
\begin{array}{rcl@{\qquad}rcl}
P_1 &=& \De_8 + \quarter\De_1\,,  &  P_2\cdots P_9 &=& \pm \eighth\De_1\,, 
\\[1ex]
Q_1 &=& \De_1\,,  &  Q_2\cdots Q_9 &=& \pm \De_8\,. \\
\end{array}
\eeq
That is, eight of the nine quarks in the theory have
a gap parameter given by $\Delta_8$, while the remaining
linear combination of the quarks has a gap parameter $\Delta_1$.

The Hamiltonian can be rewritten
in this color-flavor basis in terms of particle/antiparticle 
creation/annihilation
operators $\ab_\rho,\bb_\rho$. (This time, we keep the
antiparticles throughout the derivation.)
We also expand in momentum modes (see Refs.~7,79 for
our spinor conventions) and
now explicitly include the
form factors $F(p)$,
\beq\label{Col:H}
\begin{array}{rcl}
H &=& \dsp\sum_{\rho,k>\mu} (k-\mu)\ab^\ad_\rho(\bk) \ab_\rho(\bk)
+ \dsp\sum_{\rho,k<\mu} (\mu-k) \ab^\ad_\rho(\bk) \ab_\rho(\bk)\\[4ex]
& &\hspace{1.5cm} +\dsp\sum_{\rho,\bk} (k+\mu) \bb^\ad_\rho(\bk) \bb_\rho(\bk) \\[4ex]
& &+ \dsp \half \sum_{\rho,\bp}
F(p)^2 Q_\rho \e^{-i\phi(\bp)}\Bigl(
   \ab_\rho(\bp) \ab_\rho(-\bp)
+  \bb^\ad_\rho(\bp) \bb^\ad _\rho(-\bp)
\Bigr) + \cc\ ,
\end{array}
\eeq
where the perturbative ground state, annihilated by $\ab_\rho$ and $\bb_\rho$
is the Fermi sea, with states up to $p_F=\mu$ occupied.
Finally, we change basis to creation/annihilation operators
$y$ and $z$ for quasiparticles,
\beq\label{Col:yz}
\begin{array}{rcl}
y_\rho(\bk) &=& \cos(\th^y_\rho(\bk))\ab_\rho(\bk) + \sin(\th^y_\rho(\bk))
\exp(i\xiy_\rho(\bk)) \ab^\ad_\rho(-\bk)\,, \\[1ex]
z_\rho(\bk) &=&  \cos(\th^z_\rho(\bk))\bb_\rho(\bk) + \sin(\th^z_\rho(\bk))
\exp(i\xiz_\rho(\bk)) \bb^\ad_\rho(-\bk)\,,
\end{array}
\eeq
where
\beq\label{Col:th}
\begin{array}{rcl@{\qquad}rcl}
\cos(2\th^y_\rho(\bk))
&=& \dsp {|k-\mu| \over \sqrt{ (k-\mu)^2 + F(k)^4 Q_\rho^2 }}\,,
& \quad \xiy_\rho(\bk) &=& \phi(\bk)+\pi\,, \\[1ex]
\cos(2\th^z_\rho(\bk)) &=&
\dsp {k+\mu \over \sqrt{ (k+\mu)^2 + F(k)^4 Q_\rho^2 }}\,,
& \quad \xiz_\rho(\bk) &=& -\phi(\bk) \,.
\end{array}
\eeq
These values are chosen so that $H$ has the form of a free Hamiltonian for
quasiparticles:
\beq\label{Col:quasiH}
\begin{array}{rl}
H = \dsp\sum_{\bk,\rho}\Biggl\{ \phantom{+} &
    \sqrt{ (k-\mu)^2 + F(k)^4 Q_\rho^2} \,\, y^\ad_\rho(\bk)y_\rho(\bk) \\
+&  \sqrt{ (k+\mu)^2 + F(k)^4 Q_\rho^2} \,\, z^\ad_\rho(\bk)z_
\rho(\bk)\Biggr\}\ .
\end{array}
\eeq
Clearly then
the ground state $|\psi\>$ contains no quasiparticles:
\beq\label{Col:vac}
y_\rho(\bk) |\psi\> = z_\rho(\bk) |\psi\> = 0\,.
\eeq

The gap equations follow from requiring that
the mean field ansatz (\ref{Col:mf}) hold in the quasiparticle basis.
In other words, we use (\ref{Col:yz}) and (\ref{Col:th}) to rewrite 
(\ref{Col:mf})
in terms of quasiparticle creation/annihilation operators, and then
evaluate the expectation value using (\ref{Col:vac}).
We get two gap equations:
\beq\label{Col:gap}
\begin{array}{rcl}
 \De_8 + \quarter \De_1  &=& \dsp G K(\De_1)\,, \\[2ex]
 \eighth \De_1 &=& \dsp G K(\De_8)\,,
\end{array}
\eeq
where
\beq
K(\Delta) = -{1\over 2}\sum_{\bk}\Biggl\{
  {F(k)^4 \Delta\over \sqrt{ (k-\mu)^2 + F(k)^4 \Delta^2}}
+ {F(k)^4 \Delta\over \sqrt{ (k+\mu)^2 + F(k)^4 \Delta^2} }\Biggr\}\ .
\eeq
{}From (\ref{Col:quasiH}) we see that the physical gap, namely the
minimum energy of the quasiparticles, is $F(\mu)^2 |\De|$.
Creating a quasiparticle-quasihole pair requires at least
twice this energy. If we take $F(p)$ to be a step function
as above, then the gaps are simply $|\Delta_1|$ and $|\Delta_8|$.

In this derivation, the
variational nature of the
procedure is somewhat hidden.   The point is that the $\theta$ angles were
chosen to guarantee the
absence of linear terms in the effective Hamiltonian.  Such terms would
indicate the instability of the
no-``particle'' state, where particles are defined by the action of the
operators (\ref{Col:yz}).   This state
varies with the $\theta$ angles.   Thus in searching for a
$\theta$-dependent Hamiltonian free of linear
terms we are in effect looking for the stationary point of the expectaton
value of the energy, within a
continuous manifold of candidate states.

For the same choice of parameters as we used in the 2SC
case (namely $\Lambda=800$ MeV, $G$ chosen so that the
vacuum constituent quark mass is $M=400$~MeV, and $\mu=400$~MeV)
solving the coupled gap equations (\ref{Col:gap}) yields
$\Delta_8=80$~MeV and $\Delta_1=-176$~MeV. As promised, this
is not far from the ratio $\Delta_1=-2\Delta_8$ required
if the condensate were entirely in the color ${\bf\bar 3}$ 
channel.  The color ${\bf 6}$ condensate is nonzero 
but small, as is  also the case in the CFL phase at asymptotically
high densities.\cite{SchaeferPatterns,ShovWij}
If $\mu$
is increased to 500 MeV, we find $\Delta_8=109$~MeV 
and $\Delta_1=-249$~MeV.  These estimates of $\Delta$
may be a bit of an overestimate, because it is
likely that the instanton interaction contributes significantly
to $M$ in vacuum but, because it is a six-fermion interaction,
it does not contribute significantly to $\Delta$. For this
reason, we are overestimating $G$ by assuming that $M$ in vacuum is due
entirely to the one-gluon exchange interaction.  (This
uncertainty did not arise with two flavors because in that 
case, the instanton and one-gluon exchange four-fermion interactions
both contribute to $M$ and to $\Delta$ in the same way.)
Reducing $G$ by a factor of two results in $\Delta_8=16$~MeV 
and $\Delta_1=-34$~MeV at $\mu=500$~MeV.  
The magnitude of the gap {\it is} sensitive to the
strength of the interaction, even though it is insensitive
to details of the form factor as we described
above. This reminds us that we
should use these models as a qualitative guide only.

In three-flavor QCD, the instanton interaction is
a six-fermion interaction.  As we have discussed in Section 2,
its effects are small but are nevertheless important
because they are the leading source of $U(1)_A$ breaking.
It is also interesting to note that the four-fermion interaction
introduced by instanton--anti-instanton pairs contributes
to both $M$ and $\Delta$. In the analysis of Ref.~9,
this interaction is employed in an analogous fashion 
to the way we have employed single-gluon exchange, and
quite similar gaps are obtained.

We can now outline the argument that the CFL phase is indeed the favored
pattern of condensation.  First, the interaction favors
antisymmetry in color.  Second, pairing with zero
angular momentum is favored because it allows condensation
which utilizes the entire Fermi surface. This has
been confirmed in a variety of explicit calculations.
Pauli then requires flavor antisymmetry.  These
arguments favor a condensate of the form
\begin{equation}
\langle\psi^\alpha_i C \gamma_5 \psi^\beta_j\rangle 
\sim \epsilon^{\alpha\beta A}\epsilon_{ijB}\equiv \phi^A_B\ .
\end{equation}
Color-flavor locking corresponds to $\phi^A_B\sim {\rm diag}(1,1,1)$,
in which all nine quarks pair. The antithetical possibility
is $\phi^A_B\sim {\rm diag}(0,0,1)$, but this is just the 2SC phase,
whose gap equation we have also solved! For $\mu=400$~MeV and
$G$ fixed so that $M=400$~MeV in vacuum,
we found that in the CFL phase (with $\phi^A_B\sim {\rm diag}(1,1,1)$)
there are 8 quasiparticles with gap 
80 MeV and one with gap 176 MeV.  For $\phi^A_B\sim {\rm diag}(0,0,1)$,
on the other hand, we find four quasiparticles with 
gap 106 MeV and 5 quarks (strange quarks; blue up and blue down quarks)
with zero gap.  Clearly, the condensation energy (i.e.
$\langle \Omega \rangle$ relative to that in the unpaired state)
will be
greater with $\phi^A_B\sim {\rm diag}(1,1,1)$. Explicit evaluation
of $\langle \Psi|\Omega|\Psi \rangle$ using the variational
wave functions for the 2SC and CFL phases confirms this.

These calculations are strongly  
suggestive that
$\phi^A_B\sim {\rm diag}(1,1,1)$ is the 
choice with the lowest free energy, although they do
also demonstrate that a small admixture of the color $\bf 6$
condensate which is symmetric in both color and flavor is
mandatory. This means that a complete demonstration
is more complicated.  Sch\"afer has, however, constructed
an effective potential for $\phi^A_B$ (and the color ${\bf 6}$
condensate) within a particular NJL model but making
the most general assumptions for the form of the condensate. 
He finds that the CFL phase, with 
$\phi^A_B\sim {\rm diag}(1,1,1)$, minimizes $\langle \Omega \rangle$.
The authors of Refs.~62,12 have addressed this question 
in analyses done using 
the interaction generated by the exchange of a propagating
gluon, rather than in a model with point-like interactions.
They confirm that $\phi^A_B\sim {\rm diag}(1,1,1)$ is favored,
thus explicitly confirming that the CFL phase is favored
at arbitrarily high density. The essence of all of these
different analyses is that in the CFL phase all quarks are
able to pair, and the condensation energy is therefore
larger than that for any less symmetric condensate.
This simple argument could in principle be outweighed
if a less symmetric condensate were found
whose gaps were much larger in magnitude than those
of the CFL phase.  With all parameters fixed, though,
different choices for  $\phi^A_B$ turn out to yield comparable
gaps. Although $\phi^A_B\sim {\rm diag}(0,0,1)$ does yield
somewhat larger gaps, this advantage is not close to outweighing
the $9/4$ numerical advantage enjoyed by the CFL phase.

\subsection{Model Hamiltonian by Diagrammatic Methods}

In this brief subsection, we use diagrammatic
methods, sometimes named after Dyson and Schwinger or,
in this context, Nambu and Gorkov, to
rederive the gap
equation (\ref{BCSgap}) for the 2SC phase, still in
a model in which the QCD interaction between
quarks has been replaced by a point-like four-fermion
interaction.
We do so both because these methods may be more familiar
to some readers and because they serve as a warmup
for the weak-coupling calculations of the next subsection.
To that end, we also discuss the $G\rightarrow 0$ limit
of the model.

We follow the standard Nambu-Gorkov formalism and introduce
an eight-component field $(\psi,\bar\psi^T)$.  In this
basis, the inverse quark propagator takes the form
\beq
\label{sinv}
S^{-1}(p) = \left(\begin{array}{cc}
 \pslash+\mu \gamma_0 &  \bar\Delta \\
 \Delta  & (\pslash-\mu\gamma_0)^T
\end{array}\right),
\eeq
where $\bar \Delta=\gamma_0\Delta^\dagger \gamma_0$ and
where in this expression $\Delta$ is a matrix with color, flavor
and Dirac indices which have all been suppressed.  The diagonal
blocks correspond to ordinary propagation of massless quarks
and the off-diagonal
blocks reflect the possibility for ``anomalous propagation''
in the presence of a diquark condensate. 
The 2SC ansatz
for the form of the gap matrix is
\beq\label{2SCansatz}
\Delta_{ij}^{\alpha\beta}(p)=\epsilon^{\alpha\beta 3}\epsilon_{ij}
C\gamma_5 \Delta\ ,
\eeq
where $\Delta$ on the right-hand side is now the gap
parameter, and all matrix
structure has been written explicitly.

The full propagator of the interacting fermion system
determines the proper self energy $\Sigma$ via
\begin{equation}\label{properselfenergy}
S=\left( S_0^{-1} +\Sigma \right)^{-1}
\end{equation}
where $S_0^{-1}$ is the inverse propagator in the
absence of any interactions (given by 
(\ref{sinv}) with $\Delta=0$).  
The gap $\Delta$ is then determined
by solving a self-consistent Dyson-Schwinger 
equation for $\Sigma$.  Assuming a four-fermion
interaction, 
this is given schematically by
\begin{center}
$\displaystyle{{ \Sigma} \quad = \quad }$
\parbox{1.in}{
\epsfxsize=0.8in
\epsffile{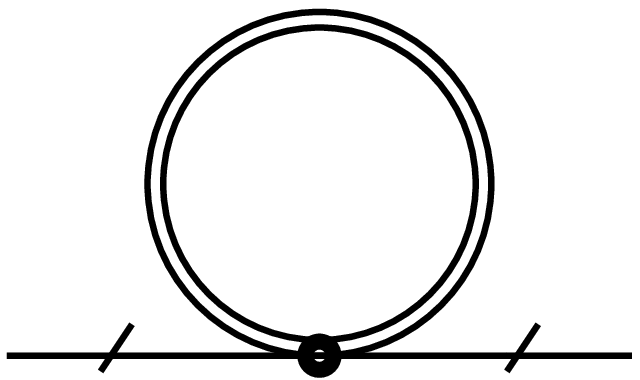}}\vspace{0.1in}
\end{center}
where the loop denotes a momentum integration
over the full propagator $S$ and where external
legs have been amputated. Note that this equation
sums an infinite class of diagrams, because $\Sigma$
occurs within the propagator on the right-hand side.
It is nevertheless (for obvious reasons) called
the one-loop Dyson-Schwinger equation.  As we
shall see, it is precisely equivalent to the 
variational approximation we described in Section 4.2,
in which we wrote an ansatz which admitted pairing
but neglected other correlations.  It is also
often called a mean-field approximation. The
Bogoliubov-Valatin derivation, also presented
in Section 4.2 and also completely equivalent, makes this
nomenclature clear.

Using the four-fermion interaction (\ref{Hinteraction})
of Section 4.2 which models single-gluon exchange, 
the Dyson-Schwinger equation is given explicitly by
\begin{equation}
\Sigma=-6G\int \frac{d^4p}{(2\pi)^4} \Gamma^A_\mu S(p) \Gamma^{A\mu}
\end{equation}
where $\Gamma^A_\mu$ is the quark-gluon vertex
\beq
\label{vert_0}
\Gamma^A_\mu = \left(\begin{array}{cc}
 \gamma_\mu\lambda^A/2 & 0 \\
 0 & -(\gamma_\mu\lambda^A/2)^T \end{array}\right)\ ,
\eeq
and where we have chosen to work in Euclidean space.
After some algebra (essentially the
determination of $S$ given the $S^{-1}$ specified
above), and upon suitable projection, this matrix
equation reduces to a gap equation for the gap parameter $\Delta$
given by
\begin{equation}\label{d4pBCSgap}
\Delta=4G\int \frac{d^4 p}{(2\pi)^4}\left(\frac{\Delta}{p_0^2+
(|\vec p|-\mu)^2 +\Delta^2}+\frac{\Delta}{p_0^2+
(|\vec p|+\mu)^2 +\Delta^2}\right)\ .
\end{equation}
Whether the $p_0$ integration contour is closed in the upper-half
$p_0$-plane or in the lower-half plane, one pole from
each of the two terms on the right-hand side contributes
and one obtains the 2SC gap equation (\ref{BCSgap}).
The second term on the right-hand side of  
(\ref{d4pBCSgap})  describes the contribution of antiparticles
to the gap equation. The first term, due to particles
and holes, yields a logarithmic divergence in the $|\vec p|$
integral if $\Delta\rightarrow 0$. As we have seen, this
divergence at the Fermi surface is the
hallmark of the BCS phenomenon.   

In the variational approach, once we have solved the gap
equation we know the variational wave function and with
this in hand we can evaluate the expectation
value of the free energy $\Omega$.  If, instead, we have
obtained the gap equation diagrammatically, 
we can still obtain
the free energy upon realizing that the gap equation
can be seen as the statement that the free energy
is stationary with respect to the gap parameter.
That is, we integrate the gap equation:
\begin{equation}\label{Omegafromgapeq}
\Omega = \Omega_{\rm free} + V\int_0^{\Delta_{\rm solution}} d\Delta 
\left(-\frac{2 \Delta}{G} + 8\int \frac{d^4p}{(2\pi)^4} 
{\rm ~integrand}\right)\,,
\end{equation}
where the ``integrand'' is that on the 
right hand side of the gap equation (\ref{d4pBCSgap}), 
where $\Delta_{\rm solution}$ is the value
of $\Delta$ which solves the gap equation, and where $\Omega_{\rm free}$
is that for noninteracting fermions.

Our next goal is the study of color superconductivity
in the limit of asymptotically high density and thus, because
of asymptotic freedom, at weak coupling.   It goes against
the spirit of the phenomenologically motivated models which
we have used in Sections 4.2 and 4.3 to apply them
at these densities. Our  approach has been to 
use such models to investigate
how the quarks resolve the instability revealed by the
renormalization group analysis of Section 4.1 (they pair)
and then to use the models as devices which allow us to
make a phenomenological normalization
of the strength of the attraction between quarks,
and thus to obtain a qualitative measure of the magnitude
of the gap.  At asymptotically high densities, instead of
attempting to normalize the interaction via vacuum phenomenology,
we should use the weak-coupling QCD interaction itself. This
is the subject of Section 4.4. Before beginning, and
in order to provide a contrast,
let us see what does happen if we take 
the four-fermion coupling $G\rightarrow 0$, and thus
take $\Delta\ll \mu,\Lambda$.
Solving the gap equation (\ref{BCSgap}) explicitly in 
this limit yields
\beq\label{weakcouplingresult}
\Delta=2\sqrt{\Lambda^2-\mu^2}\exp\left(\frac{\Lambda^2-3\mu^2}{2\mu^2}\right)
\exp\left(-\frac{\pi^2}{2\mu^2 G}\right)\ .
\eeq
As expected from BCS theory, 
the factor $2\mu^2/\pi^2$ which multiplies $G$ in the exponent 
is the density of states (for red and green $u$ and $d$
quarks of a single chirality) at the Fermi surface
in the absence of interactions. The $\Lambda$-dependent prefactor
is clearly model-dependent.
However, since the four-fermion interaction with coupling $G$  
models single gluon exchange,
the result (\ref{weakcouplingresult}) 
could suggest that for $\mu\rightarrow\infty$ in QCD, we
may find $\Delta\sim\exp(-{\rm const}/g^2)$.  We shall
see in the next section that this is {\it not} correct.

The free energy can be obtained explicitly
in the $G\rightarrow 0$ limit from (\ref{Omegafromgapeq}),
and, after using (\ref{weakcouplingresult}) to eliminate $G$,
turns out to be
\beq\label{weakcouplingfreeenergy}
\frac{\Omega-\Omega_{\rm free}}{V} = -\frac{1}{\pi^2}\mu^2 \Delta^2\ .
\eeq
It is nice that $G$ and $\Lambda$, which pertain to a specific
model, have disappeared.  Only the physical observable $\Delta$
remains.  It is reasonable 
to expect that in weak-coupling QCD, once $\Delta$
has been determined, the result (\ref{weakcouplingfreeenergy}) is --- except
perhaps for the prefactor --- a good estimate of the 
asymptotic condensation energy. Indeed, this can be argued
for by dimensional analysis.  In a similar spirit, we expect
that the conclusions we have drawn in the previous section
for what condensation pattern is favored in two- and three-flavor
QCD are correct at asymptotic densities.  
Both these expectations are borne out.\cite{EHHS} What remains
to be determined is the asymptotic, weak-coupling, 
value of $\Delta$.  Given its
model dependence, it is not surprising that 
the $G\rightarrow 0$ result (\ref{weakcouplingresult}) for $\Delta$
fails to describe QCD at asymptotic densities; 
what was unexpected until the work of Son\cite{Son}
was that it fails qualitatively.

\subsection{Asymptotic Analysis}

Using diagrammatic methods, we now address the weak-coupling behavior 
of the gap $\Delta$ in QCD, which we expect will correctly
reproduce the behavior of QCD at asymptotically high densities.
We replace the point-like caricature of single-gluon exchange
by single-gluon exchange itself. 
When retardation or relativistic effects are important, a Hamiltonian
treatment  (as in Section 4.2)
is no longer appropriate.  One must pass to Lagrangian and
diagrammatic methods, as in Section 4.3.
The Dyson-Schwinger equation becomes 
\begin{center}
\vspace{0.15in}
$\displaystyle{{ \Sigma} \quad = \quad }$
\parbox{1.in}{
\epsfxsize=1.0in
\epsffile{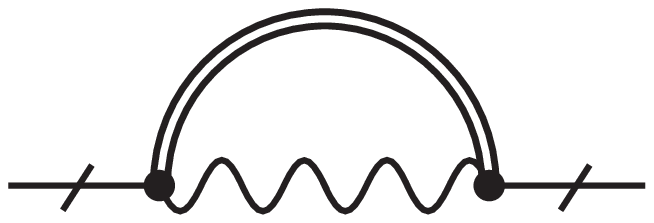}}\ .\vspace{0.15in}
\end{center}
A proper discussion of the fully microscopic 
calculation\cite{Son,PisarskiRischke,Hong,HMSW,SW3,rockefeller,Hsu2,BBS,RajagopalShuster}
is necessarily quite
technical.  Several subtleties arise, fundamentally
because (as we have already seen in previous sections)
ordinary perturbation theory is based on an unstable
and qualitatively incorrect ground state and also because
even if the 
ground state were valid there
would be many charged low-energy excitations whose response produces
electric screening and magnetic
Landau damping, which drastically affect the nature of the interaction, and
must be incorporated from the start.
Before launching in,
we present a schematic calculation, that conveys the spirit of the
thing --- and one of the most striking results --- in easily digestible form.

Upon replacing gluon exchange by a contact interaction,
we have already seen that we obtain a gap equation of the form
\begin{equation}
\Delta \propto  g^2 \int d\epsilon {\Delta \over \sqrt {\epsilon^2 +
\Delta^2}}\ .
\end{equation}
where $\epsilon=(|\vec q|-\mu)$ is the distance to the Fermi surface.
The integral           
diverges at small $\epsilon$, so that as long as the proportionality
constant is positive one will have non-trivial solutions for $\Delta$,
no matter how small is $g$.  Indeed, for small $g$ the solution
is $\Delta \sim e^{-{\rm const}/g^2}$.

At asymptotically large $\mu$, models with short-range interactions
are bound to fail because the dominant interaction is
that due to the long-range magnetic interaction coming from
single gluon exchange.\cite{PisarskiRischke1OPT,Son}  
When we now
restore the gluon propagator, we find a nontrivial angular
integral which diverges logarithmically for forward scattering in the absence
of any mechanism which screens the long-range magnetic interaction.
We suspect, however, that in the superconducting phase
this divergence will at the very least be precluded by
the generation of a gluon mass $\propto \Delta$
by the Meissner-Higgs mechanism.\footnote{The Meissner
mass is of order 
$g\mu$,\protect\cite{SonStephMesons,TwoFlavorMeissner1,TwoFlavorMeissner2,RischkeMeissner,CasalbuoniGattoNardulli} but
the gluon wave function renormalization is of order
$g\mu/\Delta$ and the physical screening length, which is the ratio
of the two,  
is therefore of order $1/\Delta$.\cite{CasalbuoniGattoNardulli}}
We therefore expect a gap
equation of the form
\begin{equation}\label{wrongequation}
\Delta \propto g^2 \int d\epsilon {\Delta \over \sqrt {\epsilon^2 +
\Delta^2}} d\theta {\mu^2 \over \theta\mu^2 + \Delta^2 }\ ,
\end{equation}
where $\theta$ is the angle between the external momentum
and the loop momentum and where
the integral is dominated by small $\epsilon$
and small $\theta$.  
This gap equation
yields $\Delta\sim g^2\Delta (\log \Delta)^2$
and hence $\Delta \sim e^{-{\rm const}/g}$!  

The proper discussion of the microscopic
gap equation is considerably more involved than (\ref{wrongequation}).
For example, it turns out that the collinear logarithmic
divergence is cutoff by dynamical screening (Landau damping) 
and not by the Meissner
effect.\cite{Son}  As a result, the prefactor of $1/g$ in the exponent
turns out to differ from that which one would derive from 
(\ref{wrongequation}). 
It nevertheless turns out that 
the conclusion that at weak coupling the gap goes
exponentially in the inverse coupling (rather than its square) 
still emerges.\cite{Son}  As a result, the gap is parametrically
larger at $\mu\rightarrow\infty$ than it would be for any point-like
four-fermion interaction. 
This has the amusing consequence, that at
asymptotically high densities the gap becomes arbitrarily large!
This is because asymptotic freedom ensures that it is the microscopic
coupling $1/g(\mu)^2$ which vanishes logarithmically, so that
$e^{-{\rm const}/g(\mu ) }$ does not shrink as fast as $1/\mu$.  Since
the ``dimensional analysis'' scale of the gap is set by $\mu$,
$\Delta$ grows asymptotically even though $\Delta/\mu$ shrinks.

We shall discuss the asymptotic analysis by example,
focussing on the 2SC phase and following
the presentation of Ref.~64.  
After describing the results, we shall
quote the analogous results for the CFL phase.  We use
the 2SC phase as our example only 
because the algebra is simpler with only two flavors. 
The physics is under much better control in the CFL phase.
In the calculation we outline, we determine the dominant gap
in the 2SC phase, describing the pairing of red and green
$u$ and $d$ quarks.  This leaves
the blue and strange quarks
unpaired, and thus leaves open the possibility of further
infrared divergences at the Fermi surface, leading to small
gaps via the formation of Cooper pairs with nonzero angular momentum.
Also, three gluons are left unscreened raising the spectre of
further infrared complications.
In the CFL phase, the dominant condensate gives a gap
to all nine quarks and gives a mass to all
eight gluons, and there are no further infrared
difficulties to be resolved. 

As in Section 4.3, we introduce an eight-component field   
$(\psi,\bar\psi^T)$ and  the inverse quark propagator (\ref{sinv}).
We assume that the gap matrix
is anti-symmetric in both color and flavor, which is the
channel in which single-gluon exchange is attractive.
We also assume that the condensation is in the channel
with total angular momentum $J=0$.   
In the case of short range interactions, all these 
assumptions can be justified by reference to the
renormalization group equations for a general four-fermion
interaction,\cite{Hsu1,SW0} as described in Section 4.1.
The single-gluon exchange
interaction is long range, however,
and other forms of pairing
might take place in addition.  In particular, since the interaction
is dominated by almost collinear scattering, it includes substantial
strength in all
partial waves, and one might
expect some condensation in the higher partial 
waves.\cite{Son,PisarskiRischke}  It turns out
that these condensates have the same leading $g$-dependence
as the $J=0$ 
condensate,\cite{Son,PisarskiRischke,rockefeller}
but they are suppressed by numerical factors of order 
$1/100$ to $1/1000$.\cite{rockefeller}
We therefore neglect them, and concentrate on the $J=0$ gap.

We also assume that the gap has positive parity.
As we discussed in Sections 4.1 and 4.2,
single-gluon exchange does not distinguish between the
positive parity $\langle\psi C \gamma_5 \psi\rangle$ condensate
and the negative parity $\langle\psi C \psi\rangle$ condensate.
This degeneracy
is lifted by instantons, which favor the positive parity
channel.\cite{ARW1,RappETC,SW0}. At large chemical
potential instanton effects are exponentially suppressed.
In the following, we will therefore assume that the only
instanton effect is to determine the parity of the gap.
Finally, we neglect quark mass effects and
chiral symmetry breaking $LR$ condensates. As shown in
Ref.~58 and described in Section 4.1,
there is no BCS instability in the case
of pairing between left and right handed quarks. The
formation of $LR$ condensates is therefore suppressed
by $m/\mu$.\cite{PisarskiRischke}

Given all these assumptions and specializations,
the gap matrix takes the form
\cite{BailinLove,PisarskiRischke,SW3}
\beq
\label{gap_ans}
 \Delta^{\alpha\beta}_{ij}(p) = (\lambda_2)^{\alpha\beta}(\tau_2)_{ij}
  C\gamma_5 \left( \Delta_1(p_0)P_+(p)
             +\Delta_2(p_0)P_-(p)\right)\ ,
\eeq
where the projection operators are given by
\begin{eqnarray}
P_+(p) &=& \half\left(1+\gamma_0\vec\gamma\cdot \hat p\right)\,,
\nonumber\\[1ex]
P_-(p) &=& \half\left(1-\gamma_0\vec\gamma\cdot \hat p\right)\,.
\end{eqnarray}
We have neglected the dependence
of the gap on the magnitude of the
momentum $|\vec p|$, but kept the dependence on frequency $p_0$.
The $|\vec p|$-dependence can be dropped because, in the
weak coupling limit, all momenta are close to the Fermi
surface. For short-range interactions, the dependence on
frequency can also be neglected.  Thus, in Section 4.3
we chose the ansatz (\ref{2SCansatz})
in which $\Delta_1(p_0)=\Delta_2(p_0)$ and both were
given by the constant $\Delta$.  
This was appropriate in that context, and we had no difficulty finding
a solution to the gap equation upon making this ansatz.
Now, however, 
because long range interactions are important, retardation
effects cannot be neglected.  

$P_+$ and $P_-$ are projection
operators which project onto particles and antiparticles respectively.
This means that $\Delta_1(p_0)$ describes the modification of 
the propagator due to particle-particle and hole-hole pairing
whereas $\Delta_2(p_0)$ describes that due to antiparticle-antiparticle
pairing.  The gap in the quasiparticle spectrum at the
Fermi surface is given by $\Delta_1(\Delta_1)$, and we 
are therefore primarily interested in deriving a gap equation
for $\Delta_1(p_0)$.  We expect that at weak coupling, quarks
near the Fermi surface should dominate and the 
contribution of $\Delta_2$ to the gap equation for $\Delta_1$
should vanish.  (This is what we found in our analysis of
the gap equation (\ref{BCSgap}), for example.) 
This emerges explicitly as a consequence of the
fact that   
in the weak coupling limit, we can replace $\gamma_0\vec\gamma\cdot\hat p$
by the unit matrix using the equations of motion.   
Note that this does not mean that $\Delta_2$ is in any sense
smaller than $\Delta_1$. It only means that the contribution
of $\Delta_2$ to the gap equation for $\Delta_1$ becomes small.

The self energy in the Nambu-Gorkov formalism obeys the
Dyson-Schwinger equation \cite{BailinLove}
\beq
\label{ds}
 \Sigma(k) = ig^2 \int \frac{d^4p}{(2\pi)^4}
  \Gamma_\mu^A S(p)\Gamma_\nu^A D_{\mu\nu}(k-p)\,.
\eeq
Here, $\Sigma(k)$ is the proper
self energy defined via (\ref{properselfenergy}), 
$\Gamma^A_\mu$ is the quark-gluon vertex
and $D_{\mu\nu}(k-p)$ is the gluon propagator. 
We have written (\ref{ds}) in Minkowski space.
To
leading order in the perturbative expansion, we can use
the free vertex (\ref{vert_0}).
To leading order, we can also neglect the
diagonal part of the proper self energy, namely the
fermion wave function renormalization.\cite{Son}
We shall see below that although we obtain the asymptotic
form for the gap, 
the results we obtain are 
uncontrolled for $g>g_c\sim 0.8$, corresponding
to a rather large value of $\mu\sim10^8$~MeV.\cite{RajagopalShuster}
This breakdown could in principle reflect a failure
in any of our assumptions. We expect, however, that
it arises because contributions which have been truncated
in writing the one-loop Schwinger-Dyson equation (\ref{ds}) with 
(\ref{vert_0}) are large for $g>g_c$. That is, we expect
that this truncation (for example the neglect of vertex corrections),
and not any of the simplifications we introduced in making the 
ansatz (\ref{gap_ans}) for $\Delta$, is the most significant
assumption we make in this analysis.

The Dyson-Schwinger equation (\ref{ds}) reduces to an equation
for the gap matrix,
\beq
\label{gap2}
 \Delta(k) = -ig^2 \int \frac{d^4p}{(2\pi)^4}
 \left(\gamma_\mu\frac{\lambda^a}{2}\right)^T
 S_{21}(p) \left(\gamma_\nu\frac{\lambda^a}{2}\right)
 D_{\mu\nu}(k-p)\,.
\eeq
Here, $S_{21}(p)$ is the 21-component of the fermion
propagator in the Nambu-Gorkov representation. $S_{21}(p)$
is determined from the inverse of (\ref{sinv}). We
have
\beq
\label{S21}
 S_{21}(p) = -\frac{1}{(\pslash-\muslash)^T}\Delta
   \frac{1}{(\pslash+\muslash) -\bar\Delta
     [(\pslash-\muslash)^T]^{-1}\Delta}\,.
\eeq
Inserting the ansatz (\ref{gap_ans}) for the gap gives
\beq
\label{S21_2}
 S_{21}(p) = -\lambda_2\tau_2C\gamma_5 \left(
 \frac{\Delta_1(p_0)P_-(p)}{p_0^2-(|\vec{p}|-\mu)^2-\Delta_1(p_0)^2}
 + \frac{\Delta_2(p_0)P_+(p)}{p_0^2-(|\vec{p}|+\mu)^2-
\Delta_2(p_0)^2}
 \right).
\eeq
Both the RHS and the LHS of the gap equation are proportional
to $\tau_2$, so the flavor structure simply drops out. The
color coefficient is given by
\beq
\label{color}
\frac{1}{4}(\lambda_a)^T\lambda_2\lambda_a
  = -\frac{N_c+1}{2N_c}\lambda_2 = -\frac{2}{3}\lambda_2
 \hspace{0.5cm}(N_c=3)\,,
\eeq
where we have used the Fierz identity $(\lambda^a)_{ij}
(\lambda^a)_{kl}=-(2/N_c)\delta_{ij}\delta_{kl}+2\delta_{il}
\delta_{jk}$ and the factor 1/4 comes from the color generators
$t^a=\lambda^a/2$. Projecting (\ref{gap2}) on $\Delta_{1,2}$ gives
two coupled gap equations
\bea
\label{gap3}
\Delta_{1,2}(k_0)\! &=&\! \frac{2ig^2}{3} \int \!\frac{d^4p}{(2\pi)^4}
\!  \left\{ \frac{1}{2}{\rm tr}
  \left(\gamma_\mu P_-(p)
        \gamma_\nu P_\pm(k)\right)
  \frac{\Delta_1(p_0)}{p_0^2-(|\vec{p}|-\mu)^2-\Delta_1(p_0)^2}
  \right. \nonumber \\[2ex]
  & & \hspace{-0.8cm}\left.\!\!  +\frac{1}{2}{\rm tr}
 \left(\gamma_\mu P_+(p)
        \gamma_\nu P_\mp(k)\right)
  \frac{\Delta_2(p_0)}{p_0^2-(|\vec{p}|+\mu)^2-\Delta_2(p_0)^2}
  \right\} D_{\mu\nu}(k-p)\,,
\eea
where the upper and lower signs on the RHS 
correspond to $\Delta_1$ and $\Delta_2$ on the LHS.

We now must specify the gluon propagator. The gluon propagator
in a general covariant gauge is given by
\beq
\label{D_dec}
 D_{\mu\nu}(q) = \frac{P_{\mu\nu}^T}{q^2-G(q)}
 + \frac{P_{\mu\nu}^L}{q^2-F(q)} - \xi\frac{q_\mu q_\nu}{q^4}\ ,
\eeq
where 
the projectors $P_{\mu\nu}^{T,L}$ are defined by
\beq
\label{proj}
 P_{ij}^T = \delta_{ij}-\hat{q}_i\hat{q}_j \,, \hspace{0.7cm}
 P_{00}^T = P_{0i}^T = 0\,,  \hspace{0.7cm}
P_{\mu\nu}^L = -g_{\mu\nu}+\frac{q_\mu q_\nu}{q^2}
   -P_{\mu\nu}^T \,.
\eeq
The functions $F(q)$ and $G(q)$ describe the effects of the medium
on gluon propagation. If we neglect the Meissner effect (that
is if we neglect the modification of $F(q)$ and $G(q)$ due to the gap $\Delta$
in the fermion propagator) then $F(q)$ describes Thomas-Fermi 
screening, $G(q)$ describes dynamical screening (Landau damping), and 
they are given in the hard dense loop approximation 
by \cite{LeBellac} 
\bea\label{HDLapprox}
F(q)&=&2m^2\frac{q^2}{\vec{q}^{\,2}}\Bigg( 1-\frac{iq_0}{|\vec{q}|}
     Q_0\Bigg(\frac{iq_0}{|\vec{q}|}\Bigg)\Bigg)\,,
\hspace{0.5cm} Q_0(x)=\frac{1}{2}\log\left(\frac{x+1}{x-1}\right)\,, \\[1ex]
G(q)&=&m^2\frac{iq_0}{|\vec{q}|}\Bigg[
    \Bigg(1-\Bigg(\frac{iq_0}{|\vec{q}|}\Bigg)^2\Bigg)
   Q_0\Bigg(\frac{iq_0}{|\vec{q}|}\Bigg) +
   \frac{iq_0}{|\vec{q}|} \Bigg]\ ,
\eea
where $m^2=N_f g^2 \mu^2/(4\pi^2)$ and we are working with $N_f=2$ flavors.
Note that $G\rightarrow 0$ for $q_0\rightarrow 0$:
static magnetic modes are not screened. Nonstatic modes
are dynamically screened due to Landau damping. 
Note that the gluon propagator contains
the gauge parameter $\xi$, which must not appear
in physical results. 

In the weak coupling limit, $q_0$ is small as compared to
$|\vec{q}|$. In this case we can expand the projectors
$P_{\mu\nu}^L\simeq \delta_{\mu 0}\delta_{\nu 0}$ and
$q_i q_j/q^2\simeq \hat{q}_i\hat{q}_j$. (See Ref.~69
for an analysis
in which these simplifying 
kinematic approximations and others which follow are not made.)
The gap equation
now becomes
\bea
\label{gap_4}
\hspace{-0.5cm}\Delta_{1}(k_0) &=& -\frac{2ig^2}{3} \int \frac{d^4p}{(2\pi)^4}
  \Biggl\{
  \frac{\Delta_1(p_0)}{p_0^2-(|\vec{p}|-\mu)^2-\Delta_1(p_0)^2}
\nonumber\\[1ex]
& & \hspace{2cm}  
\times\left(
  \frac{\frac{3}{2}-\frac{1}{2}\hat{k}\cdot\hat{p}}{(k-p)^2-G(k-p)}
 +\frac{\frac{1}{2}+\frac{1}{2}\hat{k}\cdot\hat{p}}{(k-p)^2-F(k-p)}
  \right)
  \nonumber \\
  & & \hspace{-1cm} +
  \frac{\Delta_2(p_0)}{p_0^2-(|\vec{p}|+\mu)^2-\Delta_2(p_0)^2}
\nonumber\\[1ex]
& & \hspace{-0.5cm}  
\times\left(
  \frac{\frac{1}{2}+\frac{1}{2}\hat{k}\cdot\hat{p}}{(k-p)^2-G(k-p)}
 +\frac{\frac{1}{2}-\frac{1}{2}\hat{k}\cdot\hat{p}}{(k-p)^2-F(k-p)}
 +\frac{\xi}{(k-p)^2}
  \right)
  \Biggr\}\,.
\eea
There is a similar equation for $\Delta_2$ in which the two
terms in the round brackets are interchanged. Only the first
term in (\ref{gap_4}) has a singularity on the Fermi
surface.  In the weak coupling limit, we can therefore drop
the second term, and we are left with an equation for
$\Delta(k_0)\equiv \Delta_1(k_0)$. This equation is
independent of the gauge parameter $\xi$.  The $\xi$-independence
of the $\Delta_1$ term in (\ref{gap_4}) 
arises because of the kinematic approximations made in deriving
it.  If one goes beyond these approximations, $\Delta_1$ 
is $\xi$-dependent.\cite{RajagopalShuster}
The second
gap parameter $\Delta_2$ is not suppressed in magnitude
and its gauge-dependence is not kinematically suppressed.
However, $\Delta_2$ does not lead to a gap on the Fermi
surface.

The fact that the gap is gauge
independent in the present
weak-coupling approximation is a consequence
of the fact that the gap is determined by the scattering
of quarks that are almost on shell.  For on-shell quarks,
the fact that the gauge dependent part of the propagator
does not contribute follows directly from the equations
of motion for the quark fields.   This argument is adequate
only to lowest order, as appeared in the specific diagrammatic
cancellation (and non-cancellation) above.

For large chemical potential the integral over $p$ is
dominated by momenta in the vicinity of the Fermi surface,
$|\vec{p}|\simeq\mu$ and $p_0\ll\mu$. We can expand all
momenta as $\vec{p}=\vec{p}_F+\vec{l}$, where $\vec{p}_F$
is on the Fermi surface, and $\vec{l}$ is orthogonal to it.
Asymptotically, $|\vec{l}|\ll|\vec{p}_F|$ and the integration
measure becomes $dp_0\,\mu^2 dl\,d\cos\theta\,d\phi$. We
also have $|\vec{k}-\vec{p}|\simeq\sqrt{2}\mu (1-\cos\theta)$.
The integral over $\phi$ is performed trivially. We analytically
continue to imaginary $p_0$, and perform the integral over
$\vec{l}$ by picking up the pole in the diquark propagator.
We find
\bea
\label{gap_5}
\Delta(k_0) \hspace{-0.2cm}&=&\hspace{-0.2cm} 
\frac{g^2}{12\pi^2} \int dp_0\int d\cos\theta\,
 \left(\frac{\frac{3}{2}-\frac{1}{2}\cos\theta}
            {1-\cos\theta+(G+(k_0-p_0)^2)/(2\mu^2)}\right.\nonumber \\[1ex]
 & & \left.    +\frac{\frac{1}{2}+\frac{1}{2}\cos\theta}
            {1-\cos\theta+(F+(k_0-p_0)^2)/(2\mu^2)} \right)
 \frac{\Delta(p_0)}{\sqrt{p_0^2+\Delta(p_0)^2}}\ . 
\eea
The integral over $\cos\theta$ is dominated by small $\theta$,
corresponding to almost collinear scattering.  We must, therefore,
take careful account of the medium modifications of the 
gluon propagator described by $F(q)$ and $G(q)$ given
in (\ref{HDLapprox}).   
In the gap equation (\ref{gap_5}) 
$F(q)$ and $G(q)$ are to be evaluated at  $q_0=k_0-p_0$ and 
$|\vec q|=|\vec k - \vec p|\simeq\sqrt{2}\mu (1-\cos\theta)$,
and are therefore functions of $k_0-p_0$ and $\cos\theta$.
For $q_0\ll\vec{q}\to 0$ the  expressions (\ref{HDLapprox}) for $F(q)$ 
and $G(q)$ 
simplify, yielding
\beq
 F(q) = 2m^2\,, \hspace{1cm}
 G(q) = \frac{\pi}{2}m^2\frac{q_0}{|\vec{q}|}\,.
\eeq
In the longitudinal part, 
$m_D^2=2m^2$ is the familiar Debye screening mass. In the
transverse part, 
nonstatic modes are dynamically screened. In our case, typical frequencies
are on the order of the gap, $q_0\simeq \Delta$. This means
that the electric part of the interaction is screened at
$q_E\simeq m_D^{1/2}$ whereas the magnetic interaction
is screened at $q_M\simeq (\pi/4\cdot m_D^2\Delta)^{1/3}$.


Asymptotically, $q_M\ll q_E$, and magnetic gluon exchange
dominates over electric gluon exchange. We therefore begin
by analyzing the gap equation taking into account the
magnetic part of the interaction only. We will also
approximate $\cos\theta\simeq 1$ in the denominator
and drop $(k_0-p_0)^2$ in the denominator. All of these
terms will be reinstated later. The integration over
$\cos\theta$ is now straightforward. We have
\beq
\label{eliash}
\Delta(k_0) = \frac{g^2}{18\pi^2} \int dp_0
 \log\left(1 + \frac{64\pi\mu}{N_fg^2|k_0-p_0|}\right)
 \frac{\Delta(p_0)}{\sqrt{p_0^2+\Delta(p_0)^2}}\ .
\eeq
If we are only interested in the leading exponential
behavior of the gap we can drop the numerical factors
and the powers of $g$ in the logarithm. We then arrive
at
\beq
\label{eliash1}
\Delta(k_0) = \frac{g^2}{18\pi^2} \int dp_0
 \log\left(\frac{c\mu}{|k_0-p_0|}\right)
 \frac{\Delta(p_0)}{\sqrt{p_0^2+\Delta(p_0)^2}}\ ,
\eeq
where $c$ is
a factor to be determined more fully below.
This equation was first derived by Son.\cite{Son}
(In ordinary 
superconductivity, the corresponding equation
was first derived by Eliashberg.\cite{Eliashberg})
What we have
shown here (following Ref.~64)
is that one can derive this equation directly
from the Dyson-Schwinger equation in the weak coupling
limit, and that upon making
the approximations we have made the result is independent of the gauge
parameter.  The integral equation (\ref{eliash1}) can
be converted to a differential equation,\cite{Son} and
in the weak coupling limit an approximate solution is
given by\cite{Son,PisarskiRischke,SW3,rockefeller}
\begin{equation}\label{Deltaapprox}
\Delta(k_0)\simeq \Delta_0 \sin\left(\frac{g}{3\sqrt{2}\pi}
\log\left(\frac{c\mu}{k_0}\right)\right)\ , \qquad k_0>\Delta_0\ ,
\end{equation}
where
\begin{equation}\label{resultforDelta0}
\Delta_0 = 2c\mu\exp\left(-\frac{3\pi^2}{\sqrt{2}g}\right)\ .
\end{equation}

At this point a few comments are in order. First, we note that
the use of perturbation theory to determine the dynamic screening
is self-consistent. Since $\Delta\sim\mu\exp(-{\rm constant}/g)$, the gap
grows as $\mu\to\infty$ and $q_M\gg\Lambda_{QCD}$. Second, we note
that it is essential to keep the frequency dependence of the gap.
For small frequencies, $\Delta(k_0)$ varies over scales on the
order of $k_0\sim\Delta_0$ itself. Therefore, $\Delta(k_0)$ cannot
be replaced by a constant.  Were we to approximate $\Delta(k_0)
\simeq\Delta_0$, 
we would obtain a gap equation for $\Delta_0$ that has
the correct double logarithmic structure and gives $\Delta_0
\simeq\mu\exp(-{\rm constant}/g)$, but the constant in the exponent
would  not be correct.

We now come to the role of electric gluon exchanges. We include
the second term in (\ref{gap_5}) with $F=m_D^2$. We again use the
approximation $\cos\theta\simeq 1$ in the numerator and drop
the $(k_0-p_0)^2$ term in the denominator. Let us note that
in the forward direction, electric and magnetic gluon exchanges
have the same overall factor. Performing the integral over
$\cos\theta$, we find
\bea
\label{eliash_mel}
\Delta(k_0) &=& \frac{g^2}{18\pi^2} \int dp_0
 \left\{ \log\left(1+\frac{64\pi\mu}{N_fg^2|k_0-p_0|}\right)
     +  \frac{3}{2}\log\left(1+\frac{8\pi^2}{N_fg^2} \right) \right\}
\nonumber\\[1ex]
& &\hspace{5.5cm}\times\frac{\Delta(p_0)}{\sqrt{p_0^2+\Delta(p_0)^2}}\ ,
\eea
where the factor $3/2$ in front of the second term comes from the
difference between dynamic screening, $q_M\sim |\vec{q}|^{1/3}$, and
static screening, $q_E\sim |\vec{q}|$. In the weak coupling limit,
the gap equation (\ref{eliash_mel}) reduces to the 
form (\ref{eliash1}) with approximate solution (\ref{Deltaapprox}),
where we now see that we can  
estimate the constant $2c$ to be
\begin{equation}\label{bresult}
2c = 2048\sqrt{2}\pi^4 N_f^{-5/2}\, g^{-5}
   = 512\pi^4 g^{-5}\simeq 5.0\times 10^4 g^{-5}
   \hspace{0.5cm} (N_f=2)\ .
\eeq

We have obtained the result (\ref{bresult}) for the prefactor $2c$
in the expression for 
the gap $\Delta_0$
by collecting the leading logarithms
from both electric and magnetic gluon exchange.  To do better,
Sch\"afer and Wilczek solve the gap equation (\ref{gap_5}), which
incorporates both electric and magnetic exchange,
numerically.\cite{SW3}  
They also keep the
$\cos\theta$ dependence in the numerator, and the terms $(k_0-p_0)^2$
in the denominator. Finally, they use the exact forms of $G(q)$ and $F(q)$
in the hard dense loop approximation, (\ref{HDLapprox}). 
This takes into account that there is no dynamic screening
for $|\vec{q}|<q_0$.  Asymptotically (i.e. for small $g$)
the numerical solution to the gap equation is well
described by (\ref{Deltaapprox},\ref{resultforDelta0}) 
with $2c\simeq 1.4\times 10^4 g^{-5}$,
at least for $k_0<\sqrt{N_f/(8\pi)}g\mu$.  At larger $k_0$,
the retardation terms $\sim(k_0-p_0)^2$ dominate over screening
and $\Delta$ falls off more quickly with increasing $k_0$ than
in (\ref{Deltaapprox}).
We are interested in $k_0=\Delta_0$, however, and in this
regime (\ref{Deltaapprox}) describes the shape of the numerical
solution quite well.

In order to use the weak-coupling results, we must choose an energy
scale at which to evaluate the running QCD coupling.  
Sch\"afer and Wilczek chose to use
the one-loop running coupling constant evaluated at the Fermi
momentum $p_F=\mu$, which is an average over the momenta of the
exchanged gluons, which are in the range $[q_M,2\mu]$. 
Strictly speaking, only a 
higher order calculation can fix the scale in the running
coupling.  It turns out, however, that with $g$ evaluated at $\mu$,
the gap $\Delta_0$ only depends rather weakly on $\mu$. We return
to this below.
The numerical solution yields 
$\Delta_0\simeq 40$~MeV at $\mu=10^{10}$~MeV, corresponding
to $\Delta_0/\mu=4\times 10^{-9}$ at $g=0.67$ and a value
of $2c$ which is about $2/5$ that in (\ref{bresult}),
and 
$\Delta_0\simeq 90$~MeV at $\mu=400$~MeV, corresponding
to $\Delta_0/\mu=0.23$ at $g=3.43$ and a value of $2c$ which happens
to be almost exactly that in (\ref{bresult}).   As $\mu$ increases from 
400~MeV to about $10^6$~MeV, $\Delta$ drops from 90~MeV
to about 10~MeV; as $\mu$ increases further, $\Delta$ then
begins its inexorable, but logarithmic, rise.

Clearly, with $\mu=400$~MeV and
$g=3.43$, the calculation of the gap is determined by momenta which
are not large compared to $\Lambda_{\rm QCD}$, and 
all
of the approximations we have used have broken down.
More on this below, but before plunging into caveats let us note that
it is gratifying to see
that the order of magnitude of the result agrees with
that obtained in the calculations described in Sections 4.2 and 4.3,
based on more phenomenological
effective interactions, which were normalized
to the strength of chiral symmetry
breaking at zero density, rather than to the calculable asymptotics of
the running coupling.

Son's result \cite{Son} 
\begin{equation}\label{eq:SonResult}
\frac{\Delta}{\mu}\sim \frac{b}{g^5} \exp\left(-\frac{3\pi^2}{\sqrt{2}g}
\right) 
\end{equation}
has now been confirmed using a variety of 
methods.\cite{PisarskiRischke,Hong,HMSW,SW3,rockefeller,Hsu2,EHHS,BBS,RajagopalShuster}
However, even the 
${\cal O}(g^0)$ contribution to the prefactor $b$ in 
(\ref{eq:SonResult}) is not yet fully understood,
and certainly nobody has proposed a controlled calculational scheme
which could be used to push the calculation of $b$ to arbitrarily
high accuracy.  
We have worked through the analysis of Sch\"afer and Wilczek,\cite{SW3}
which leads to the estimate that $b \sim 512 \pi^4$ in the 2SC phase,
a result which is supported by their numerical work and
by the work of other authors.\cite{PisarskiRischke,rockefeller,Hsu2,EHHS,BBS}
A similar analysis yields the estimate that 
$b\sim 512 \pi^4 2^{-1/3} (2/3)^{5/2}$ in the CFL phase,
with the $(2/3)^{5/2}$ coming from the $N_f$-dependence of the
Debye mass and the $2^{-1/3}$ coming from an analysis of
the color-flavor structure of the CFL gap 
equation,\cite{SchaeferPatterns} 
along
the lines of that we have presented in Section 4.2 in the 
context of models with four-fermion interactions.
Numerical solutions of the weak-coupling CFL gap equation 
by Evans {\it et al.} are in good agreement
with this estimate, and these authors also confirm
that including Meissner effects in the gluon propagator
makes little difference to the final result.\cite{EHHS}

The argument that the CFL phase
is favored at asymptotically high density in QCD with 
three flavors of quarks is similar to 
that given at the end of Section 4.2.
2SC pairing would
yield a gap given by (\ref{eq:SonResult}) with $b\sim 512 \pi^4 (2/3)^{5/2}$,
whereas in the CFL phase the gap is smaller by a factor
of $2^{-1/3}$. 
In the asymptotic regime, the condensation energy 
is proportional to $\Delta^2\mu^2$ up to
logarithmic corrections.\cite{SchaeferPatterns,EHHS}
As expected, the expression (\ref{Omegafromgapeq}) 
derived by taking $G\rightarrow 0$ in the analysis done using a four-fermion
interaction is a good guide.
In the 2SC phase, four quarks
pair with gap $\Delta$ and the condensation
energy energy is $\sim 4(\Delta^2\mu^2/4\pi^2)$.  
In the CFL phase at asymptotically high density,
the color ${\bf 6}$ condensate is suppressed relative
to the color ${\bf \bar 3}$ condensate by a factor 
$g\sqrt{2}\log(2)/36\pi$,\cite{SchaeferPatterns}
and so to leading order 
one has eight quarks with gap $\Delta$ and one 
with gap $2\Delta$, and a 
condensation energy which is therefore $\sim 12(\Delta^2\mu^2/4\pi^2)$.
Although the CFL gap is smaller than the 2SC gap by a factor of $2^{-1/3}$,
this does not outweigh the 
advantage the CFL phase enjoys because it allows all nine quarks
to pair. As a result, the CFL phase is favored.  As we
saw in Section 3, this conclusion continues to hold even
when the strange quark mass $m_s$ is nonzero, as long
as $m_s^2/2\mu$ is small compared to $\Delta$.

There
are several contributions to $b$ which we have not 
taken into account above. They modify the magnitude of the gap, 
but do so in the same way for CFL or 2SC pairing.
For example, modifications to the quasiparticle dispersion
relations in the normal (nonsuperconducting; high temperature)
phase~\cite{rockefeller,ManuelDispersion,BoyanovskydeVega}
and quasiparticle damping effects in the superconducting
phase~\cite{Manuel} do affect the gap, and both tend to reduce $b$. 
In the normal phase, the effects
of wave function renormalization 
are only
important within $\sim\exp(-1/g^2)$ of the Fermi 
surface,\cite{BoyanovskydeVega}
which supports Son's conclusion that in 
the superconducting phase these effects
result only in a subleading correction to $\Delta$.\cite{Son}  This
correction does, however, reduce $b$ by about
a factor of 5.\cite{rockefeller}

The value
of $b$ is also
affected by the choice of the scale at which $g$ is
evaluated in (\ref{eq:SonResult}).  The consequent modifications
to $\Delta$ have been considered by
Beane {\it et al.}\cite{BBS}
They do a renormalization
group analysis within the effective theory constructed
by Hong,\cite{Hong} who noted that the antiparticles can safely
be integrated out.  Beane {\it et al} use the renormalization
group to compute and resum contributions of the form 
$\alpha^{n+1}\beta_0^n\log^n(\Delta/\mu)$ to the gap equation,
where $\beta_0$ is the first coefficient in the QCD $\beta$-function.
Their results
demonstrate that $g$ should be evaluated at a $\mu$-dependent scale which is
much lower than $\mu$.\cite{BBS} Upon assuming that the QCD
coupling runs according to the vacuum $\beta$-function
all the way down to the magnetic scale $q_M$ --- an assumption
which is not justified but which has not yet been superseded ---
they find that if, 
by convention,
$g$ is taken as $g(\mu)$, then $b$ is enhanced by a factor of about 20.

Finally, we return to the question of the gauge dependence
of $b$.  Both the analysis of Sch\"afer and Wilczek presented above
and that of Pisarski and Rischke~\cite{PisarskiRischke}
demonstrate that the ${\cal O}(g^0)$ contribution to $b$
is gauge-independent.  However, Rajagopal and Shuster have
gone back to (\ref{gap3}), eliminated $\Delta_2$ in the 
gap equation for $\Delta_1$, but made no further kinematic
simplifications.  The resulting gap equation for $\Delta_1$
is gauge-dependent.
Examination of the gauge-dependent (and $g$-dependent)
contributions to $b$ obtained by solving this equation 
reveals that these  do decrease with decreasing $g$, but that
they
only begin to decrease 
for $g< 0.8$.\cite{RajagopalShuster}
This means that effects like vertex corrections which have to date been
neglected in all calculations 
based on
the one-loop Schwinger-Dyson equation (e.g. 
those of Refs.~61,63,64,66)
are small corrections to $b$ only for $\mu\gg 10^8$~MeV.  

All these caveats should by now have made it clear
that although we are confident of the leading $g$-dependence
of the gap $\Delta$ at asymptotically large $\mu$,
there remains physics which has not been satisfactorily
treated which contributes at order $g^0$ to 
the prefactor $b$.  Furthermore, effects which are higher
order in $g$ are large enough that they can only be neglected
for $\mu\gg 10^8$~MeV.  

The asymptotic results nevertheless have very important
consequences.  We have seen in Section 3 that the outcome of the 
competition between the CFL and 2SC phases depends on
the relative magnitudes of $\Delta$ and $m_s^2/2\mu$.
The latter decreases at large $\mu$, while the 
result (\ref{eq:SonResult}) demonstrates  
that $\Delta$
increases logarithmically as $\mu\rightarrow\infty$. 
This means
that the CFL phase is favored over the 2SC phase 
for $\mu\rightarrow\infty$ for any $m_s\neq \infty$.\cite{ABR2+1}
Even though the asymptotic regime where $\Delta$ can
be calculated from first principles with confidence is
not accessed in nature, it is of great theoretical 
interest.
The weak-coupling calculation of the gap in the CFL phase 
is the
first step toward the weak-coupling calculation of other
properties of this phase, in which chiral symmetry is broken
and the spectrum of excitations is as in a confined phase.
As we have described in Section 2, for example, the masses and decay constants
of the pseudoscalar mesons can be calculated from first
principles once $\Delta$ is known.

Although the value of $\Delta$ 
is under control asymptotically, it seems fair 
to say that applying these asymptotic results 
at $\mu=400$~MeV is currently at least as uncertain a proposition
as applying estimates made using phenomenologically
normalized models with point-like interactions.
Nevertheless, 
if we take the estimates for the prefactor $b$
provided by Sch\"afer and Wilczek's numerical results
described above and apply them at 
$\mu\sim 400$~MeV, they predict gaps of order
$100$~MeV.  The consequent critical temperatures
are related to the zero temperature gap 
$\Delta$ by the standard weak-coupling BCS
result $T_c=0.57\Delta$,\cite{PisarskiRischke,rockefeller}
and are therefore of order 50~MeV.
We have seen that some known corrections
push this estimate up while others push it down, and
that the calculation whence it came is, regardless, of  
quantitative value only for $\mu\gg 10^8$~MeV.
It is nevertheless satisfying that two very different approaches,
one using zero density phenomenology to normalize models, the
other using weak-coupling methods valid at asymptotically
high density, yield predictions for the
gaps and critical temperatures at accessible
densities
which are in qualitative agreement.  
Neither can be trusted quantitatively
for quark number chemical potentials $\mu\sim 400-500$~MeV,
as appropriate for the quark matter which may occur in
compact stars.  Still, both methods agree that the
gaps at the Fermi surface are of order tens to 100~MeV, with
critical temperatures about half as large.

$T_c\sim 50$~MeV is much larger relative to the
Fermi momentum (say $\mu\sim 400-500$~MeV) than in 
low temperature superconductivity in metals.
This reflects the fact that color superconductivity
is induced by an attraction due to the primary,
strong, interaction in the theory, rather
than having to rely on much weaker secondary interactions,
as in phonon mediated superconductivity in metals.
Quark matter is a high-$T_c$ superconductor by any reasonable
definition. It is unfortunate
that its $T_c$ is nevertheless low enough that
it is unlikely the phenomenon can be realized in heavy ion
collisions.

\subsection{Challenges for the Future}

While the rigorous asymptotic justification of color superconductivity
is a marvelous result, there are compelling physical and mathematical
motivations to go further.

For physical applications, we are interested in densities not much beyond
nuclear, which (as we have
seen) appears to be quite far from asymptopia.   It is not at all clear, in
particular, that the form of the
interaction which dominates asymptotically, namely near-forward magnetic
scattering, is the most
important at physically achievable densities.   Indeed, electric 
scattering --- or binding through electric
flux tubes! --- and instantons more nearly resemble the interactions used
successfully in phenomenological
models of hadrons at zero density and applied to nonzero densities
as described in Sections 4.2 and 4.3.

On the mathematical side, it clearly would be very desirable to have
a systematic approximation framework that did not depend on selective
diagram resummation, and whose
results were manifestly gauge invariant.   Similar problems arise in the
treatment of
ordinary superconductors in condensed matter physics,  albeit in a simpler
form since the gauge group is
abelian.  However,  
in that context the lowest-order approximation is adequate in
practice since
electrodynamic radiative corrections are overwhelmed, quantitatively, by
many other uncertainties.

In formulating an appropriate perturbation theory, the difficulties arise
due to the gauge dependence of
the primary order parameter and to the complicated (in particular,
non-instantaneous) nature of the
microscopic interaction.    The whole notion of the primary order parameter
depends on fixing a
gauge, and we assume that is done.  On physical grounds, as we have
emphasized already, one expects a
weak-coupling approach to work only when it is based on perturbing around
the correct, gapped ground
state; but of course this begs the question of how to construct the ground
state.  The BCS-inspired variational
procedure works well for instanteous interactions, where one can define a
simple Hamiltonian, but is very
awkward for retarded interactions.  As we saw in Section 4.2, 
in one of its many implementations (that due originally 
to Bogoliubov and Valatin)
it takes the form of
a search among
``trial Hamiltonians'' incorporating quadratic terms mixing particles and
holes, with coefficients
depending on the momentum.   This suggests that one should employ trial
Lagrangians with similar terms,
allowing also for energy dependence, using the action variational principle
popularized by Feynman in
connection with the polaron problem.   Such a procedure has the great
advantage that one need only
do perturbation theory around the ground state of the trial Lagrangian,
which can be designed to
incorporate the correct physics (i.e. gaps) to begin with.   In principle,
the choice of gauge can simply be
treated as an additional variational parameter.

Ultimately, truly quantitative results on QCD at moderately ultra-nuclear
densities will probably have to
await numerical work using lattice gauge theory, as has been the case for
zero density.   In principle, we
can bypass all the difficulties involved in defining the primary condensate
by working directly with the
distinctive gauge-invariant parameters predicted for color-flavor locking.
Specifically, one should
encounter diverging susceptibilities for chiral symmetry breaking,
especially in the diquark-antidiquark channel (\ref{chiralOP}),
and for 
$U(1)_B$ breaking, in the  
baryon number 2 channel.
The corresponding vacuum
expectation values, signaling 
chiral symmetry breaking and superfluidity,
could be stabilized
by adding the corresponding infinitesimal source terms.   These features,
and the associated appearance of
Nambu-Goldstone bosons,  would serve as unmistakable signatures of the
color-flavor locked phase.  In
practice, unfortunately, all presently known numerical algorithms for QCD
at finite density and zero
temperature converge too slowly to be useful.    The discovery of suitable
algorithms for this problem
stands as perhaps the greatest theoretical challenge in QCD today.

\section{The Phase Diagram}

In this article we focus largely on 
zero temperature quark matter, but it is important to understand
where the 2SC and CFL phases fit on the phase
diagram of QCD as a function of both temperature
and density.   Because QCD is asymptotically
free, its high temperature and high baryon
density phases are more simply and more appropriately
described in terms of quarks and gluons as degrees
of freedom, rather than hadrons.  At high
temperatures, in the resulting quark-gluon plasma
(QGP) phase all of the symmetries of the QCD
Lagrangian are unbroken and the excitations have
the quantum numbers of quarks and gluons. 
At high densities, on the other hand, we have
seen that quarks form Cooper pairs and new condensates
develop.  
At high enough density, chiral symmetry
is broken by color-flavor locking. At densities 
which are high enough that nucleons overlap
and the matter is in a quark matter phase but
which are not high enough for color-flavor locking,
we expect to find the 2SC phase.

In this section, we describe the qualitative
features of the QCD phase diagram.
We choose to describe the entire phase diagram,
and not just the region of low temperatures and 
high densities.  The exploration of the higher temperature regions
of the diagram is the object of extensive 
experimental efforts   
in heavy ion collision experiments at CERN
and Brookhaven. Due to its topical interest,
we describe the high temperature
regime of the phase diagram in some detail, in addition
to explaining how the dense quark matter phases which
occupy us in the remainder of this article fit in.
In Section 6 below, we refocus
on cold dense quark matter, as we describe current
efforts to understand how to
use observed phenomena occuring 
in compact stars to map this region of the phase diagram.

Let us begin with a brief review of the phase changes wich
occur as a function of temperature at zero baryon number 
density.\cite{rajreview} 
That is, we begin by restricting ourselves to 
the vertical axis in Figures~1 through 4.  This slice of
the phase diagram was explored throughout the early universe
during the first tens of microseconds after the big bang.
It can also be studied in lattice simulations.  As heavy ion collisions
are performed at higher and higher energies, they create plasmas
with a lower and lower baryon number to entropy ratio and therefore
explore regions of the phase diagram 
closer and closer to the vertical axis.

\begin{figure}[t]
\begin{center}
\vspace{-0.1in}
\hspace*{0in}
\epsfysize=2.in
\epsffile{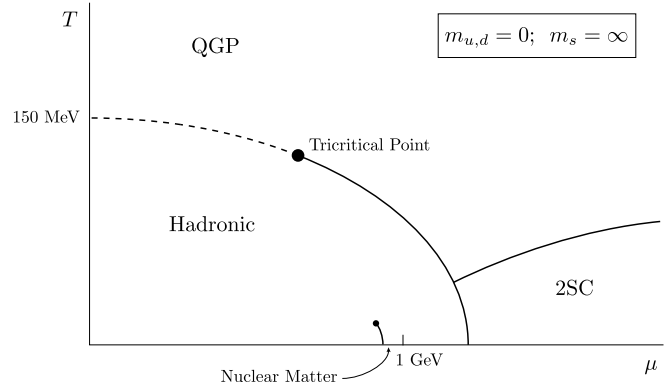}
\end{center} \label{fig1}
\caption{QCD Phase diagram for two massless quarks. Chiral symmetry
is broken in the hadronic phase and is restored elsewhere in the 
diagram. The chiral phase transition changes from second to
first order at a tricritical point. The phase at high density and low
temperature is a color superconductor in which up and down
quarks with two out of three colors pair and form a condensate.
The transition between this 2SC phase and the QGP phase 
is likely first order. The transition on the horizontal
axis between the hadronic and 2SC phases is first 
order. The transition between a nuclear matter ``liquid''
and a gas of individual nucleons is also marked.  At $T=0$,
it separates the vacuum phase from the nuclear matter
phase; Lorentz-boost symmetry
is broken to its right but unbroken to its left.
At nonzero temperature, Lorentz-boost symmetry is broken
in both the nuclear gas and nuclear liquid, and this
line of phase transitions may therefore end.
It is thought to end
at a critical point at a 
temperature of order 10~MeV, characteristic of the forces
which bind nucleons into nuclei.}
\vspace{-0.1in}
\end{figure}
In QCD with
two massless quarks ($m_{u,d}=0$; $m_s=\infty$; Figure~1)
the vacuum phase, with hadrons as excitations, is characterized
by a chiral condensate $\langle \bar\psi_{L}\psi_R\rangle$.
Whereas
the QCD Lagrangian is invariant under separate global flavor
rotations of the left-handed and right-handed
quarks, the presence of the chiral condensate spontaneously
breaks $SU(2)_L \times SU(2)_R$ to the subgroup $SU(2)_{L+R}$,
in which only  simultaneous flavor rotations of $L$ and $R$ quarks
are allowed.  Locking left- and right-handed rotations 
in this way breaks global symmetries and 
results in three massless Goldstone bosons, the pions.
The chiral order parameter, a $2\times 2$ matrix $M^{ab}$ in
flavor space,
can be written in terms of
four real fields $\sigma$ and $\vec\pi$ as
\begin{equation}
\langle \bar\psi_{L\alpha}^i\psi_{R}^{\alpha j}\rangle = M^{ij} = 
\sigma \delta^{ij} + \vec \pi \cdot \left(\vec\tau\right)^{ij} \ ,
\end{equation}
where the $\vec\tau$ are the three Pauli matrices.
$SU(2)_L$ and $SU(2)_R$ rotations act on $M^{ij}$ from the
left and right, respectively.  
The order parameter
can also be written as a four component scalar 
field $\phi =(\sigma,\vec\pi)$
and the  $SU(2)_L \times SU(2)_R$ rotations are then simply
$O(4)$ rotations of $\phi$.  In this language, the symmetry breaking
pattern $SU(2)_L \times SU(2)_R\rightarrow SU(2)_{L+R}$ is
described as
$O(4)\rightarrow O(3)$: in the vacuum, $\langle\phi\rangle\neq 0$ 
and this condensate picks a direction in $O(4)$-space.
The direction in which the condensate points is
conventionally taken
to be the $\sigma$ direction.  In the presence 
of $\langle\sigma\rangle\neq 0$, the $\vec\pi$ excitations
are excitations of the direction in which $\langle\phi\rangle$
is pointing, and are therefore
massless goldstone modes.

At nonzero but low temperature, one finds a gas of pions,
the analogue of a gas of spin waves, but $\langle\phi\rangle$
is still nonzero.  Above some temperature $T_c$, entropy
wins over order (the direction in which $\phi$
points is scrambled) and $\langle\phi\rangle=0$.
The phase transition at which chiral symmetry is restored 
is likely second order and belongs to the universality
class of $O(4)$ spin models in three dimensions.\cite{piswil}
Below $T_c$, chiral symmetry is broken and there are three
massless pions.  At $T=T_c$, there are four massless degrees
of freedom: the pions and the sigma. Above $T=T_c$, the pion
and sigma correlation lengths are degenerate and finite.

\begin{figure}[t]
\begin{center}
\vspace{-0.1in}
\epsfysize=2.in
\hspace*{0in}
\epsffile{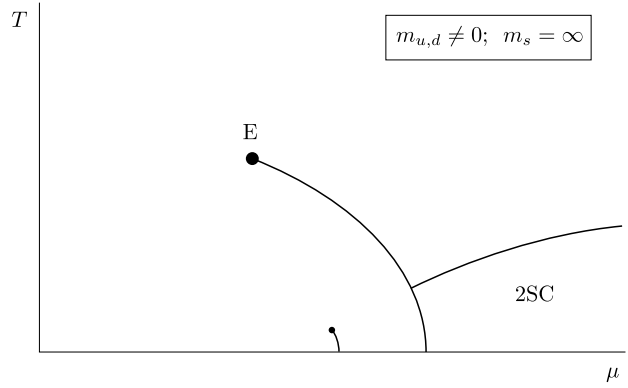}
\end{center} \label{fig2}
\caption{QCD phase diagram for two light quarks. Qualitatively as in Figure~1,
except that the introduction of light quark masses  turns the
second order phase transition into a smooth crossover. 
The tricritical point becomes the critical endpoint $E$, which
can be found in heavy ion collision experiments.}
\vspace{-0.1in}
\end{figure}
In nature, the light quarks are not massless.  Because
of this explicit chiral symmetry breaking,
the second order phase transition is replaced by an 
analytical crossover: physics changes dramatically but smoothly 
in the 
crossover region, and no correlation length diverges.
Thus, in Figure~2, there is no sharp boundary on the 
vertical axis separating the low temperature hadronic world
from the high temperature quark-gluon plasma.  
This picture is consistent with present lattice 
simulations,\cite{latticereview,latestlattice}
which suggest $T_c\sim 140-190$~MeV.\cite{latticeTc,latestlattice}

Arguments based on a variety of 
models\cite{NJL,steph,ARW1,RappETC,bergesraj,stephetal}
indicate that the chiral symmetry restoration
transition is first order at 
large $\mu$.  The 2SC quark matter phase that 
arises in two-flavor QCD at values of $\mu$ 
above this first order transition features
new condensates, different from those in the hadronic
phase, and these condensates do not break chiral symmetry.
The fact that this is a 
transition in which two different condensates compete 
strengthens previous model-based arguments that this transition
is first order.\cite{bergesraj,CarterDiakonov}  
This suggests that the
phase diagram features a critical point $E$ at which
the line of first order phase transitions present for 
$\mu>\mu_E$ ends, as shown in Figure~2.\footnote{If
the up and down quarks were massless, $E$ would
be a tricritical point,\cite{lawrie} at which the first
order transition becomes second order. See Figure~1.}
At $\mu_E$, the phase transition is second order
and is in the Ising universality class.\cite{bergesraj,stephetal}
Although the
pions remain massive, the correlation length in the $\sigma$ channel
diverges due to universal long wavelength fluctuations
of the order parameter.
This results in characteristic signatures,
analogues of critical opalescence in the sense that they
are unique to collisions which freeze out near the
critical point, which
can be used to discover $E$.\cite{SRS1,SRS2}

\begin{figure}[t]
\begin{center}
\vspace{-0.1in}
\epsfysize=2.in
\hspace*{0in}
\epsffile{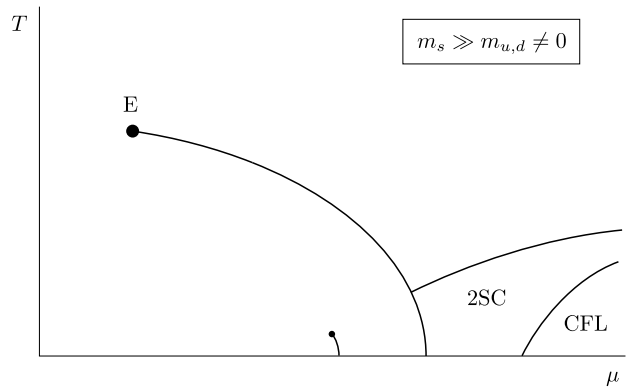}
\end{center} \label{fig3}
\caption{QCD phase diagram for two light quarks and a strange quark
with a mass comparable to that in nature.
The presence of the strange quark shifts
$E$ to the left,  as can be seen by comparing with Figure~2. 
At sufficiently high density, cold quark matter is necessarily 
in the CFL phase in which quarks of all three colors and
all three flavors form Cooper pairs. The diquark condensate in
the CFL phase breaks chiral symmetry, and this
phase has the same symmetries as baryonic matter which is
dense enough that the nucleon and hyperon densities are 
comparable. The unlocking phase transition
between the CFL and 2SC phases is first order.}
\vspace{-0.15in}
\end{figure}
Returning to the $\mu=0$ axis,
universal arguments,\cite{piswil} again backed by lattice 
simulation,\cite{latticereview}
tell us that if the strange quark were as light as the
up and down quarks, the transition would be first order,
rather than a smooth crossover.  
This means that if one could dial
the strange quark mass $m_s$, one would find a critical
$m_s^c$ at which the transition as a function of temperature
is second order.\cite{rajwil,rajreview} 
Figures~2, 3 and 4 are drawn
for a sequence of decreasing strange quark masses. Somewhere
between Figures~3 and 4, $m_s$ is decreased below $m_s^c$ and
the transition on the vertical axis becomes first order.
The value of $m_s^c$ is an open question,
but lattice simulations suggest that it is about half the
physical strange quark mass.\cite{columbia,kanaya} 
These results are not yet conclusive\cite{oldkanaya}  but
if they are correct then the phase
diagram in nature is as shown in Figure~3,  and the phase transition
at low $\mu$ 
is a smooth crossover.  
\begin{figure}[t]
\begin{center}
\vspace{-0.1in}
\epsfysize=2.in
\hspace*{0in}
\epsffile{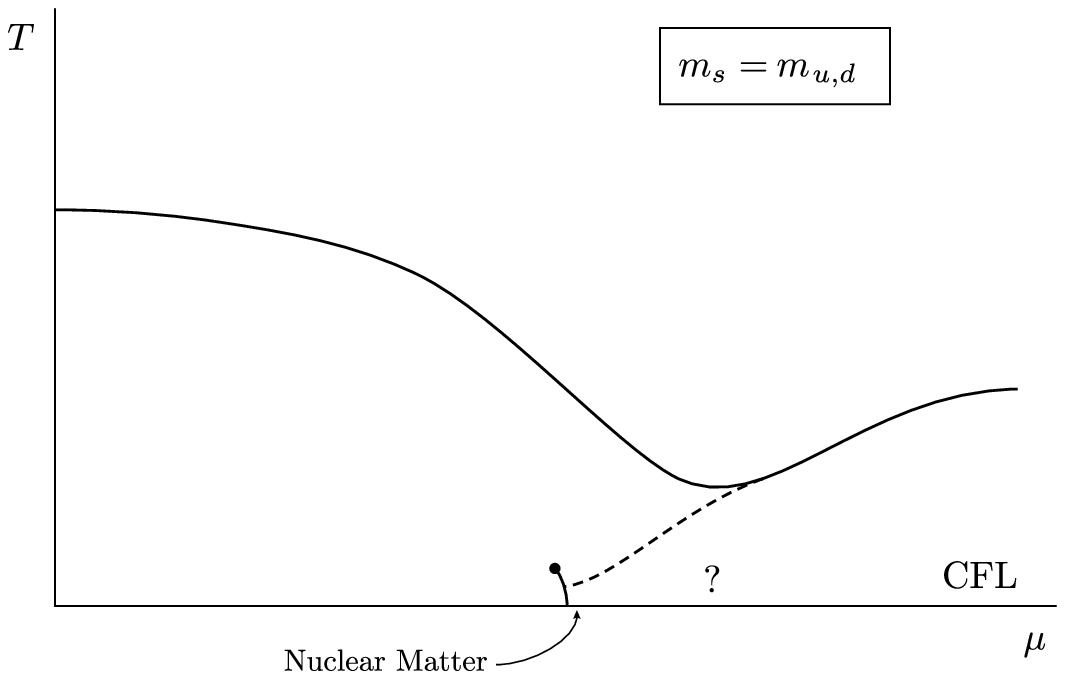}
\end{center} \label{fig4}
\caption{QCD phase diagram for three quarks which
are degenerate in mass and which are either massless or light.
The CFL phase
and the baryonic phase have the same symmetries and 
may be continuously connected. The dashed line 
denotes the critical temperature at which baryon-baryon (or
quark-quark) pairing vanishes; the region below the dashed line 
is superfluid.
Chiral symmetry is 
broken everywhere below the solid line, which is a first order
phase transition. 
The question mark serves to remind
us that although no transition is required in this region,
transition(s) may nevertheless arise as the magnitude of the gap
increases qualitatively in going from the hypernuclear to the
CFL phase.  For quark
masses as in nature, the high density region of the map
may be as shown in Figure~3 or may be closer to that shown here, albeit
with transition(s) in the vicinity of the question mark
associated with the onset
of nonzero hyperon density and the breaking 
of $U(1)_S$.\protect\cite{ABR2+1}}
\vspace{-0.15in}
\end{figure}

These observations fit together in a simple
and elegant fashion.
If we
could vary  $m_s$, we would find that as $m_s$
is reduced from infinity to $m_s^c$, the critical
point $E$ in the $(T,\mu)$ plane moves toward the $\mu=0$
axis.\cite{SRS1}  This is shown in Figures~2-4.
In nature, $E$ is at some nonzero $T_E$ and $\mu_E$.
When $m_s$ is reduced to $m_s^c$, between Figure~3 and Figure~4,
$\mu_E$ reaches zero.
Of course, experimentalists cannot vary $m_s$.  They
can, however, vary $\mu$.  AGS collisions with
center of mass energy 
$\sqrt{s}=5$~AGeV create fireballs which freeze out
near $\mu\sim 500-600$~MeV.\cite{PBM}  
SPS collisions with $\sqrt{s}=17$~AGeV
create fireballs which freeze out near $\mu\sim 200-300$~MeV.\cite{PBM}
In time, we will also have data from SPS collisions
with $\sqrt{s}=9$~AGeV and $\sqrt{s}=12$~AGeV
and from RHIC collisions with
$\sqrt{s}=56$, 130 and 200~AGeV and other energies.\footnote{The 
first data from RHIC collisions at $\sqrt{s}=56$~AGeV and 
$\sqrt{s}=130$~AGeV
have already appeared.\cite{PHOBOS,STAR} 
This bodes well for the analyses to come.} 
By dialing $\sqrt{s}$ and thus $\mu$, experimenters can find
the critical point $E$.  

We hope that the study of heavy ion collisions will, in the
end, lead both to a quantitative study of the 
properties of the quark-gluon plasma
phase at temperatures well above the transition and to a 
quantitative understanding of how to draw the phase transition
region of the phase diagram.  Probing the partonic matter
created early in the collision relies on a suite of signatures
including: the use of $J/\Psi$ mesons, charmed mesons, and perhaps
the $\Upsilon$ as probes; the energy loss of high momentum
partons and consequent effects on the high-$p_T$ hadron spectrum;
and the detection of photons and dileptons over and above those 
emitted in the later hadronic stages of the collision.
We will not review this program here.  Instead, we focus
on signatures of the critical point.
The map of the QCD phase diagram which we have sketched
so far is simple, coherent and consistent with all we know
theoretically; the discovery of the critical point would
provide an experimental foundation for the central
qualitative feature of the landscape.
This discovery would in addition
confirm that in higher energy heavy ion collisions and in
the big bang, the QCD phase transition is a smooth crossover.
Furthermore, the 
discovery of collisions
which create matter that
freezes out near $E$ would imply that 
conditions above the transition existed prior to freezeout,
and would thus make it much easier to interpret the results of other
experiments which study those observables which can probe the 
partonic matter created early in the collision. 

We theorists must clearly do as much as we can
to tell experimentalists {\it where} and {\it how} to find $E$.
The ``where'' question, namely the question of predicting
the value of $\mu_E$ and thus suggesting the $\sqrt{s}$ to
use to find $E$, is much harder for us to answer.   
First, as we have stressed in the previous Section, 
{\it ab initio} analysis of QCD in its full glory --- i.e.
lattice calculations --- are at present impossible at nonzero $\mu$.
We must therefore rely on models.
Second, an intrinsic feature of the picture we have described 
is that $\mu_E$ is sensitive to the mass of the strange quark,
and therefore particularly hard to predict.
Crude models suggest that $\mu_E$ could be $\sim 600-800$~MeV
in the absence of the strange quark;\cite{bergesraj,stephetal} 
this in turn suggests that
in nature $\mu_E$ may have of order half this value, and may therefore
be accessible at the SPS if the SPS 
runs with $\sqrt{s}<17$~AGeV.   However, at present theorists cannot
predict the value of $\mu_E$ even to within a factor of two.
The SPS can search a significant fraction of the parameter
space; if it does not find $E$, it will then be up to 
the RHIC experiments to map the $\mu< 250 $~MeV region.

Although we are trying
to be helpful with the ``where'' question, we are not very
good at answering it quantitatively.  This question can only
be answered convincingly by an experimental discovery.
What we theorists {\it can} do reasonably
well is to answer the ``how'' question, thus
enabling experimenters to answer ``where''.  
This is the goal of Ref.~95.
The signatures proposed there are based
on the fact that $E$ is a genuine thermodynamic singularity
at which susceptibilities diverge and the order parameter
fluctuates on long wavelengths. The resulting signatures
are {\it nonmonotonic} as a function of $\sqrt{s}$: as
this control parameter is varied, we should see the signatures
strengthen and then weaken again as the critical point is
approached and then passed.   

The critical point $E$ can also be sought by varying control
parameters other than $\sqrt{s}$. 
Ion size, centrality selection
and rapidity selection can all be varied.
The advantage of using $\sqrt{s}$  
is that we already know (by comparing
results from the AGS and SPS)
that dialing it changes 
the freeze out chemical potential $\mu$, which is the goal
in a search for $E$.

The event-by-event fluctuations of the 
mean transverse momentum of the charged particles
in an event, $p_T$, will be enhanced in collisions
which freeze out near $E$.\cite{SRS2}
The fluctuations 
measured by NA49 at $\sqrt{s}=17$~AGeV 
are as perfect Gaussians as the data statistics
allow,\cite{NA49} 
as expected for freeze-out from a system in thermal equilibrium.
The width of the event-by-event distribution\footnote{This width 
can be measured even
if one observes only two pions per event;~\cite{bialaskoch}
large acceptance data as from NA49 is required in order
to learn that the distribution is Gaussian, that 
thermodynamic predictions may be valid, and that
the width is therefore the only interesting quantity to measure.} 
of mean $p_T$
is in good agreement with predictions based on noncritical thermodynamic 
fluctuations in an equilibrated resonance gas.\cite{SRS2}
That is, NA49 data are consistent with the hypothesis
that almost all the observed event-by-event fluctuation in 
mean $p_T$, an intensive quantity,
is thermodynamic in origin.
This bodes well for the detectability of systematic
changes in thermodynamic fluctuations near $E$.

One 
analysis described in detail in Ref.~95 is based on the ratio
of the width of the true event-by-event distribution of the mean $p_T$
to the width of the distribution in a sample of mixed events. This
ratio was called $\sqrt{F}$. NA49 has measured $\sqrt{F}=1.002\pm
0.002$,\cite{NA49,SRS2} which is consistent with expectations for
noncritical thermodynamic fluctuations.\footnote{In an infinite system
made of classical particles which is in thermal equilibrium,
$\sqrt{F}=1$.  Bose effects increase $\sqrt{F}$ by $1-2\%$;~\cite{Mrow,SRS2} 
an anticorrelation introduced by energy conservation
in a finite system --- when one mode fluctuates up it is more likely
for other modes to fluctuate down --- decreases $\sqrt{F}$ by 
$1-2\%$;~\cite{SRS2} two-track resolution also decreases $\sqrt{F}$ 
by $1-2\%$.\cite{NA49} The contributions due to correlations introduced by
resonance decays and due to fluctuations in the flow velocity are each
much smaller than $1\%$.\cite{SRS2}}  Critical fluctuations
of the $\sigma$ field, i.e. the characteristic long wavelength
fluctuations of the order parameter near $E$, influence pion momenta
via the (large) $\sigma\pi\pi$ coupling and increase $\sqrt{F}$.\cite{SRS2}
The effect is proportional to $\xi_{\rm freezeout}^2$,
where $\xi_{\rm freezeout}$ is the $\sigma$-field correlation length
of the long-wavelength fluctuations at freezeout.\cite{SRS2}  If
$\xi_{\rm freezeout}\sim 3$ fm (a reasonable estimate, as we
describe below)
the ratio $\sqrt{F}$ increases by
$\sim 3-5\%$, ten to twenty times the statistical error in the present
measurement.\cite{SRS2}  This observable is valuable because data on
it has been analyzed and presented by NA49, and it can therefore be
used to learn that Pb+Pb collisions at 158~AGeV do {\it not} freeze
out near~$E$. The $3-5\%$ nonmonotonic
variation in $\sqrt{F}$ as a function
of $\sqrt{s}$  which we predict is easily detectable but is
not so large as to make one confident of using this alone
as a signature of $E$.

Once $E$ is located, however, other observables which 
are more sensitive to critical effects will be more useful.
For example, a $\sqrt{F_{\rm soft}}$,
defined using only the softest $10\%$ of the pions in each event, 
will be much more sensitive to the critical long wavelength 
fluctuations.  The higher $p_T$ pions are less affected
by the $\sigma$ fluctuations,\cite{SRS2} 
and these relatively unaffected pions
dominate the mean $p_T$ of all the pions in the
event.  This is why the increase in $\sqrt{F}$ near the critical point 
will be much less than that of $\sqrt{F_{\rm soft}}$. 
Depending on the details of the
cut used to define it,  $\sqrt{F_{\rm soft}}$ should be enhanced by many tens
of percent in collisions passing near $E$.
Ref.~95 suggests other such observables,
and more can surely be found.

The multiplicity of soft pions is an 
example of an observable which may
be used to detect the critical fluctuations 
without an event-by-event analysis.
The post-freezeout decay of sigma mesons, which are copious
and light at freezeout near $E$ and which
decay subsequently when their mass increases above
twice the pion mass, should result in a population of pions 
with $p_T\sim m_\pi/2$ which appears only for freezeout
near the critical point~\cite{SRS2}.  
If $\xi_{\rm freezeout}> 1/m_\pi$, this population
of unusually low momentum pions will be comparable in
number to that of the ``direct'' pions (i.e. those which
were pions at freezeout) and will result in a large
signature.  This signature is therefore certainly
large for $\xi_{\rm freezeout}\sim 3$~fm and would
not increase much further if $\xi_{\rm freezeout}$ were larger still.

The variety of observables
which should {\it all} vary nonmonotonically with $\sqrt{s}$
(and should all peak at the same $\sqrt{s}$)
is sufficiently great that if it were to turn out that 
$\mu_E<200$~MeV, making $E$ inaccessible to the SPS, all four
RHIC experiments could play a role in the study of the critical
point.

The purpose of Ref.~106 
is to estimate how large 
$\xi_{\rm freezeout}$ can become, thus making the predictions
of Ref.~95 for the magnitude of various signatures
more quantitative.  
The nonequilibrium dynamics analyzed in Ref.~106
is guaranteed
to occur in a heavy ion collision which passes near $E$,
even if local thermal equilibrium is achieved 
at a higher temperature during the earlier evolution
of the plasma created in the collision.  
If this plasma were to cool arbitrarily slowly, $\xi$ would
diverge at $T_E$.  However, it would take an infinite
time for $\xi$ to grow infinitely large.  Indeed, near
a critical point, the longer the correlation length, the
longer the equilibration time, and the slower the 
correlation length can grow. This critical slowing
down means that the
correlation length cannot grow sufficiently fast for the 
system to stay in equilibrium.  
We use the theory of dynamical critical phenomena\cite{HoHa} to 
describe the effects of critical slowing down
of the long wavelength dynamics near $E$
on the time development of the correlation length.
The correlation length does not have time
to grow as large as it would in equilibrium:  
we find 
$\xi_{\rm freezeout}\sim 2/T_E \sim 3$ fm for trajectories
passing near $E$. 
Although critical slowing down hinders the growth of $\xi$, it 
also slows the decrease of $\xi$  as the system continues to cool 
below the critical point.  As 
a result, $\xi$ does not
decrease significantly between the phase transition and 
freezeout.  

We have learned much from the beautiful gaussian event-by-event
fluctuations observed by NA49.  The magnitude of these fluctuations
are consistent with the hypothesis that the hadronic system
at freezeout is in approximate thermal equilibrium. These and
other data show none of the non-gaussian features that would 
signal that the system had
been driven far from equilibrium either by a rapid
traversal of the transition region or by the bubbling
that would occur near a strong first order
phase transition.  There is also no sign of
the enhanced, but still gaussian, fluctuations which would signal
freezeout near the critical point $E$.  Combining these 
observations with the observation of tantalizing indications
that the matter created in SPS collisions is not well described
at early times by hadronic models\cite{HeinzJacob} suggests
that collisions at the SPS may be exploring the crossover region
to the left of the critical point $E$, in which
the matter is not well-described as a hadron gas but
is also not well-described as a quark-gluon plasma.  This speculation
could be confirmed in two ways.   First, if the SPS is probing
the crossover region then the coming experiments
at RHIC may discover direct signatures of an early partonic phase, 
which are well-described by theoretical calculations beginning from 
an equilibrated quark-gluon plasma.  
Second, 
if $\sqrt{s}=17$~AGeV collisions
are probing the
crossover region not far to the left of the critical point $E$, 
then SPS data taken at lower energies
would result in the discovery of $E$. 
If, instead, RHIC were to discover
$E$ with $\mu_E<200$~MeV, that would indicate that the SPS 
experiments have probed the weakly first order region 
just to the right of $E$. Regardless, 
discovering $E$ would take all the speculation
out of mapping this part of the QCD phase diagram.

Let us now return to the high density, low temperature
regions of the phase diagram.  In this regime, as 
we have seen in previous sections, the most symmetric
starting point with which to begin our analysis 
is the case of three quarks with degenerate masses.
In QCD with $m_s=m_{u,d}$ as in Figure 4, cold dense quark matter is
in the CFL phase which has the same symmetries
as the hypernuclear matter
phase, characterized by a condensate of 
Cooper pairs of baryons.\cite{SW1}  
Furthermore,
many non-universal features of these two phases 
correspond.\cite{SW1}
This raises the possibility that quark matter and baryonic
matter may be continuously connected,\cite{SW1} 
as shown in Figure~4.

Both CFL quark matter and hypernuclear matter
are superfluids, characterized by a spontaneously
broken  global $U(1)$ symmetry
associated with baryon number.
Chiral symmetry is broken in these (or this!) phase.
This means that as we increase the temperature from zero
in Figure 4, there may be two distinct phase transitions
between the CFL/hypernuclear phase and the quark-gluon
plasma, as superfluidity is lost and chiral symmetry is
restored.  As hypernuclear matter is heated, we expect
superfluidity to be lost at a lower temperature than
that at which chiral symmetry is restored.
At high densities, on the other hand, we expect only a single
transition because both chiral symmetry breaking and
superfluidity are caused by the diquark condensate
which locks color to flavor. Figure 4 is drawn accordingly.
The transition at which chiral symmetry is restored
is first order, for the same reasons as at 
$\mu=0$: there is no infrared fixed point which could
describe a second order chiral symmetry restoration transition
in QCD with three massless quarks.\cite{piswil}

Nature chooses two light quarks and one middle-weight
strange quark, rather than three
degenerate quarks as in Figure~4. 
As we have
discussed in Section 3.2, a nonzero $m_s$ weakens those condensates which
involve pairing between light and strange quarks.
The CFL phase requires
nonzero $\langle us \rangle$ and $\langle ds \rangle$
condensates which 
can only exist if the associated gaps are 
larger than of order $m_s^2/2\mu$.
If one imagines increasing $m_s$ at fixed $\mu$, one finds a first order
unlocking transition:~\cite{ABR2+1,SW2} for larger
$m_s$ only $u$ and $d$
quarks pair and the 2SC phase is obtained.  
Conversely, as $m_s$
is reduced in going from Figure~2 to 3 to 4, the 
region occupied by the CFL phase expands to encompass 
regions with smaller
and smaller $\mu$.\cite{ABR2+1,SW2}  For  
any $m_s\neq \infty$,
the CFL phase is the ground state at arbitrarily
high density.\cite{ABR2+1}  For larger values of $m_s$,
there is a 2SC interlude on the horizontal axis,
in which chiral symmetry is restored, before
the CFL phase breaks it again at high densities.
For smaller values of $m_s$, the possibility of
quark-hadron continuity\cite{SW1} as shown in Figure~4 arises.
It should be noted that when the strange and light quarks 
are not degenerate,
the CFL phase may be continuous
with a baryonic phase in which the densities of
all the nucleons and hyperons are comparable; there
are, however, phase transition(s) between this 
hypernuclear phase and ordinary nuclear matter
made of neutrons and protons only.\cite{ABR2+1}

We can now describe what happens in QCD
with $m_{u,d}< m_s<\infty$ as in Figure~3 
to CFL quark matter as it is heated.  
Because the $\langle us \rangle$ and $\langle ds \rangle$
condensates are smaller than the $\langle ud \rangle$
condensate, they vanish first.  At this temperature,
we find a first order unlocking phase transition at
which chiral symmetry is restored.\footnote{That this transition  
is first order can be seen both via the same argument 
which demonstrates that the unlocking
transition between the 2SC and CFL phases at $T=0$ is first
order\cite{ABR2+1,SW2} and via the fact that it is a finite
temperature phase transition at which chiral symmetry is restored
in a three-flavor theory.}
Above this
transition, we find the 2SC phase in which only 
red and green $u$ and $d$ quarks pair.\footnote{As we have seen in Section 3, 
in the 2SC phase at zero temperature, there may
be small $J=1$ 
$\langle ss \rangle$ and blue $\langle ud \rangle$ condensates; these
condensates would not persist at the temperatures we are 
discussing here.}
The Cooper pairs are $ud-du$
flavor singlets and the global flavor symmetry 
$SU(2)_L\times SU(2)_R$ is intact. There
is also an unbroken global symmetry which plays the
role of $U(1)_B$. Thus, no global symmetries are broken
in this 2SC phase.  
There need therefore be no transition between
the 2SC and quark-gluon plasma phases in Figure~3 
(or in Figures~1 and 2) because neither phase breaks
any global symmetries.
However, this transition, which
is second order in mean field theory, is likely first
order in QCD due to gauge field fluctuations,\cite{bergesraj} 
at least at high enough 
density.\cite{PisarskiPhaseDiagram}

Throughout our description of the phase diagram, we have
assumed that the chemical potentials $\mu_u=\mu_d=\mu_s=\mu$
are degenerate.  Different values
of the strange quark $m_s$ lead to differences between
the Fermi momentum for $s$ quarks relative to that
for $u$ and $d$ quarks.   This analysis neglects
electromagnetic and weak effects. In particular, looking
ahead to the study of color superconductivity
in compact stars, it neglects the requirement
that the quark matter be electrically neutral and in beta
equilibrium. If the up, down and strange quarks were
degenerate, quark matter with $\mu_u=\mu_d=\mu_s=\mu$
would be electrically neutral and in beta equilibrium.
Instead, because $m_s\gg m_{u,d}$, the quark matter
in a compact star must have $\mu_{d,s} - \mu_u = \mu_e > 0$,
where $\mu_e$ is the electron chemical 
potential.\cite{RajagopalWilczekNeutrality}
In quark matter with average quark 
chemical potential $\mu\sim 400 - 500$~MeV,
as may occur in a compact star, reasonable
estimates for $\mu_e$ are in the few tens of MeV.
The presence of a small but nonzero $\mu_e$ would
certainly have quantitative effects in Figures 1 to 4;
we shall see in Section 6.6 that it can have qualitative
effects also.

\section{Color Superconductivity in Compact Stars}

Our current understanding of the color superconducting
state of quark matter leads us to believe that it
may occur naturally in compact stars. 
The critical temperature $T_c$ below which quark matter 
is a color superconductor is high enough that
any quark matter which occurs within
neutron stars that are more than a few seconds old
is in a color superconducting state.
In the absence of lattice simulations, present theoretical
methods are not accurate enough to determine whether 
neutron star cores are made of hadronic matter or quark
matter.  They also cannot determine whether any quark
matter which arises will be in the CFL or 2SC phase: 
the difference between the $u$, $d$ and $s$ Fermi momenta
will be a few tens of MeV which is comparable to estimates
of the gap $\Delta$; the CFL phase occurs when $\Delta$ is
large compared to all differences between Fermi momenta.
Just as the higher temperature regions of the QCD
phase diagram are being mapped out in heavy ion collisions,
we need to learn how to use neutron star phenomena to 
determine whether they feature cores made of 2SC quark matter,
CFL quark matter or hadronic matter, thus teaching us
about the high density region of the QCD phase diagram.
It is therefore important to look for astrophysical consequences of
color superconductivity.

\subsection{Equation of State} 

Much of the work on the consequences 
of quark matter within a compact star has focussed on
the effects of quark matter on the equation of state,
and hence on the radius of the star.  As a Fermi surface
phenomenon, color superconductivity has little effect on
the equation of state: the pressure is an integral over
the whole Fermi volume.  Color superconductivity 
modifies the equation of state at the $\sim (\Delta/\mu)^2$
level, typically by a few percent.\cite{ARW1}  Such small effects
can be neglected in present calculations, and for
this reason we will not attempt to survey
the many ways in which observations of neutron stars
are being used to constrain the equation of state.\cite{Henning}

We will describe one current idea, however.
As a neutron star in a low mass X-ray binary (LMXB)
is spun up by accretion from its companion, it becomes
more oblate and its central density decreases. If it contains
a quark matter core, the volume fraction occupied by this
core decreases, the star expands, and its moment of inertia
increases.  This raises the possibility\cite{GlendenningWeberSpinup}
of a period during the spin-up history of an LMXB when
the neutron star is gaining angular momentum via accretion,
but is gaining sufficient moment of inertia that its angular
frequency is hardly increasing.  In their modelling of this effect,
Glendenning and Weber\cite{GlendenningWeberSpinup} discover 
that LMXB's should spend a significant fraction
of their history with a frequency of around 200~Hz,
while their quark cores are being spun out of existence,
before eventually spinning up to higher frequencies.  
This may explain the observation that 
LMXB frequencies are clustered around 250-350~Hz,\cite{vanderKlis}
which is otherwise puzzling in that it is thought that LMXB's provide
the link between canonical pulsars and millisecond pulsars,
which have frequencies as large as 600~Hz.\cite{ChakrabartyMorgan}
It will be interesting to see how robust the result of 
Ref.~111 is to changes in model
assumptions and also how 
its predictions fare when compared to 
those of other 
explanations which posit upper bounds on LMXB 
frequencies,\cite{Bildsten2}
rather than a most probable frequency range with no 
associated upper bound.\cite{GlendenningWeberSpinup} 
We note here that because Glendenning
and Weber's  effect depends only 
on the equation of state and not on other
properties of quark matter, the fact that the quark
matter must in fact be a color superconductor
will not affect the results in any significant way.
If Glendenning and Weber's explanation for the observed clustering
of LMXB frequencies proves robust, it would imply
that pulsars with lower rotational frequencies feature quark matter
cores.  

\subsection{Cooling by Neutrino Emission}

We turn now to neutron star phenomena which {\it are} affected
by Fermi surface physics.  For the first $10^{5-6}$ years of its
life, the cooling of a neutron star is governed by the balance
between heat capacity and the loss of heat by neutrino emission.
How are these quantities affected by the presence of a
quark matter core? This has been addressed recently in 
Refs.~115,116, following earlier work in Ref.~117.
Both the specific heat $C_V$  and the neutrino emission rate 
$L_\nu$ are dominated
by physics within $T$ of the Fermi surface.  If, as 
in the CFL phase,  all quarks have a gap $\Delta\gg T$ then
the contribution of quark quasiparticles to $C_V$ and $L_\nu$ 
is suppressed by $\sim \exp(-\Delta/T)$.  There may be other
contributions to $L_\nu$,\cite{Blaschke} but these are also
very small.  In the CFL phase, the specific heat is  
dominated by that of the electrons.\cite{RajagopalWilczekNeutrality} 
(There is an additional contribution
from the superfluid
mode in the CFL phase --- i.e. the Goldstone boson
associated with the spontaneous breaking of $U(1)_B$ ---
and there 
may also be small contributions from the light but not
massless pseudo-Goldstone
bosons associated with chiral symmetry breaking.)  
Although further work is required, it is already
clear that both $C_V$ and $L_\nu$ are much smaller than in
the nuclear matter outside the quark matter core. This
means that the total heat capacity and
the total neutrino emission rate (and hence
the cooling rate) of a neutron star with a CFL core will 
be determined completely by the nuclear matter outside 
the core.  The quark matter core is ``inert'':
with its small heat capacity and emission rate it
has little influence on the temperature of the star as a whole.
As the rest of the star emits neutrinos and cools, the core
cools by conduction, because the electrons keep it in good thermal
contact with the rest of the star.   These qualitative expectations
are nicely borne out in the calculations presented
by Page et al.\cite{Page}

The analysis of the cooling history of a neutron star with 
a quark matter core in the 2SC phase is more complicated.
The red and green up and down quarks pair with a gap
many orders of magnitude larger than the temperature, which is
of order 10 keV, and
are therefore inert as described above.  
Any strange quarks present will form an
$\langle ss \rangle$ condensate with
angular momentum $J=1$ which locks to color
in such a way that rotational invariance is not
broken.\cite{Schaefer1Flavor}
The resulting gap has been estimated to be of order
hundreds of keV,\cite{Schaefer1Flavor} although applying
results of Ref.~49 suggests a somewhat smaller gap, around
10 keV.  The blue up and down quarks also pair, forming
a $J=1$ condensate which breaks rotational invariance.\cite{ARW1}
The related gap was estimated to be a few keV,\cite{ARW1} but this 
estimate was not robust and should be revisited in light of more
recent developments given its importance
in the following.  The critical temperature $T_c$ above
which no condensate forms is of order the  zero-temperature gap
$\Delta$. ($T_c=0.57 \Delta$ for $J=0$ condensates.\cite{PisarskiRischke}) 
Therefore, 
if there are quarks for which $\Delta\sim T$ or smaller, these quarks
do not pair at temperature $T$. Such quark quasiparticles 
will radiate neutrinos rapidly (via direct URCA
reactions like $d\rightarrow u+e+\bar\nu$, 
$u\rightarrow d+e^+ +\nu$, etc.)
and the quark matter core will cool rapidly and determine the
cooling history of the star as a whole.\cite{Schaab,Page}
The star
will cool rapidly until its interior temperature is
$T<T_c\sim\Delta$, at which time the quark matter core will become
inert and the further cooling history will be dominated by
neutrino emission from the nuclear matter fraction of the star. 
If future data were to show that neutron
stars first cool rapidly (direct URCA) and then cool more
slowly, such data would allow an estimate of the smallest 
quark matter gap. We are unlikely to be so lucky.
The simple observation of rapid cooling would {\it not} be an unambiguous
discovery of quark matter with small gaps; there are other
circumstances in which the direct URCA processes occur.
However, if as data on neutron star temperatures improves in coming
years the standard cooling scenario proves correct,
indicating the absence of the direct URCA processes, 
this {\it would} rule out the presence
of quark matter with gaps in the 10 keV range or smaller.  
The presence of a quark matter core
in which {\it all} gaps are $\gg T$ can never be revealed by
an analysis of the cooling history.

\subsection{Supernova Neutrinos}

We now turn from neutrino emission from a neutron star which
is many years old to that from the protoneutron star 
during the first seconds of  a supernova.
Carter and Reddy\cite{CarterReddy}
have pointed out that when this protoneutron
star is at its maximum temperature of order 30-50~MeV,
it may have a quark matter core which is too hot for color
superconductivity.  As such a  protoneutron star core cools
over the next few seconds, this quark
matter will cool through $T_c$, entering the color superconducting
regime of the QCD phase diagram.  For $T\sim T_c$, the
specific heat rises and the cooling slows. Then, as $T$ drops
further and $\Delta$ increases to become greater than $T$,
the specific heat drops rapidly. Furthermore, as the number
density of quark quasiparticles becomes suppressed by $\exp(-\Delta/T)$,
the neutrino transport mean free path rapidly 
becomes very long.\cite{CarterReddy}
This means that all the neutrinos previously trapped
in the now color superconducting
core are able to escape in a sudden burst.  If 
a terrestrial neutrino detector
sees thousands of neutrinos from a future supernova, Carter
and Reddy's results suggest that there may be a signature of the
transition to color superconductivity present in the time distribution
of these neutrinos.  Neutrinos from the core of the protoneutron
star will lose energy as they scatter on their way out, but because
they will be the last to reach the surface of last scattering, they
will be the final neutrinos received at the earth.  If they are released
from the quark matter core in a sudden burst, they may therefore
result in a bump at late times in the temporal distribution of
the detected neutrinos.  More detailed study remains to be done
in order to understand how Carter and Reddy's signature, dramatic
when the neutrinos escape from the core, is processed as the neutrinos
traverse the rest of the protoneutron star and reach their
surface of last scattering.

\subsection{R-mode Instabilities}
  
Another arena in which color superconductivity comes into play
is the physics of r-mode instabilities.  A neutron star whose
angular rotation frequency $\Omega$ is large enough is unstable
to the growth of r-mode oscillations which radiate 
away angular momentum via gravitational waves, reducing $\Omega$.
What does ``large enough'' mean?  The answer depends on 
the damping mechanisms which act to prevent the growth of
the relevant modes.  Both shear viscosity and bulk viscosity
act to damp the r-modes, preventing them from going unstable.
The bulk viscosity and the quark contribution
to the shear viscosity both become exponentially
small in quark matter with $\Delta>T$ and as a result,
as Madsen\cite{Madsen} has shown, 
a compact star made {\it entirely} of quark matter with
gaps $\Delta=1$~MeV or greater is 
unstable if its spin frequency is greater than tens to 100~Hz.
Many compact stars spin faster than this, and Madsen therefore
argues that compact stars cannot be strange quark stars
unless some quarks remain ungapped.  Alas, this powerful argument
becomes much less powerful in the context of a neutron star
with a quark matter core.  First, the r-mode oscillations 
have a wave form whose amplitude is largest at large radius,
outside the core. Second, in an ordinary neutron star there
is a new source of damping: friction at the boundary between
the crust and the neutron superfluid ``mantle'' keeps the 
r-modes stable regardless of the properties of a quark matter 
core.\cite{Bildsten,Madsen}

\subsection{Magnetic Field Evolution}

Next, we turn to the physics of magnetic fields within
color superconducting neutron star cores.\cite{Blaschkeflux,ABRflux}  
The interior
of a conventional neutron star is a superfluid (because of neutron-neutron
pairing) and is an electromagnetic superconductor
(because of proton-proton pairing).  Ordinary magnetic fields
penetrate it only in the cores of magnetic flux tubes.
A color superconductor behaves differently. At first
glance, it seems that because a diquark Cooper pair has nonzero
electric charge, a diquark condensate
must exhibit the standard Meissner effect, expelling
ordinary magnetic fields or restricting them to flux tubes
within whose cores the condensate vanishes.  This is not
the case.\cite{ABRflux}
In both the 2SC and CFL phase, a linear combination
of the $U(1)$ gauge transformation of ordinary electromagnetism
and one (the eighth) color gauge transformation remains unbroken 
even in the presence of the condensate.  This means that 
the ordinary photon $A_\mu$ and the eighth gluon $G_\mu^8$
are replaced by new linear combinations
\begin{eqnarray}
A_\mu^{\tilde Q} &=& \cos\alpha_0 \,A_\mu + \sin\alpha_0\,G_\mu^8\,,
\nonumber\\[1ex]
A_\mu^{X} &=& -\sin\alpha_0\,A_\mu + \cos\alpha_0\,G_\mu^8\,,
\end{eqnarray}
where $A_\mu^{\tilde Q}$ is massless and $A_\mu^{X}$ is massive.
That is, $B_{\tilde Q}$ satisfies the ordinary Maxwell
equations while $B_X$ experiences a Meissner effect.
$\sin(\alpha_0)$ is proportional to $e/g$ and turns
out to be about $1/20$ in the 2SC phase and $1/40$ in the CFL
phase.\cite{ABRflux}  This means that the 
$\tilde Q$-photon which propagates in color superconducting
quark matter is mostly photon with only 
a small gluon admixture. If a color superconducting neutron star core 
is subjected to an ordinary magnetic field, it will either
expel the $X$ component of the flux
or restrict it to flux tubes, but it can
(and does\cite{ABRflux}) admit the great majority of the flux
in the form of a $B_{\tilde Q}$ magnetic field satisfying
Maxwell's equations.   
The decay in time of this ``free field'' (i.e. not in flux tubes) 
is limited by the $\tilde Q$-conductivity of the quark matter.
A color superconductor is not a $\tilde Q$-superconductor --- 
that is the whole point --- but it may turn out 
to be a very good
$\tilde Q$-conductor due to the presence of electrons:
if a nonzero density of electrons is required in order to maintain
charge neutrality,  
the $B_{\tilde Q}$ magnetic field likely 
decays only on a time scale 
which is much longer than the age of the universe.\cite{ABRflux}
This means that a quark matter core within a neutron
star can serve as an ``anchor'' for the magnetic field:
whereas in ordinary nuclear matter the magnetic flux
tubes can be dragged outward by the neutron superfluid
vortices as the star spins down,\cite{Srinivasan} 
the magnetic flux within the 
color superconducting core simply cannot decay.
Even though this distinction is a qualitative one, it
will be 
difficult to confront it with data since what is
observed is the total dipole moment of the neutron star.
A color superconducting
core anchors those magnetic flux lines which pass through
the core, while in a neutron star with no quark matter core
the entire internal magnetic field can decay over time. 
In both cases, however, the total dipole moment can change
since the magnetic flux lines which do not pass through
the core can move.

\subsection{Crystalline Color Superconductivity and Glitches in Quark Matter}

The final consequence of color superconductivity 
we wish to discuss is the possibility that (some)
glitches may originate within quark matter regions of a 
compact star.\cite{BowersLOFF}
In any context in which color superconductivity arises
in nature, it is likely to involve pairing between species of quarks
with differing chemical potentials.   
If the chemical potential difference  
is small enough, BCS pairing
occurs as we have been discussing. 
If the Fermi surfaces are too far apart, no pairing between the species is
possible. The transition between the BCS and unpaired states as the
splitting between Fermi momenta increases has been studied in
electron superconductors,\cite{Clogston} 
nuclear superfluids\cite{Sedrakian} and QCD
superconductors,\cite{ABR2+1,SW2,Bedaque} assuming 
that no other state intervenes.  However,
there is good reason to think that another state can occur.  This is
the ``LOFF'' state, first explored by Larkin and Ovchinnikov\cite{LO}
and Fulde and Ferrell\cite{FF} in the context of electron
superconductivity in the presence of magnetic impurities.
They found that near the
unpairing transition, 
it is favorable to form a state in
which the Cooper pairs have nonzero momentum. This is favored because
it gives rise to a region of phase space where each of the two quarks
in a pair can be close to its Fermi surface,
and such pairs can be created at low cost in free energy.
Condensates of this sort spontaneously
break translational and rotational invariance, leading to
gaps which vary periodically in a crystalline pattern.
If in some shell within the quark matter core
of a neutron star (or within a strange quark star)  
the quark number densities are
such that crystalline color superconductivity arises,
rotational vortices may be pinned in this shell, making
it a locus for glitch phenomena.

The authors of Ref.~49, 
have explored the range of parameters
for which crystalline color superconductivity occurs in
the QCD phase diagram, upon making various simplifying assumptions.
We focus primarily on a toy model in which the
quarks interact via a
four-fermion interaction
with the quantum numbers of single gluon exchange.
Also, we only consider pairing between $u$ and $d$ quarks, with
$\mu_d=\bar\mu+\delta\mu$ and $\mu_u=\bar\mu-\delta\mu$, whereas
we expect a LOFF state wherever the difference between the Fermi momenta
of any two quark flavors is near an unpairing transition, including,
for example, near the unlocking phase transition between the 2SC and
CFL phases.  

In the LOFF state, each Cooper pair carries 
momentum $2{\bf q}$ with $|{\bf q}|\approx 1.2 \delta\mu$.
The condensate and gap parameter vary in space with wavelength
$\pi/|{\bf q}|$.  In Ref.~49, we simplify
the calculation by assuming that the condensate varies in space
like a plane wave, leaving the determination of the crystal
structure of the QCD LOFF phase to future work. 
We give an ansatz for the LOFF wave function,
and by variation obtain a gap equation which allows
us to solve for the gap parameter $\Delta_A$, the free energy and
the values of the diquark condensates which characterize
the LOFF state at a given $\delta\mu$ and $|{\bf q}|$. 
We then vary $|{\bf q}|$, to find the preferred (lowest
free energy) LOFF state at a given $\delta\mu$, and compare
the free energy of the LOFF state to that of the BCS state with
which it competes. 
The LOFF state is characterized by a gap parameter $\Delta_A$ and a 
diquark condensate, but not by an energy gap in the dispersion
relation: we obtain the quasiparticle dispersion 
relations\cite{BowersLOFF} and find that they vary
with the direction of the momentum, yielding gaps that vary from zero
up to
a maximum of $\Delta_A$.  The condensate is dominated by
the regions in momentum space in which a quark pair
with total momentum $2{\bf q}$ has both members of
the pair within $\sim \Delta_A$ of their respective 
Fermi surfaces.

The LOFF state is favored
for values of $\delta\mu$ which satisfy 
$\delta\mu_1 < \delta\mu < \delta\mu_2$, 
with $\delta\mu_1/\Delta_0=0.707$ and $\delta\mu_2/\Delta_0=0.754$ in the 
weak coupling limit in 
which $\Delta_0\ll \mu$.  (For $\delta\mu<\delta\mu_1$, we
have the 2SC phase with gap $\Delta_0$.)
At weak coupling, the LOFF gap parameter decreases from $0.23 \Delta_0$
at $\delta\mu=\delta\mu_1$ (where there is a first order BCS-LOFF
phase transition)
to zero at $\delta\mu=\delta\mu_2$ (where there is a second order
LOFF-normal transition).  
Except for very close to $\delta\mu_2$, the critical
temperature above which the LOFF state melts will be much
higher than typical neutron star temperatures.
At stronger coupling the LOFF gap parameter decreases relative
to $\Delta_0$ and 
the window of $\delta\mu/\Delta_0$ within which the LOFF state
is favored shrinks. 
The window grows if the interaction is changed to
weight electric gluon exchange more heavily than
magnetic gluon exchange.


Near the second-order critical point $\delta\mu_2$, we can describe the
phase transition with a Ginzburg-Landau effective potential.
The order parameter for the LOFF-to-normal phase transition is
\begin{equation}\label{lofforderparam}
\Phi({\bf r}) = -\frac{1}{2} \langle \epsilon_{ab} \epsilon_{\alpha\beta3} 
\psi^{a\alpha}({\bf r}) C \gamma_5 \psi^{b\beta}({\bf r}) \rangle 
\end{equation}
so that in the normal phase $\Phi({\bf r}) = 0$, while in the LOFF phase
$\Phi({\bf r}) = \Gamma_A e^{i 2 {\bf q} \cdot {\bf r}}$.  (The gap parameter
is related to the order parameter by $\Delta_A=G\Gamma_A$.)
Expressing the order
parameter in terms of its Fourier modes $\tilde\Phi({\bf k})$, we write
the LOFF free energy (relative to the normal state) as
\begin{equation}\label{ginzland}
F(\{\tilde\Phi({\bf k})\}) = \sum_{{\bf k}} \left( C_2(k^2) | 
\tilde\Phi({\bf k}) |^2 
+ C_4(k^2) | \tilde\Phi({\bf k}) |^4 + {\mathcal O}(|\tilde\Phi|^6) \right).
\end{equation}
For $\delta\mu > \delta\mu_2$, 
all of the $C_2(k^2)$ are positive and the normal
state is stable.  Just below the critical point, all of the modes
$\tilde\Phi({\bf k})$ are stable except those on the sphere $|{\bf k}| =
2q_2$, where $q_2$ is the value of $|{\bf q}|$ at $\delta\mu_2$ 
(so that $q_2\simeq 1.2 \delta\mu_2 \simeq 0.9 \Delta_0$ 
at weak coupling).  In general,
many modes on this sphere can become nonzero, giving a
condensate with a complex crystal structure.  We consider the simplest
case of a plane wave condensate where only the one mode
$\tilde\Phi({\bf k} = 2{\bf q}_2) = \Gamma_A$ is nonvanishing.  Dropping all
other modes, we have
\begin{equation}\label{ginzland2}
F(\Gamma_A) = a(\delta\mu - \delta\mu_2) (\Gamma_A)^2 + b (\Gamma_A)^4 \,,
\end{equation}
where $a$ and $b$ are positive constants.  Finding the minimum-energy
solution for $\delta\mu < \delta\mu_2$, we obtain simple power-law relations
for the condensate and the free energy:
\begin{equation}\label{powerlaws}
\Gamma_A(\delta\mu) = K_{\Gamma} (\delta\mu_2 - \delta\mu)^{1/2}, 
\hspace{0.3in} F(\delta\mu) = - K_F ( \delta\mu_2- \delta\mu)^2.
\end{equation}
These expressions agree well with the numerical results obtained
by solving the gap equation.\cite{BowersLOFF}  
The Ginzburg-Landau
method does not specify the proportionality factors $K_\Gamma$ and
$K_F$, but analytical expressions for these coefficients can be
obtained in the weak coupling limit by explicitly solving the gap
equation,\cite{Takada1,BowersLOFF} yielding
\begin{equation}\label{keqns}
\begin{array}{rclcl}
G K_\Gamma &=& 2 \sqrt{\delta\mu_2} \sqrt{(q_2/\delta\mu_2)^2  - 1}  
&\simeq& 1.15 \sqrt{\Delta_0}\ , \\[0.5ex]
K_F &=& (4\bar\mu^2/\pi^2)((q_2/\delta\mu_2)^2-1) &\simeq& 0.178 \bar\mu^2\ .
\end{array}
\end{equation}
Notice that because $(\delta\mu_2-\delta\mu_1)/\delta\mu_2$ is small, the
power-law relations (\ref{powerlaws}) are a good model of the system
throughout the entire LOFF 
interval $\delta\mu_1 < \delta\mu < \delta\mu_2$ where the
LOFF phase is favored over the BCS phase.  The Ginzburg-Landau
expression (\ref{ginzland2}) gives the free energy of the LOFF phase
near $\delta\mu_2$, but it cannot be used to determine the location
$\delta\mu_1$ of the first-order phase transition where the LOFF window
terminates. (Locating the first-order point requires a comparison of
LOFF and BCS free energies.)

The quark matter which may be 
present within a compact star will be in
the crystalline color superconductor (LOFF) state 
if $\delta\mu/\Delta_0$ is in the requisite range.  
For a reasonable value of $\delta\mu$, say 25~MeV,
this
occurs if the gap $\Delta_0$ which characterizes the uniform
color superconductor present at smaller values of $\delta\mu$ is 
about 40~MeV. This is in the middle of the range of present
estimates.  Both $\delta\mu$ and $\Delta_0$ vary as a function
of density and hence as a function of radius in a compact star.
Although it is too early to make quantitative predictions,
the numbers are such that crystalline color superconducting
quark matter may very well occur in a range of radii within a compact 
star. It is therefore worthwhile to consider the consequences.

Many pulsars have been observed to glitch.  Glitches are sudden
jumps in rotation frequency $\Omega$ which may
be as large as $\Delta\Omega/\Omega\sim 10^{-6}$, but may also
be several orders of magnitude smaller. The frequency of observed
glitches is statistically consistent with the hypothesis that 
all radio pulsars experience glitches.\cite{AlparHo}
Glitches are thought to originate from interactions
between the rigid neutron star crust, typically somewhat 
more than a kilometer thick, and rotational vortices in a
neutron superfluid. 
The inner kilometer of crust
consists of a crystal lattice of nuclei immersed in 
a neutron superfluid.\cite{NegeleVautherin}
Because the pulsar is spinning, the neutron superfluid 
(both within the inner crust and deeper inside the star) 
is threaded with
a regular array of rotational vortices.  As the pulsar's spin
gradually slows,
these vortices must gradually move outwards since the rotation frequency
of a superfluid is proportional to the density of vortices. 
Deep within the star, the vortices are free to move outwards.
In the crust, however, the vortices are pinned by their interaction
with the nuclear lattice.  
Models\cite{GlitchModels} differ
in important respects as to how the stress associated
with pinned vortices is released in a glitch: for example,
the vortices may break and rearrange the crust, or a cluster
of vortices may suddenly overcome the pinning force and 
move macroscopically outward, with
the sudden decrease in the angular momentum
of the superfluid within the crust resulting in a sudden increase
in angular momentum of the rigid crust itself and hence a glitch.
All the models agree that the fundamental requirements
are the presence of rotational vortices in a superfluid 
and the presence
of a rigid structure which impedes the motion of vortices and
which encompasses enough of the volume of the pulsar to contribute
significantly to the total moment of inertia.

Although it is 
premature to draw quantitative conclusions,
it is interesting to speculate that some glitches may originate 
deep within a pulsar which features
a quark matter core, in a region of that core 
which is in
a LOFF crystalline color superconductor phase.
A three flavor analysis is required 
to estimate over what range
of densities LOFF phases may arise, as either 
$\langle ud \rangle$, $\langle us \rangle$ or $\langle ds \rangle$
condensates approach their unpairing transitions.  Comparison
to existing models which describe how $p_F^u$, $p_F^d$ and $p_F^s$
vary within a quark matter core in a neutron star\cite{Glendenning} 
would then
permit an estimate of how much the LOFF region contributes to
the moment of inertia of the pulsar.  Furthermore, a three 
flavor analysis is required
to determine whether the LOFF
phase is a superfluid.   If the only pairing is between $u$
and $d$ quarks, this 2SC phase is not a superfluid,\cite{ARW1,ABR2+1}
whereas if all three
quarks pair in some way, a superfluid {\it is} 
obtained.\cite{CFL,ABR2+1}
Henceforth, we suppose  that the LOFF phase is a superfluid, 
which means that if it occurs within a pulsar it will be threaded
by an array of rotational vortices.
It is reasonable to expect that these vortices will
be pinned in a LOFF crystal, in which the
diquark condensate varies periodically in space.
Indeed, one of the suggestions for how to look for a LOFF phase in
terrestrial electron superconductors relies on the fact that
the pinning of magnetic flux tubes (which, like the rotational vortices
of interest to us, have normal cores)
is expected to be much stronger
in a LOFF phase than in a uniform BCS superconductor.\cite{Modler}

A real calculation of the pinning force experienced by a vortex in a
crystalline color superconductor must await the determination of the
crystal structure of the LOFF phase. We can, however, attempt an order
of magnitude estimate along the same lines as that done by Anderson
and Itoh\cite{AndersonItoh} for neutron vortices in the inner crust
of a neutron star. In that context, this estimate has since been made
quantitative.\cite{Alpar77,AAPS3,GlitchModels}  
For one specific choice of parameters,\cite{BowersLOFF} the LOFF phase
is favored over the normal phase by a free energy 
$F_{\rm LOFF}\sim 5 \times (10 {\rm ~MeV})^4$ 
and the spacing between nodes in the LOFF
crystal is $b=\pi/(2|{\bf q}|)\sim 9$ fm.
The thickness of a rotational vortex is
given by the correlation length $\xi\sim 1/\Delta \sim 25$ fm.  
The pinning energy
is the difference between the energy of a section of vortex of length 
$b$ which is centered on a node of the LOFF crystal vs. one which
is centered on a maximum of the LOFF crystal. It 
is of order $E_p \sim F_{\rm LOFF}\, b^3 \sim 4 {\rm \ MeV}$.
The resulting pinning force per unit length of vortex is of order
$f_p \sim E_p/b^2 \sim  (4 {\rm \ MeV})/(80 {\rm \ fm}^2)$.
A complete calculation will be challenging because
$b<\xi$, and is likely to yield an $f_p$
which is somewhat less than that we have obtained by dimensional 
analysis.
Note that our estimate of $f_p$ is
quite uncertain both because it is
only based on dimensional analysis and because the values
of $\Delta$, $b$ and $F_{\rm LOFF}$ are 
uncertain.  (We have a good understanding of 
all the ratios $\Delta/\Delta_0$, $\delta\mu/\Delta_0$, $q/\Delta_0$ 
and consequently $b\Delta_0$ in the LOFF phase.  It is 
of course the value of the BCS gap $\Delta_0$ which is uncertain.) 
It is premature to compare our crude result 
to the results of serious calculations 
of the pinning of crustal neutron vortices as in 
Refs.~135,136,131.  It is nevertheless
remarkable that they prove to be similar: the pinning
energy of neutron vortices in the inner crust 
is $E_p \approx 1-3  {\rm \ MeV}$
and the pinning force per unit length is
$f_p\approx(1-3 {\rm ~MeV})/(200-400 {\rm ~fm}^2)$.

The reader
may be concerned that a glitch deep within the quark
matter core of a neutron star may not be observable:  the
vortices within the crystalline
color superconductor region suddenly unpin and leap
outward;  this loss of angular momentum is compensated
by a gain in angular momentum 
of the layer outside the
LOFF region; how quickly, then, does this increase
in angular momentum manifest itself at the {\em surface} of
the star as a glitch? 
The important point here is that the rotation of any superfluid
region within which the vortices are able to move freely is
coupled to the rotation of the outer 
crust on very short time scales.\cite{AlparLangerSauls}
This rapid coupling, due to electron scattering off vortices
and the fact that the electron fluid penetrates throughout the 
star, is usually invoked to explain that the core
nucleon superfluid speeds up quickly after a crustal glitch:
the only long relaxation time is that of the vortices within
the inner crust.\cite{AlparLangerSauls}
Here, we invoke it to explain that the outer crust speeds
up rapidly after a LOFF glitch has accelerated the quark matter
at the base of the nucleon superfluid. 
After a glitch in the LOFF region, the only
long relaxation times are those of the vortices in the LOFF
region and in the inner crust.

A quantitative theory of glitches originating within
quark matter in a LOFF phase must await further 
calculations, in particular a three flavor analysis and
the determination of the crystal structure of the QCD LOFF phase.
However, our rough estimate of the pinning force on 
rotational vortices in a LOFF region suggests that this force may be 
comparable to that on vortices in the inner crust of a conventional
neutron star.
Perhaps, therefore, glitches occurring in a region of crystalline
color superconducting quark matter may yield similar phenomenology
to those occurring in the inner crust.  This is surely
strong motivation for further investigation.

Perhaps the most interesting consequence of these speculations
arises in the context of compact stars made entirely of 
strange quark matter.  The work of Witten\cite{Witten}
and Farhi and Jaffe\cite{FarhiJaffe} raised the possibility
that strange quark matter may be stable relative
to nuclear matter even at zero 
pressure.
If this is the
case it raises the question whether observed compact stars---pulsars,
for example---are strange quark stars\cite{HZS,AFO} rather than
neutron stars.  
A conventional neutron star may feature
a core made of strange quark matter, as we have been discussing 
above.\footnote{Note that a convincing discovery of
a quark matter core within an otherwise hadronic
neutron star would demonstrate
conclusively that strange quark matter is {\it not} stable
at zero pressure, thus ruling out the existence of strange
quark stars.  It is not possible for 
neutron stars with quark matter cores and strange quark
stars to both be stable.}
Strange quark stars, on the other hand, are made (almost)
entirely of quark
matter with either no hadronic matter content at all or
with a thin crust, of order one hundred meters thick, which contains
no neutron superfluid.\cite{AFO,GlendenningWeber}
The nuclei in this thin crust
are supported above the quark matter by electrostatic forces;
these forces cannot support a neutron fluid.  Because
of the absence of superfluid neutrons, and because of the thinness of
the crust, no successful models of glitches in the crust
of a strange quark star have been proposed.  
Since pulsars are observed to glitch, the apparent lack of a 
glitch mechanism for strange
quark stars  has been the 
strongest argument that pulsars cannot be strange quark 
stars.\cite{Alpar,OldMadsen,Caldwell}
This conclusion must now be revisited.  

Madsen's conclusion\cite{Madsen} that a strange
quark star is prone to r-mode instability due to
the absence of damping must
also be revisited, since the relevant oscillations
may be damped within or at the boundary of a 
crystalline color superconductor region.

The quark 
matter in a strange quark star, should
one exist, would be a color superconductor.
Depending on the mass of the star, the 
quark number densities increase by a factor of about two to ten
in going from the surface to the center.\cite{AFO} This means
that the chemical potential differences among the three
quarks will vary also, and there could be a range of radii
within which the quark matter is in a crystalline
color superconductor phase.  This raises the 
possibility of glitches in strange quark stars.
Because the
variation in density with radius is gradual, if a shell
of LOFF quark matter exists it need not be particularly thin.
And, we have seen, the pinning forces may be comparable
in magnitude to those in the inner crust of a conventional
neutron star.
It has recently been suggested (for reasons unrelated to our considerations)
that certain accreting compact stars
may be strange quark stars,\cite{Bombaci} although the
evidence is far from unambiguous.\cite{ChakrabartyPsaltis}
In contrast, 
it has been thought that, because they glitch,  
conventional radio pulsars cannot be strange
quark stars.  Our work questions this assertion
by raising the possibility that glitches
may originate within a layer of quark matter 
which is in a crystalline color superconducting state.


There has been much recent progress in
our understanding 
of how the presence of color superconducting quark matter
in a compact star would affect five different phenomena:
cooling by neutrino emission, the pattern of the arrival times of 
supernova neutrinos, the evolution of neutron
star magnetic fields, r-mode instabilities and glitches.
Nevertheless, much theoretical work remains to be done before
we can make sharp proposals for which 
astrophysical observations can teach us 
whether compact stars contain quark matter, and if so
whether it is in the 2SC or CFL phase.

\section*{Acknowledgments}
We are grateful to M. Alford, B. Berdnikov,
J. Berges, J. Bowers, J. Kundu,
T. Sch\"afer, E. Shuryak, E. Shuster and M. Stephanov for
enjoyable and fruitful collaboration. 
This work is supported in 
part  by the Department of Energy 
under cooperative research agreement \#DF-FC02-94ER40818.
The work of 
KR was supported in part by a DOE OJI Award and by the
A. P. Sloan Foundation. Preprint MIT-CTP-3049.

\end{document}